\documentclass[pdflatex, sn-nature]{sn-jnl}
\usepackage{graphicx}%
\usepackage{multirow}%
\usepackage{amsmath,amssymb,amsfonts}%
\usepackage{amsthm}%
\usepackage{mathrsfs}%
\usepackage[title]{appendix}%
\usepackage{xcolor}%
\usepackage{textcomp}%
\usepackage{manyfoot}%
\usepackage{float}%
\usepackage{placeins}%
\usepackage{bm} 
\usepackage{booktabs}%
\usepackage{algorithm}%
\usepackage{algorithmicx}%
\usepackage{algpseudocode}%
\usepackage{listings}%
\usepackage{amssymb}
\usepackage{caption}
\usepackage{tabularx}
\usepackage{makecell}
\usepackage{array}

\newcommand{\yes}{\checkmark}
\newcommand{\no}{--}
\newcommand{\limited}{Limited}
\newcommand{\RR}{\mathbb{R}}
\newcommand{\zhat}{\hat{z}}

\theoremstyle{thmstyleone}%
\theoremstyle{thmstyletwo}%

\theoremstyle{thmstylethree}%

\newcounter{offset}
\newcounter{tableoffset}
\newcommand{\beginsupplement}{%
  \setcounter{offset}{\value{figure}}%
  \renewcommand{\figurename}{Supplementary Fig.}%
  \renewcommand{\thefigure}{%
    \number\numexpr\value{figure}-\value{offset}\relax
  }%
  \setcounter{tableoffset}{\value{table}}%
  \renewcommand{\tablename}{Supplementary Table}%
  \renewcommand{\thetable}{%
    \number\numexpr\value{table}-\value{tableoffset}\relax
  }%
}

\begin{document}

\title[Article Title]{MIMIC-MJX: Neuromechanical Emulation of Animal Behavior}



\author[1]{\fnm{Charles Y.} \sur{Zhang}}
\equalcont{These authors contributed equally to this work.}
\author[2,3]{\fnm{Yuanjia} \sur{Yang}}
\equalcont{These authors contributed equally to this work.}
\author[4,5]{\fnm{Aidan} \sur{Sirbu}}
\equalcont{These authors contributed equally to this work.}

\author[6,7,8]{\fnm{Elliott T.T.}\sur{Abe}}
\author[1]{\fnm{Emil}\sur{W\"arnberg}}
\author[2]{\fnm{Eric J.}\sur{Leonardis}}
\author[1]{\fnm{Diego E.}\sur{Aldarondo}}
\author[1,2]{\fnm{Adam}\sur{Lee}}
\author[9,10]{\fnm{Aaditya}\sur{Prasad}}
\author[2]{\fnm{Jason}\sur{Foat}}
\author[2]{\fnm{Kaiwen}\sur{Bian}}
\author[2]{\fnm{Joshua}\sur{Park}}
\author[2]{\fnm{Rusham}\sur{Bhatt}}
\author[19]{\fnm{Vyom N.}\sur{Patel}}
\author[2]{\fnm{Hutton}\sur{Saunders}}
\author[2]{\fnm{Austin}\sur{Barbano}}

\author[11]{\fnm{Akira}\sur{Nagamori}}
\author[11]{\fnm{Ayesha R.}\sur{Thanawalla}}
\author[11]{\fnm{Kee Wui}\sur{Huang}}
\author[12]{\fnm{Fabian}\sur{Plum}}
\author[12]{\fnm{Hendrik K.}\sur{Beck}}

\author[9,10,13]{\fnm{Steven W.}\sur{Flavell}}
\author[12]{\fnm{David}\sur{Labonte}}
\author[4,5,14,15,16]{\fnm{Blake A.}\sur{Richards}}
\author[6,7,8]{\fnm{Bingni W.}\sur{Brunton}}
\author[11]{\fnm{Eiman} \sur{Azim}}


\author*[1,17,18]{\fnm{Bence P.} \sur{\"Olveczky}}\email{olveczky@fas.harvard.edu}
\author*[2]{\fnm{Talmo D.} \sur{Pereira}}\email{talmo@salk.edu}


\affil[1]{\orgdiv{Department of Organismic and Evolutionary Biology}, \orgname{Harvard University}, \orgaddress{\city{Cambridge}, \state{MA}, \country{USA}}}

\affil[2]{\orgdiv{Computational Neurobiology Laboratory}, \orgname{Salk Institute for Biological Studies}, \orgaddress{\city{La Jolla}, \state{CA}, \country{USA}}}

\affil[3]{\orgdiv{Neurosciences Graduate Program}, \orgname{University of California San Diego}, \orgaddress{\city{La Jolla}, \state{CA}, \country{USA}}}

\affil[4]{\orgname{Mila}, \orgaddress{\city{Montr\'eal}, \state{QC}, \country{Canada}}}

\affil[5]{\orgdiv{School of Computer Science}, \orgname{McGill University}, \orgaddress{\city{Montr\'eal}, \state{QC}, \country{Canada}}}

\affil[6]{\orgdiv{Biology Department}, \orgname{University of Washington}, \orgaddress{\city{Seattle}, \state{WA}, \country{USA}}}

\affil[7]{\orgdiv{eScience Institute}, \orgname{University of Washington}, \orgaddress{\city{Seattle}, \state{WA}, \country{USA}}}

\affil[8]{\orgdiv{Computational Neuroscience Center}, \orgname{University of Washington}, \orgaddress{\city{Seattle}, \state{WA}, \country{USA}}}

\affil[9]{\orgdiv{Department of Brain and Cognitive Sciences}, \orgname{Massachusetts Institute of Technology}, \orgaddress{\city{Cambridge}, \state{MA}, \country{USA}}}

\affil[10]{\orgdiv{Picower Institute for Learning and Memory}, \orgname{Massachusetts Institute of Technology}, \orgaddress{\city{Cambridge}, \state{MA}, \country{USA}}}

\affil[11]{\orgdiv{Molecular Neurobiology Laboratory}, \orgname{Salk Institute for Biological Studies}, \orgaddress{\city{La Jolla}, \state{CA}, \country{USA}}}

\affil[12]{\orgdiv{Department of Bioengineering}, \orgname{Imperial College London}, \orgaddress{\city{London}, \country{United Kingdom}}}

\affil[13]{\orgname{Howard Hughes Medical Institute}, \orgaddress{\city{Cambridge}, \state{MA}, \country{USA}}}

\affil[14]{\orgdiv{Department of Neurology and Neurosurgery}, \orgname{McGill University}, \orgaddress{\city{Montr\'eal}, \state{QC}, \country{Canada}}}

\affil[15]{\orgdiv{Learning in Machines and Brains Program}, \orgname{Canadian Institute for Advanced Research}, \orgaddress{\city{Toronto}, \state{ON}, \country{Canada}}}

\affil[16]{\orgdiv{Montr\'eal Neurological Institute}, \orgname{McGill University}, \orgaddress{\city{Montr\'eal}, \state{QC}, \country{Canada}}}

\affil[17]{\orgdiv{Center for Brain Science}, \orgname{Harvard University}, \orgaddress{\city{Cambridge}, \state{MA}, \country{USA}}}

\affil[18]{\orgdiv{Kempner Institute}, \orgname{Harvard University}, \orgaddress{\city{Cambridge}, \state{MA}, \country{USA}}}

\affil[19]{\orgname{Neuromatch}, \orgaddress{\city{Beaverton}, \state{OR}, \country{USA}}}

\abstract{The primary output of the nervous system is movement and behavior. While recent advances have democratized pose tracking during complex behavior, kinematic trajectories alone provide only indirect access to the underlying control processes. Here we present MIMIC-MJX, a framework for learning biomechanically grounded neural control policies from kinematics. MIMIC-MJX provides a platform for modeling the generative process of motor control by training neural controllers that learn to actuate biomechanical animal models in physics simulation to reproduce real kinematic trajectories. We demonstrate that our implementation is accurate, fast, and generalizable to diverse animal body models, and that it can be trained with modest amounts of motion data. MIMIC-MJX can be used to model motor control policies and simulate behavioral experiments, illustrating its potential as an integrative modeling framework for neuroscience.}

\maketitle
\section{Main}\label{sec1}


The nervous system evolved to control complex bodies in dynamic and uncertain environments. Studies on neural control of movement typically break the system down into manageable modules, whether functional and/or anatomic, and probe these in isolation \cite{Robinson1968-rl,Kristan2005-zj,Grillner2006-bc,Marder2007-td}. While such reductionist approaches have enabled our current understanding of complex sensorimotor control, holistic alternatives that embrace the complex interplay between the brain, body, and environment \cite{Dickinson2000-xc,Anderson2014-cq,Datta2019-qs,Pereira2020-zz} are ultimately required.

Successfully embracing this integrated view necessitates new tools for capturing, modeling, and simulating feedback interactions between neural and biomechanical systems in the context of natural behaviors. It is now possible to record the kinematics of natural behaviors in great detail using off-the-shelf cameras and easy-to-use software tools \cite{Pereira2019-np,Mathis2018-vm,Dunn2021-zc,Karashchuk2021-ys,Pereira2022-mo,Monsees2022-nh,Biderman2024-hp}. This progress enables routine measurements of detailed postural dynamics, but kinematic descriptions by themselves do not reflect the output of the brain's control system \cite{Aldarondo2024-nc}. Rather, the nervous system acts through muscles that drive the biomechanical body in rapid feedback with real-world physics, so that the body's passive dynamics directly shape the control problem the brain must solve.

For this reason, probing the neural control of movement and behavior will require models that capture how the brain controls the biomechanical body. Artificial neural networks (ANNs) offer a path toward this goal, providing tractable and expressive substrates for modeling nervous systems \textit{in silico}. Increasingly, ANNs are used as a computational language for expressing and testing theories of brain function \cite{Doerig2023-lw, Zador2023-oj, Richards2019-om}. Fully leveraging these models as accounts of neural systems, however, will require multiscale alignment to identifiable and measurable components of real circuits. For example, for organisms with mapped connectomes, such as \textit{C. elegans} \cite{Cook2019-hl} and \textit{D. melanogaster} \cite{Dorkenwald2024-ju}, whole-brain wiring diagrams can constrain network architectures at the circuit level. Just as connectomes anchor these models at the circuit level, aligning them at the level of behavior requires grounding them in the physical problem the nervous system actually solves: the biomechanical control of the body.

Connecting such models of neural control to recorded kinematic trajectories will require modeling how the biomechanics of the body, together with the physics of the environment, generate forces that result in coordinated movement. Recent work has pioneered the use of physics simulation to model the biomechanics of complex behavior in diverse animals, including in rats \cite{Merel2019-mh,Aldarondo2024-nc}, flies \cite{Lobato-Rios2022-om,Wang-Chen2024-uc,Melis2024-pn,Vaxenburg2025-mb}, worms \cite{Chen2023-zs, ModWorm-Kim2019-wr, BAAIWorm-Zhao2024-wi, NeuroSimWorm-Wang2025-yp, OpenWorm-Sarma2018-bk}, and mice \cite{Monsees2022-nh, DeWolf2024-xg, Ramalingasetty2021-rq}. Because joint torques and muscle activations are prohibitively difficult to directly measure in freely behaving animals, the control signals that actuate the body must be inferred. This inference, however, is ill-posed: many combinations of muscle activations and joint torques can produce the same kinematics, a redundancy known as Bernstein's degrees-of-freedom problem \cite{Bernstein1967-ez}. This poses a significant challenge for aligning biomechanical simulation with behavioral data.

A separate line of work in computer graphics \cite{Peng2018-di}, robotics \cite{Merel2018-my} and trajectory forecasting \cite{Eyjolfsdottir2016-ua} devised approaches that use deep reinforcement learning (DRL) to train neural controllers capable of producing motor commands that mimic reference kinematic trajectories, a task commonly referred to as ``motion imitation''. Building on this idea, a recent system called MIMIC \cite{Aldarondo2024-nc} demonstrated that motion imitation can be successfully applied to a rat body model. However, because MIMIC was not fully released as open-source software and sufficient physics simulation throughput requires access to high-performance computing clusters with substantial CPU capacity, the system's accessibility was substantially constrained. To address these deficits, we develop MIMIC-MJX, an open-source framework that builds on this motion imitation approach, aiming to democratize neuromechanical behavioral emulation from kinematics. MIMIC-MJX generalizes MIMIC's approach to body models of different species (including muscle-based actuation) and greatly simplifies its deployment by leveraging MuJoCo XLA (MJX) \cite{MuJoCo-XLA-AuthorsUnknown-il} for massively parallel physics simulation on a single GPU alongside ANN training. A comparison of MIMIC-MJX with representative musculoskeletal simulation, neuromechanical simulation, and robot-learning frameworks is provided in \hyperref[supptable:framework-comparison]{Supplementary Table~\ref*{supptable:framework-comparison}}.

MIMIC-MJX implements a two-stage pipeline in JAX \cite{Bradbury2018-ge} that takes 3D pose tracking data as input, aligns it to biomechanically realistic body models, and trains ANNs to control these bodies within physics simulation environments by learning to mimic the input kinematic trajectories.
We show that MIMIC-MJX is accurate, fast, and efficient, making it accessible to academic labs without the computing resources available to industry researchers. We demonstrate its generalizability by applying it to a diverse set of scenarios: a wide range of species, under freely moving and restrained experimental settings, and using both previously published and novel biomechanical body models (a rat, fruit fly, mouse arm, stick insect, and worm). We validate and demonstrate the utility of our trained controllers by recapitulating results on gait kinematics and behavioral representation learning, and by leveraging their reuse as low-level controllers in a new task environment to enable simulated experimentation. Finally, to facilitate its adoption and further development, we have made our framework available as open-source software, along with all datasets, trained models, and documentation on the project page at: \url{https://mimic-mjx.talmolab.org}.


\section{Results}\label{sec2}

\subsection{MIMIC-MJX is a pipeline for training neuromechanical control models from motion data.}\label{subsec1}

\begin{figure}[!htb]
  \centering
  \includegraphics[width=1\textwidth,height=0.8\textheight,keepaspectratio]{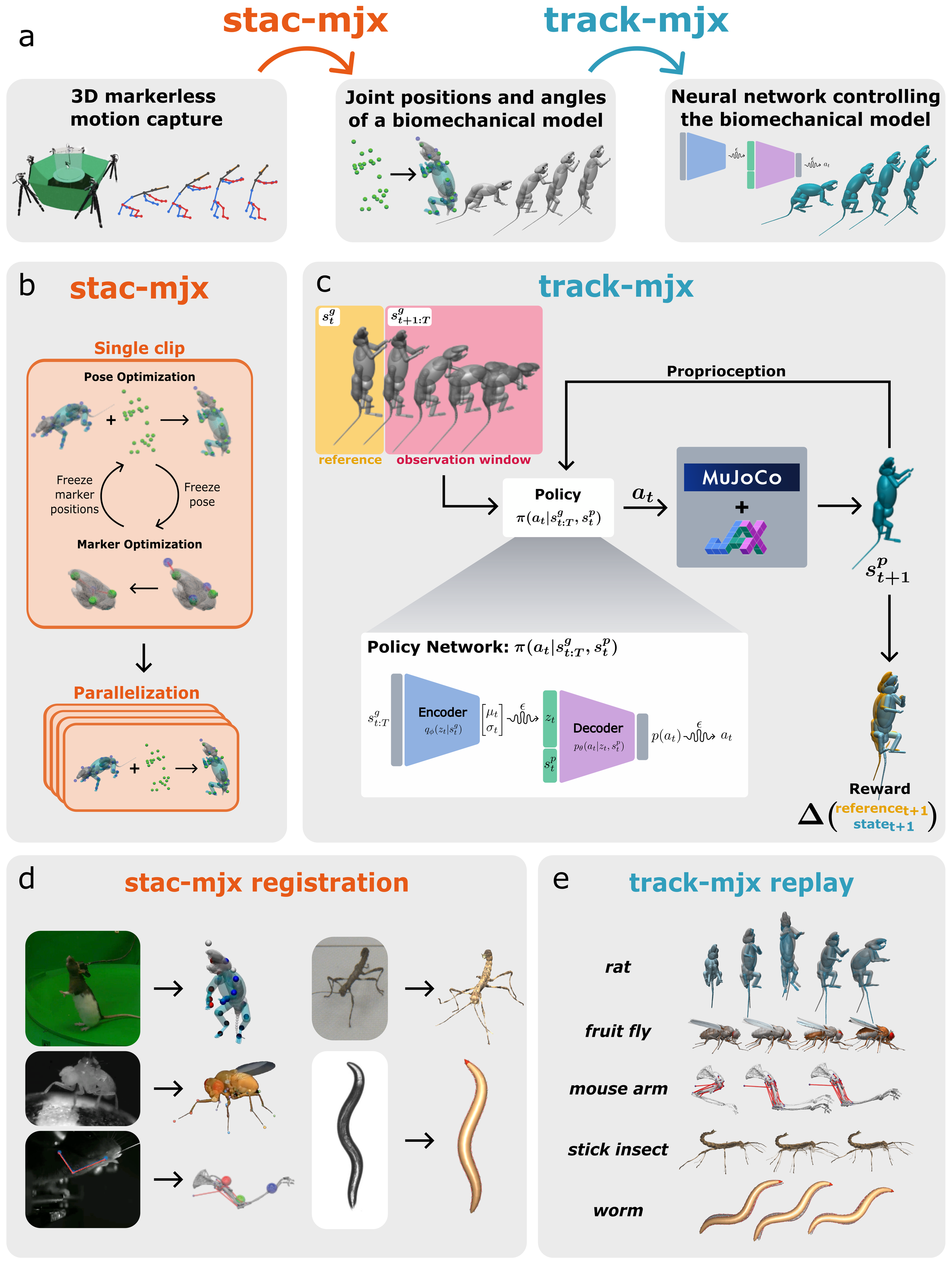}
  \caption{\textbf{MIMIC-MJX is a complete pipeline for training neuromechanical control models from pose tracking data.} \textbf{a}, Diagram of MIMIC-MJX at a high level \textbf{b}, Diagram of the \texttt{stac-mjx} module \textbf{c}, Diagram of the \texttt{track-mjx} training scheme \textbf{d}, \texttt{stac-mjx} registration is generalizable across arbitrary body models \textbf{e}, Rendered trajectories of \texttt{track-mjx} controlled agents (colored) and \texttt{stac-mjx} reference motion (gray)}
  \label{fig1}
\end{figure}

The input to MIMIC-MJX is the reference pose tracking data along with a MuJoCo-compatible biomechanical body model describing the animal's kinematic chain, actuators that can receive control signals to elicit movement, and optional sensors (e.g., touch) for richer sensory feedback. First, \texttt{stac-mjx} is used to estimate calibrated joint-angle trajectories from pose data via inverse kinematics. Then, \texttt{track-mjx} trains a neural controller that computes the motor control signals necessary to realize the tracked kinematic trajectories with the biomechanical body model (\hyperref[fig1]{Fig. 1a}).

\texttt{stac-mjx} implements a GPU-parallelized version of the Simultaneous Tracking and Calibration (STAC) algorithm for marker calibration and inverse kinematics \cite{Wu2013-ci}. It takes 3D pose tracking data as input and infers the joint angles for a given MuJoCo body model that best reproduce the keypoint trajectories while accounting for a fixed per-keypoint offset from a body part and geometric constraints imposed by the body model (\hyperref[fig1]{Fig. 1b}). Because pose tracking data are often biased towards tracking surface landmarks rather than the internal skeleton which defines the true articulation points that drive the biomechanical model, \texttt{stac-mjx} alternates between estimating marker offsets and joint angles to best fit the observed keypoints while accounting for a fixed, rigid displacement from the internal model landmarks that correspond to the video-derived keypoints. Our approach is robust, having been applied successfully to multiple species with datasets derived from different pose tracking pipelines.

To facilitate the application of this system to new body models and tracking data, we developed a graphical user interface (GUI) that enables interactive selection of correspondences between keypoints and their most appropriate counterpart in the body model, while also solving for an optimal geometric transformation (scaling, rotation, translation) to account for coordinate system differences in the data relative to the model (\hyperref[suppfig:stac-ui]{Supplementary Fig.~\ref*{suppfig:stac-ui}}).

\texttt{stac-mjx} speeds up the inverse kinematics process by solving many frames together on the GPU, leveraging MJX's vectorized forward kinematics across a batch of frames. It is capable of processing data at $\sim$400 frames per second on a single H100 GPU, enabling full inverse kinematics of an hour-long session of pose tracking in just 10 minutes (\hyperref[suppfig:stac-mjx-fps-scaling]{Supplementary Fig.~\ref*{suppfig:stac-mjx-fps-scaling}}). The output of \texttt{stac-mjx} is stored in a portable HDF5 file containing the solved body configuration for all frames and derived features, including Euclidean coordinates, quaternions defining orientation, joint angles, and velocities.

In the second stage, \texttt{track-mjx} receives the postural trajectories registered to the body model as input, and optimizes a neural controller to actuate the body model and ``track'' the movements of the input reference trajectories (\hyperref[fig1]{Fig. 1c}).

The neural controller is an ANN policy network trained via deep reinforcement learning (DRL) to maximize a composite reward function that rewards similarity between reference and reproduced trajectories and penalizes energy expenditure (see \hyperref[Methods]{Methods} for details). The ANN is structured as an encoder-decoder architecture. The encoder receives a window of future timesteps of poses in the reference trajectory relative to the agent's current pose, and compresses these into a low-dimensional Gaussian-regularized stochastic latent vector.

We interpret this latent vector as representing the ``motor intention" of the agent as it encodes a window of future target positions (the outcome that is intended) and gets decoded into the control command for the next timestep (the immediate action that realizes the intention). This motor intention vector is then concatenated with sensory inputs from the current state of the environment and passed to the decoder, which outputs the action for the next time step, driving the body to move in the direction of the reference motion trajectory. The action output parameterizes a probability distribution over actuator control signals specific to the body model and can represent any number of degrees-of-freedom (DOFs) or actuator types present in the body model (e.g., direct torque-control, position-control, or muscles). The modularity of the encoder-decoder architecture facilitates reuse of the naturalistic motor control capabilities learned from motion capture imitation for new downstream tasks, such as locomotion at a target velocity and escape from bowl-shaped terrain (\hyperref[fig5]{Fig. 5}). By keeping the decoder frozen, new task policies can learn to solve the task by producing motor intention vectors to steer the decoder, effectively decoupling low-level motor control from high-level perception and decision-making.

To train the ANN policy, \texttt{track-mjx} uses the Proximal Policy Optimization (PPO) learning algorithm. This is motivated by its reputation as an effective and widely used algorithm for continuous control, especially when paired with massively parallel GPU physics simulation \cite{Schulman2017-rt,Rudin2021-ma}. PPO optimizes the ANN parameters through batches of ``rollouts''---consecutive iterations of the agent-environment interaction loop in which the policy attempts to track a given reference trajectory. These batches are collected via parallel simulations and used to compute synchronous gradient updates.

\texttt{track-mjx} implements this approach to learning a neural controller from reference trajectories as a general strategy that can be used for any given MuJoCo-compatible body model and corresponding reference motion trajectories (\hyperref[fig1]{Fig. 1d,e}). It uses a composable configuration system through Hydra \cite{Yadan2019Hydra} for experimentation and supports extensive logging of the optimization trajectory through Weights \& Biases \cite{Biewald2020-mg}, enabling granular visibility into individual reward terms, system performance, rollout visualizations (\hyperref[fig1]{Fig. 1e}) on held-out data, and other metrics that facilitate troubleshooting and provide crucial insights when applying this system to new data.

Together, this system unifies pose registration, GPU-accelerated physics simulation, and DRL training, lowering the technical barriers to neuromechanical modeling.

\subsection{MIMIC-MJX accurately registers and reproduces keypoint sequences from real data.}\label{subsec2}

\begin{figure}[htbp]
  \centering
  \includegraphics[width=0.98\textwidth,height=0.75\textheight,keepaspectratio]{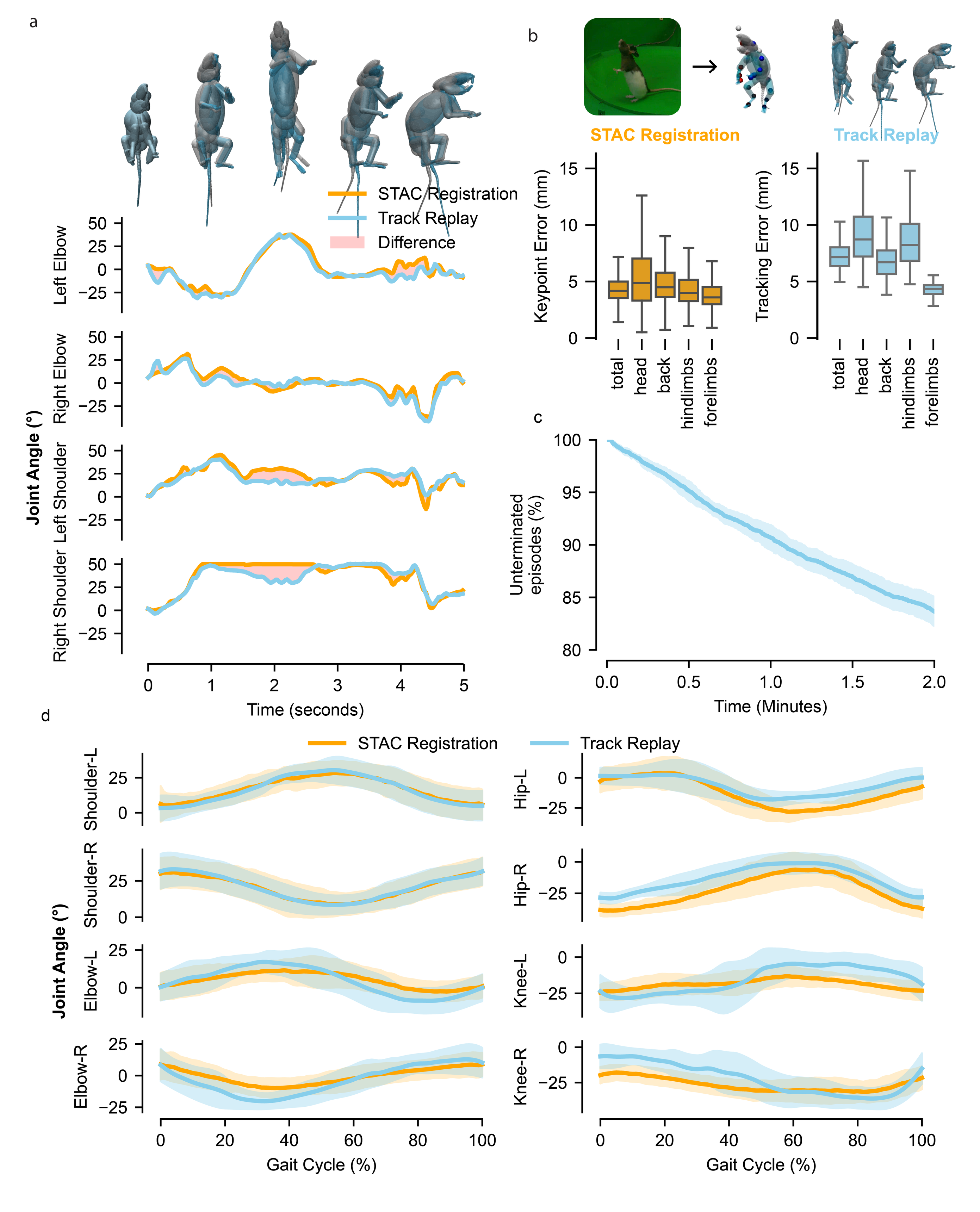}
  \caption{\textbf{MIMIC-MJX accurately registers and reproduces keypoint sequences from real data.} \textbf{a}, An example of a rollout of the rat imitating a rearing motion. Top: representative rendered frames of the biomechanical model during a roll-out showing the reference (gray) and reproduced (blue) trajectories. Bottom: joint-angle traces for the left and right shoulders and elbows over time; \texttt{stac-mjx}-registered reference (orange) and \texttt{track-mjx}-reproduced (blue) trajectories are overlaid, with their difference highlighted in the shaded region (pink).
    \textbf{b}, Keypoint error distributions between \texttt{stac-mjx}-registered reference and 3D pose tracking landmarks (orange) and between \texttt{track-mjx}-reproduced trajectories and \texttt{stac-mjx}-registered reference (blue). Each distribution represents per-frame keypoint errors of markers within each body region (forelimbs, hindlimbs, back, head) or over all 23 markers (total).
    \textbf{c}, Evaluation of continuous tracking performance, measured as the percentage of episodes that remain unterminated as a function of elapsed time (minutes). The solid line shows the mean and the shaded band $\pm 1$ standard error of the mean across 22 held-out sessions from 7 rats.
    \textbf{d}, Phase-normalized gait kinematics across multiple locomotion cycles. Joint-angle trajectories (degrees) for left/right shoulders and elbows (left panel) and hips and knees (right panel) over the gait cycle (0–100\%); \texttt{stac-mjx}-registered (orange) and \texttt{track-mjx}-reproduced (blue). Solid lines show the mean joint angle and shaded bands $\pm 1$ standard deviation across gait cycles (n = 18 bouts).
  }
  \label{fig2}
\end{figure}
We first applied MIMIC-MJX to train a neural controller for a virtual rat model, using a dataset of diverse freely-moving rat behaviors as reference trajectories. Behaviors were classified as walking, rearing, and grooming, and segmented into 5-second clips \cite{Aldarondo2024-nc}. ANNs trained with \texttt{track-mjx} reproduced trajectories accurately, closely following the observed reference sequences registered by \texttt{stac-mjx} as measured by the alignment of joint angles between reference and generated trajectories (\hyperref[fig2]{Fig. 2a}).

Evaluated over a set of held-out clips, we found that \texttt{stac-mjx} and \texttt{track-mjx} achieved low registration and tracking error, respectively. Performance was quantified as the absolute Euclidean distance between predicted and target keypoints. \texttt{stac-mjx} showed low median registration error ($\sim$5 mm) relative to the pose tracking keypoint trajectories, while \texttt{track-mjx} likewise demonstrated low median tracking error ($\sim$7 mm) when evaluated against the \texttt{stac-mjx}-registered keypoints, with errors grouped by body region (\hyperref[fig2]{Fig. 2b}); both medians were computed over a held-out dataset. This corresponds to less than $\sim$5\% of the adult rat body length, consistent with previous work \cite{Aldarondo2024-nc}, and reflects markerless keypoint localization error baselines \cite{Dunn2021-zc,Ulutas2025-ig}.

We next evaluated \texttt{track-mjx}-trained controllers on their ability to track motion sequences at longer timescales than they were trained on. Tracking errors tended to slowly accumulate when the controller failed to completely recover from mistakes in the control sequence; at a threshold deviation distance from the reference pose (following the parameterization in Aldarondo et al. \cite{Aldarondo2024-nc}), we terminated the episode and reset the environment by realigning the pose to the reference at that timestep. We tested our rat controller on full 3-hour sessions of pose-tracked data (22 sessions from 7 rats) and measured, over time, the fraction of episodes which were not yet terminated (``unterminated episode percentage''), yielding the survival curve. $\sim$83\% of episodes persisted after 2 minutes, far beyond the training clip length of 5 seconds (\hyperref[fig2]{Fig. 2c}). This reflects our system's ability to generalize to longer timescales and to faithfully simulate continuous control at the timescale of complex natural behavior sequences, a substantial improvement over previous approaches (\hyperref[suppfig:mimic-comparison]{Supplementary Fig.~\ref*{suppfig:mimic-comparison}}). Not all episodes remained stable over the full evaluation window, indicating that stable infinite-horizon imitation remains an open direction for future improvement.

Finally, we assessed whether these error rates had an impact on the kinematic features used to characterize motor behaviors. To test this, we turned to locomotion, whose stereotyped, cyclic structure allows movement to be aligned to a standard gait cycle for direct comparison. After identifying and segmenting bouts of locomotion, we phase-aligned both the registered and generated movement sequences to a standard gait cycle. We found that MIMIC-MJX accurately reproduces the postural dynamics across the entire gait cycle, an indication that it performs well across both short timescales as well as repeated instantiations of periodic motor behaviors (\hyperref[fig2]{Fig. 2d}).

Together, these results reflect MIMIC-MJX's ability to accurately reconstruct biomechanically plausible poses through motor actuation while maintaining the necessary temporal structure and joint-level dynamics for precise kinematic reproduction.

\subsection{MIMIC-MJX generalizes to diverse animal models and experimental settings.}\label{subsec3}

\begin{figure}[htbp]
  \centering
  \includegraphics[width=0.9\textwidth]{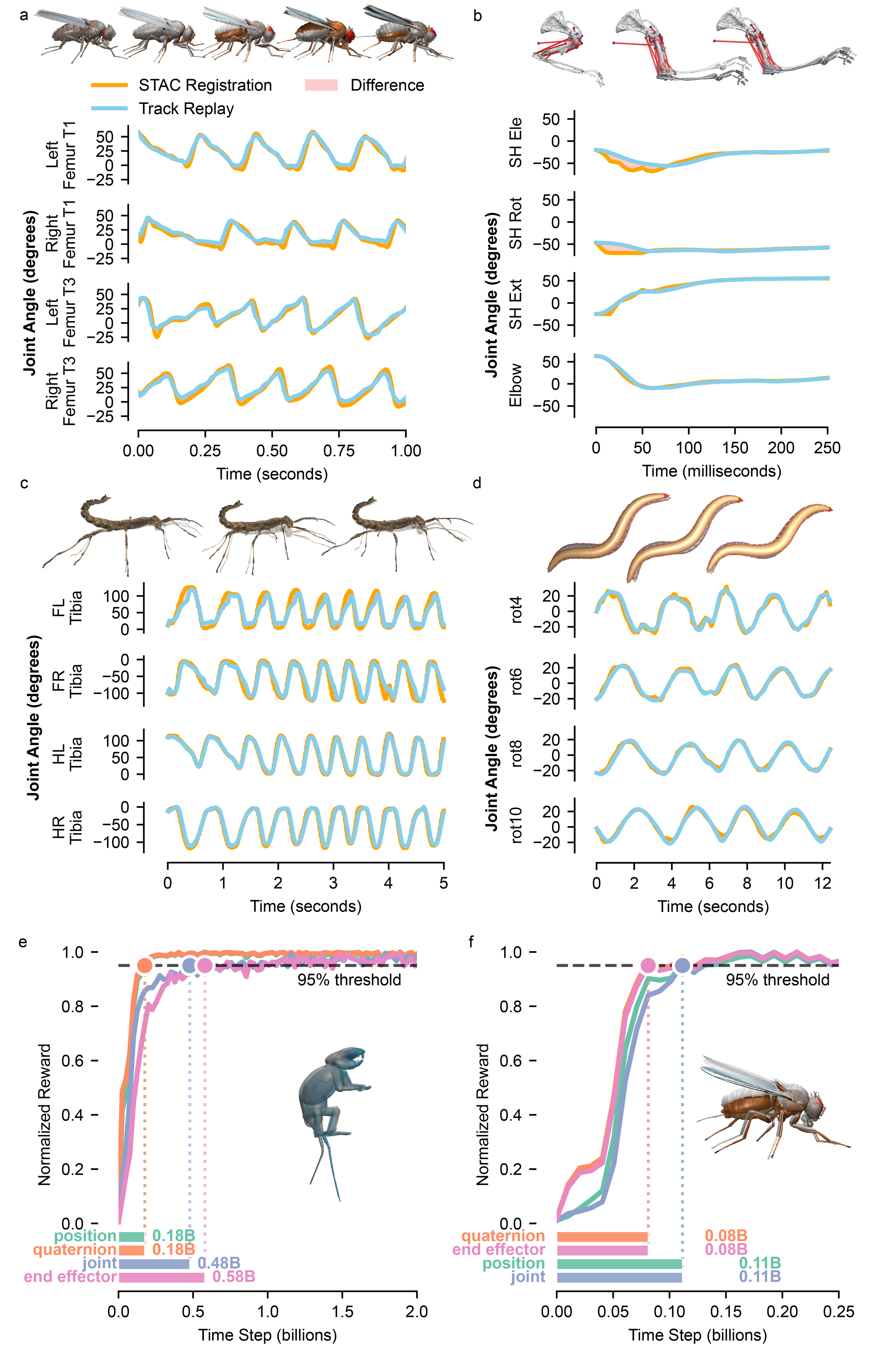}
  \caption{\textbf{MIMIC-MJX generalizes to diverse animal models and experimental settings. }\textbf{a--d}, STAC registration and Track replay with pointwise differences highlighted from rollouts of a fly (\textbf{a}), mouse arm (\textbf{b}), stick insect (\textbf{c}), and worm (\textbf{d}). \textbf{e, f}, Individual reward components of the rat (\textbf{e}) and fly model (\textbf{f}). The dashed line marks a 95\% threshold of each term's maximum observed value. The number of time steps required to reach the 95\% threshold is shown.}
  \label{fig3}
\end{figure}

We designed MIMIC-MJX to accommodate diverse animal body models. To demonstrate this, in addition to our primary focus on the rat model, we successfully applied MIMIC-MJX to biomechanical models and pose tracking data from a fruit fly, a mouse arm, a stick insect, and a worm (\hyperref[fig3]{Fig. 3a-d}). These four models were either already publicly available (fruit fly and worm) or newly developed through collaborations (mouse arm and stick insect). The actuator type (torque vs. muscle) varies between models, with muscle-actuated models being more costly to build and thus less common. Adding muscle-based actuation requires localizing insertion points and tuning force-production parameters that are difficult to measure experimentally---often fit per muscle by matching force-velocity curves through grid searches \cite{Wang-Chen2024-uc, caillet2025hill}. As such, the rat, fruit fly, and stick insect models are torque-actuated, while the mouse arm and worm are muscle-actuated. By applying MIMIC-MJX to both, we illustrate the generality of our training setup to different types of body models and its forward-compatibility as more detailed biomechanical models are developed.

Fruit fly, stick insect, and worm controllers were trained to track locomotor behavior, while the mouse arm controller tracked reaching trajectories. We demonstrate performance through the alignment of \texttt{track-mjx}-controlled joint sequences with \texttt{stac-mjx}-produced reference sequences (\hyperref[fig3]{Fig. 3a-d}). Configuration details for each implemented model are provided in the supplement (\hyperref[supptable:imitation-training-config]{Supplementary Table~\ref*{supptable:imitation-training-config}}).

Because our reinforcement learning reward is made up of multiple imitation accuracy reward components (i.e., alignment of root position, root quaternions, joint angles, and end-effector positions to the reference pose), we studied each component's learning trajectory during training for multiple body models. Root position and quaternion rewards can be thought of as ``coarse'' terms (global root alignment), and joint and end-effector rewards as ``fine'' terms (pose alignment). In the rat, composing these dense reward signals produces a consistent coarse-to-fine curriculum: the ANN first stabilizes global position and orientation, then refines the many degrees-of-freedom needed to match joint angles and end-effector positions—a more constrained, higher-dimensional objective that requires additional experience (\hyperref[fig3]{Fig. 3e}). In the fruit fly, whose behavioral repertoire here is limited to walking, all reward components converge rapidly and almost simultaneously, and this staging is largely absent (\hyperref[fig3]{Fig. 3f}). This suggests that the coarse-to-fine curriculum emerges as a function of task complexity, becoming pronounced when the target behavior is high-dimensional and diverse. We observed degraded overall tracking performance when removing tracking reward components (\hyperref[suppfig:reward-ablation]{Supplementary Fig.~\ref*{suppfig:reward-ablation}}), showing that each contributes to overall accuracy. Reward scale and weight sweeps measured sensitivity to those hand-tuned parameters, with perturbations at $\pm 1$ order of magnitude producing slightly larger held-out tracking errors for pose-alignment components, but retaining robust tracking overall (\hyperref[suppfig:reward-perturbation]{Supplementary Fig.~\ref*{suppfig:reward-perturbation}}). We also tested robustness to biomechanical mismatch by perturbing the mass of the rat's head during either training or inference and found that tracking error remained close to baseline across the tested range (\hyperref[suppfig:head-mass-sensitivity]{Supplementary Fig.~\ref*{suppfig:head-mass-sensitivity}}).

\subsection{MIMIC-MJX is fast and efficient.}\label{subsec4}

Making MIMIC-MJX an accessible framework for modeling the neuromechanical basis of behavior requires training times and data demands to be compatible with iterative experimentation. For example, adding more diverse behavioral data, varying biomechanical or neural controller configurations, or extending it to a new species all require training the system from scratch. In previous work \cite{Aldarondo2024-nc}, this form of experimentation was limited by the speed of the physics simulation, a CPU-bound component of the pipeline that does not scale well with the computational resources available to academic researchers. Using JAX-based deep reinforcement learning and physics simulation, we bypassed this limitation, as demonstrated by its performance profile across different animal models and datasets.

We trained neural controllers for the same five biomechanical models (mouse arm, fruit fly, stick insect, rat, and worm) using their corresponding pose tracking datasets. To evaluate training efficiency, we quantified the number of environment timesteps (one agent–environment interaction each) needed for a policy to attain near–asymptotic performance.

The DRL training runtime is primarily composed of two components: experience generation through the agent-environment loop and the actor/critic network gradient updates. For our physics-based environments, the dominant component by far is the experience generation. Therefore, we quantified training speed in terms of environment step throughput, measuring the number of steps required to reach \(95\%\) of the normalized test–set reward.

\begin{figure}[H]
  \centering
  \includegraphics[width=1.0\textwidth]{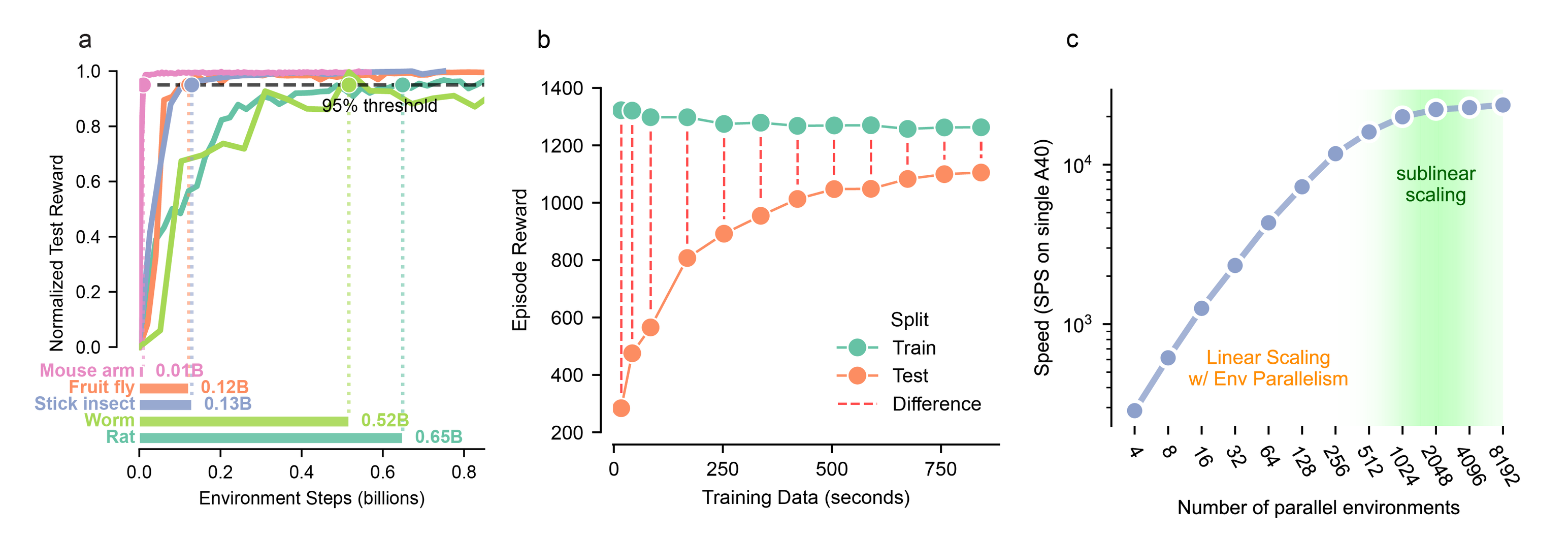}
  \caption{\textbf{MIMIC-MJX is fast and efficient.} \textbf{a}, Normalized test set reward versus environment steps for five biomechanical models (mouse arm, fly, stick insect, rat, and worm). Rewards are normalized to each model's empirical maximum (unit peak = 1). The horizontal dashed line marks the \(95\%\) threshold; colored markers and matching vertical dotted lines indicate the first crossing for each model. The bars below report the number of environment steps (in billions) required to reach the \(95\%\) threshold.
    \textbf{b}, Mean episode reward as a function of training data size. Green: training split. Orange: held-out test split of 673 five-second clips. At each \(x\) (number of 5~s clips), the paired markers come from a single training run; red dashed connectors visualize the train–test gap. The \(y\)-axis reports the average reward over evaluation rollouts (the state trajectory produced during the environment-agent loop).
  \textbf{c}, Environment throughput on a single NVIDIA A40 GPU: steps per second (SPS) versus the number of parallel environments (log scale). }
  \label{fig4}
\end{figure}

All agents reached the threshold in fewer than one billion steps: mouse arm $\sim$0.01\text{B}, fruit fly $\sim$0.12\text{B}, stick insect $\sim$0.13\text{B}, worm $\sim$0.52\text{B}, and rat $\sim$0.65\text{B} (\hyperref[fig4]{Fig. 4a}). These results show that convergence time scales with biomechanical and behavioral complexity. The mouse arm, performing simple reaching trajectories with only only four degrees of freedom and nine muscle actuators, trained very quickly---converging in tens of millions of steps. The rat took the longest to converge, consistent with it being trained on a richer behavioral repertoire compared to the more constrained reference behavior  the other animals. The worm was also comparatively slow to train despite its reference data consisting of only undulatory locomotion in both directions; this could be due to its large number of muscle-driven actuators (95), making the sensorimotor controller harder to learn \cite{Cotton2025-ni}.

To investigate generalization performance with respect to dataset size, we trained multiple controllers with varying amounts of training data. Beginning with $15$ seconds of data, we trained on increasing durations of natural behavior until the training dataset totaled $845$ seconds, or about $14$ minutes. We report mean episode reward on both the training and test splits for each of these runs (\hyperref[fig4]{Fig. 4b}). Test rewards increased monotonically with data, and the train-test gap decreased steadily (red segments), indicating task generalization rather than memorization. Notably, most of the generalization gains were achieved within $\sim$50 clips, and by $100$ clips the test curve approached the training curve. While the full training dataset contains 842 clips of similar quality, this experiment shows that a relatively small amount of data---on the order of a few minutes of behavior---is sufficient to achieve the generalization needed to reproduce unseen trajectories without retraining on them, a modest requirement that is practical to collect with modern pose-tracking pipelines.

We benchmarked environment throughput on a single A40 GPU as a function of the number of parallel environments (\hyperref[fig4]{Fig. 4c}). Steps per second (SPS, the number of total environment steps completed per second), scaled linearly until GPU compute utilization became saturated.
We noticed that for our environments, GPU memory (which constrains the total possible number of environments) is not a bottleneck for maximizing throughput on most GPUs. Detailed per-step wall-clock breakdowns are provided for the rat (\hyperref[tab:steptime-rodent]{Supplementary Table~\ref*{tab:steptime-rodent}}) and fruit fly (\hyperref[supptable:fruitfly-step-time]{Supplementary Table~\ref*{supptable:fruitfly-step-time}}) models, including a head-to-head comparison against a CPU-based fly imitation learning framework \cite{Vaxenburg2025-mb} (\hyperref[supptable:flybody-head-to-head]{Supplementary Table~\ref*{supptable:flybody-head-to-head}}). We also evaluated end-to-end throughput across different GPU models, and compared them with a previous CPU-based imitation learning pipeline for the fruitfly \cite{Vaxenburg2025-mb}, finding that MIMIC-MJX is up to 40x faster in terms of throughput (\hyperref[suppfig:hardware-throughput]{Supplementary Fig. 7a,b}). Because the previous work's approach used a different RL algorithm, we also show that MIMIC-MJX is faster in terms of total training time (time to convergence) when learning to imitate the same fly reference motion (\hyperref[suppfig:hardware-throughput]{Supplementary Fig. 7c}).
The combination of high-throughput, single-device scaling through parallelism, and portability across widely available hardware lowers the barrier to entry.

\subsection{MIMIC-MJX enables experiment simulation and neuromechanical behavioral analysis.}\label{subsec5}

\begin{figure}[!htbp]
  \centering
  \includegraphics[width=0.95\textwidth,height=0.72\textheight,keepaspectratio]{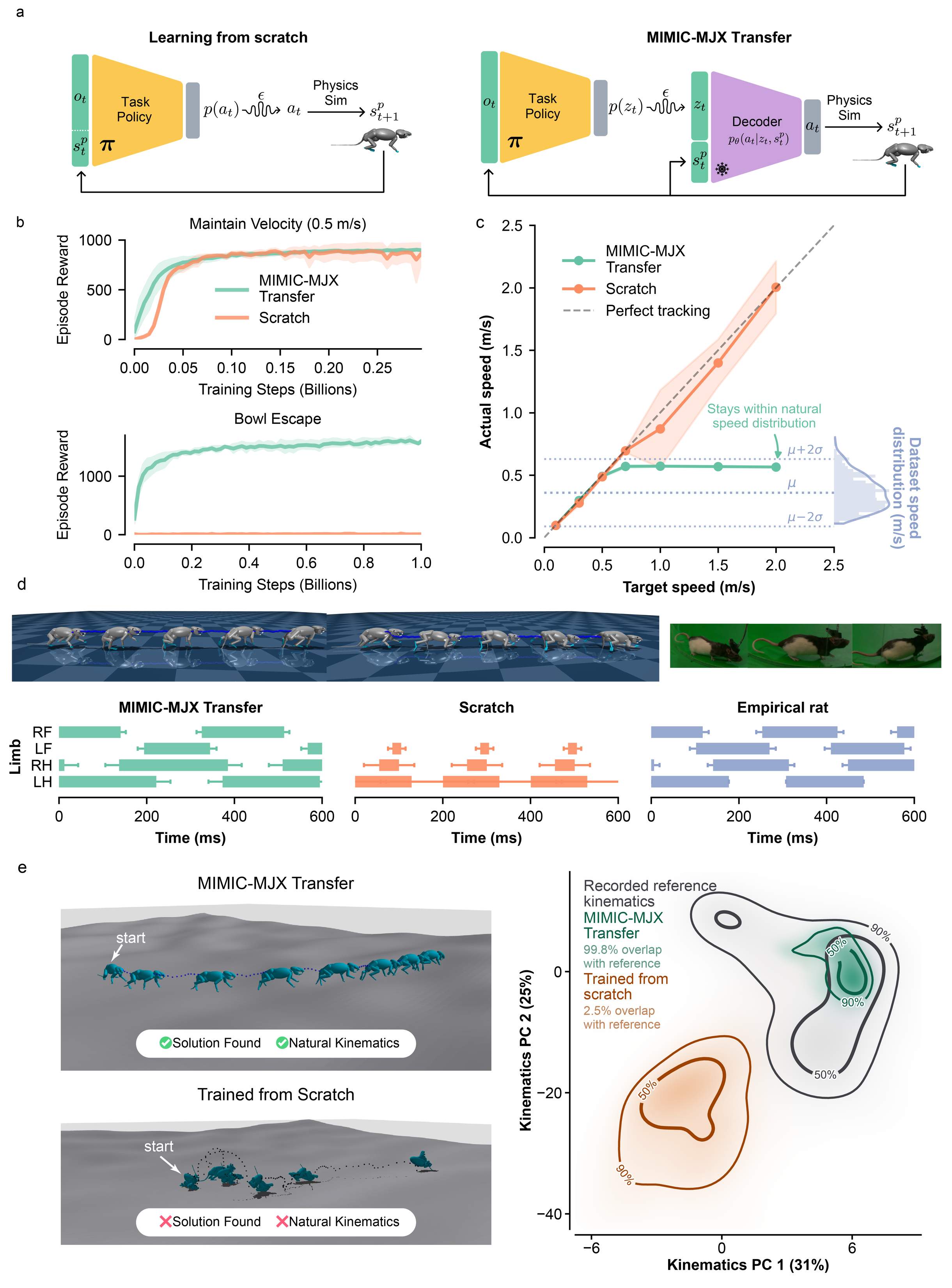}
  \caption{\textbf{MIMIC-MJX enables experiment simulation and neuromechanical behavioral analysis.} \textbf{a}, Diagrams of learning a downstream control task from scratch (left) versus MIMIC-MJX transfer (right). \textbf{b}, Episode reward during downstream training for Bowl Escape (top) and Maintain Velocity at 0.5\,m/s (bottom), comparing MIMIC-MJX transfer with learning from scratch; solid lines show the mean and shaded bands $\pm 1$ standard deviation across 5 random seeds. \textbf{c}, Actual locomotor speed as a function of target speed for transfer and scratch policies in the maintain velocity task; solid lines show the mean and shaded bands $\pm 1$ standard deviation across evaluation rollouts at each target speed. The dashed line indicates perfect target-speed tracking, and the marginal distribution shows speeds represented in the training dataset, with the dotted lines representing the mean speed in the training data and $\pm 2$ standard deviations from the mean. \textbf{d}, Representative trajectories from the maintain velocity task from the MIMIC-MJX transfer training and from-scratch training compared with rat video frames from the reference data. Footfall timing rasters below show limb-contact patterns for each condition. The empirical rat footfall timing raster was reproduced with permission from \cite{danner_spinal_2023}. \textbf{e}, Bowl escape rollouts and kinematic occupancy relative to recorded references. Left, pose montages of representative escape episodes for the MIMIC-MJX transfer policy (top) and a from-scratch policy (bottom). Arrow: start pose; dotted line: torso path. Right, kernel-density occupancy of per-frame pose features (91 pairwise landmark distances) in the leading two principal components, for reference kinematics (gray), transfer (green) and from-scratch (orange); contours enclose 50\% and 90\% of probability mass, pooled over 64 rollouts per policy. Percent overlap represents the fraction of frames inside the reference k-nearest-neighbor manifold (k = 5).}
  \label{fig5}
\end{figure}

A trained motion imitation policy yields reusable modules that encode a naturalistic behavioral prior. Transferring these modules lets a new policy learn downstream tasks while producing naturalistic motion. A randomly initialized network, by contrast, may still solve some of these tasks, but does so with unnatural motion, and fails on harder ones altogether. We leveraged the motor intention space and decoder module by making a trained decoder module part of the RL environment, having it act as a low-level controller into which a newly initialized task-specific high-level policy network inputs motor intentions (\hyperref[fig5]{Fig. 5a}). With this ``MIMIC-MJX transfer" approach, the new high-level policy's behavior was constrained to what is expressible by the decoder through the motor intention space, based on what was learned in the motion imitation task. Since the original encoder-decoder was trained with stochastic regularization, the motor intention space was easy for the task-specific policy to explore, allowing effective control in novel tasks and environments. This regularization includes injecting noise into the decoder's proprioceptive input during imitation learning, which encourages the decoder to rely more on the motor intention input rather than autocorrelation of the proprioceptive feedback signal. This strategy also has the consequence of inducing a more expressive and steerable intention space at the cost of allowing it to deviate more from the movement statistics in the reference data. A small amount of this noise in the low-level controller produced policies that were closely aligned with naturalistic motion when solving the new task, but expressive enough to generalize and solve difficult tasks (\hyperref[suppfig:maintain_vel_noise]{Supplementary Fig. 10a}).

We present the effectiveness of MIMIC-MJX transfer by comparing it against a randomly initialized policy trained from scratch on two tasks for the rat: a ``Maintain Velocity''  task and a ``Bowl Escape'' task \cite{Merel2019-mh}. In the Maintain Velocity task, the agent was placed in a flat environment and was rewarded for matching a given target velocity. In the Bowl Escape task, the agent had to navigate an environment of uneven terrain in the shape of a bowl, where the reward was based on moving away from the center of the bowl as fast as possible. These task definitions intentionally do not include additional hand-engineered reward terms or termination criteria that may improve learning or intended behavior---they are naive task setups that benefit from the pretrained low-level controller behavioral prior as an alternative to inductive biases imposed on the reinforcement learning task itself.

We found that the policy trained from scratch could learn the Maintain Velocity task but was unable to learn the Bowl Escape task, while the MIMIC-MJX transfer approach was able to solve both tasks (\hyperref[fig5]{Fig. 5b}).

We then investigated the range of locomotor velocities the two approaches could achieve by training policies on the Maintain Velocity task with varied target speeds. The policy trained from scratch was able to achieve target speeds ranging from 0.1 m/s to 2.0 m/s, but the MIMIC-MJX transfer could only achieve a maximum locomotor speed of $\sim$0.6 m/s. This speed upper bound is explained by the reference motion training data, where 0.6 m/s is approximately two standard deviations from the mean locomotor speed (\hyperref[fig5]{Fig. 5c}), reflecting the strong behavioral constraint that the low-level controller places on the high-level policy.
A closer look at the solution kinematics furthers this narrative; based on the gait footfall diagram, the MIMIC-MJX transfer produces a locomotion strategy strikingly similar to empirically gathered rodent footfall data \cite{danner_spinal_2023}, while the policy trained from scratch learned an unnatural, jittery movement pattern (\hyperref[fig5]{Fig. 5d}).
Consistent with this intuition, when high-level policies were trained on low-level controllers exposed to varying proprioceptive noise during imitation learning, higher noise levels yielded greater achieved speeds but less natural gaits (\hyperref[suppfig:maintain_vel_noise]{Supplementary Fig.~\ref*{suppfig:maintain_vel_noise}}).

In the Bowl Escape task, the policy trained from scratch completely failed to solve the task, producing neither goal achievement, as shown by task reward improvement, nor natural behavior, as shown by the kinematic trajectory's overlap with the reference kinematics (\hyperref[fig5]{Fig. 5e}). When we simplified the control problem by switching from torque to position actuators, the policy trained from scratch solved the task, yet still exhibited unnatural behavior (\hyperref[suppfig:bowl-position-actuators]{Supplementary Fig.~\ref*{suppfig:bowl-position-actuators}}). This demonstrates that pretraining on the imitation objective using \texttt{track-mjx} enables the encoding of strong priors for naturalistic behavior, which can subsequently be leveraged by higher-level controllers.

\begin{figure}[!htbp]
  \centering
  \includegraphics[width=0.95\textwidth,height=0.72\textheight,keepaspectratio]{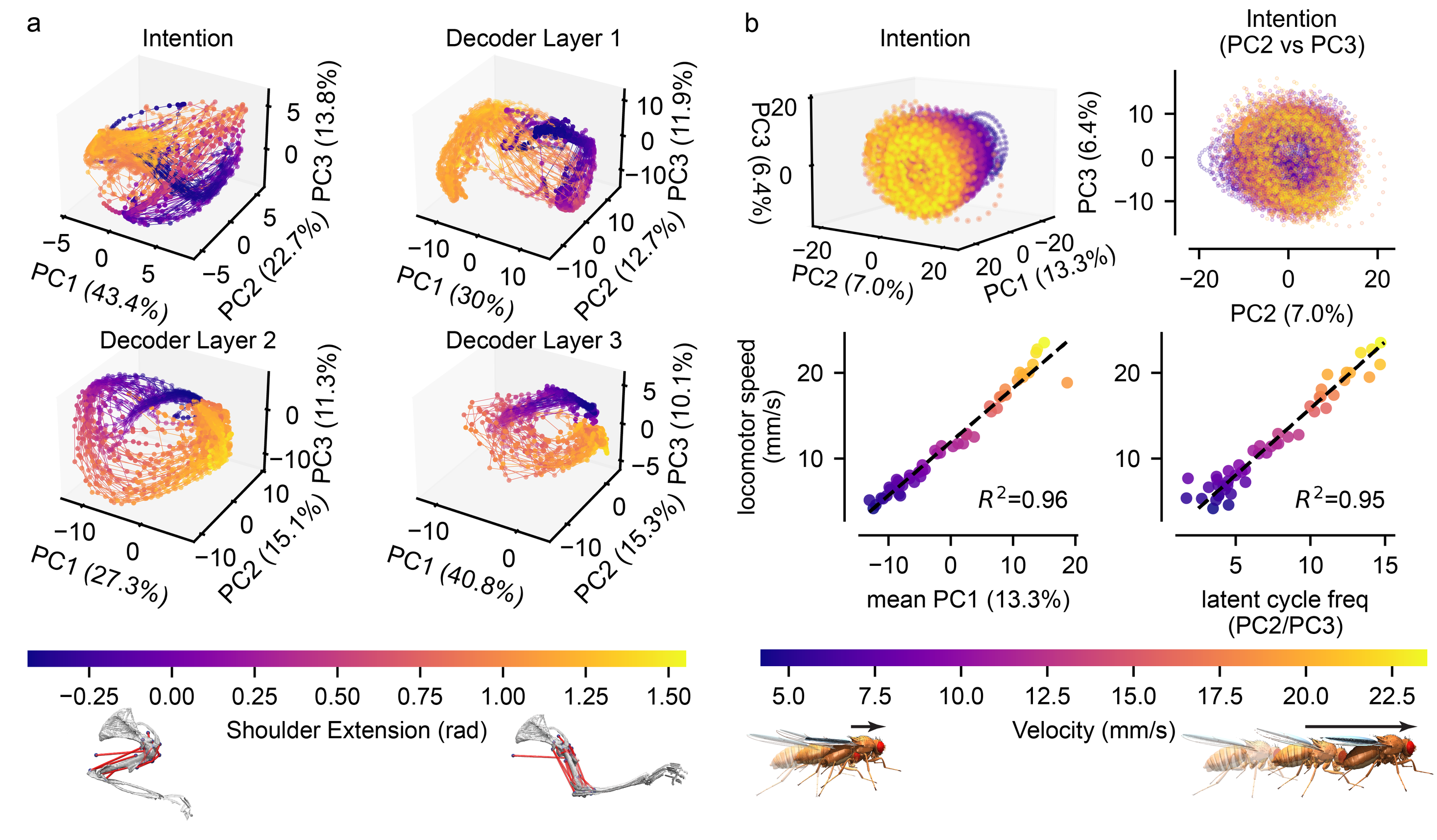}
  \caption{\textbf{MIMIC-MJX controllers enable neural-style analysis of latent control dynamics.} \textbf{a}, First three principal components (PCs) of the motor intention vectors and of successive decoder hidden-layer activations during mouse-arm reaching, colored by shoulder extension. The variance explained by each PC is indicated on the axes. \textbf{b}, PCA embeddings of the motor intention vectors (top left, top right). Fly locomotor speed as a function of the mean intention-latent PC1 (bottom left), and of the cycle frequency (bottom right). The coefficient of determination (\(R^2\)) of a linear fit is shown in each panel.}
  \label{fig6}
\end{figure}

To illustrate another way the framework can be used, we asked whether the network activity produced by MIMIC-MJX controllers can be interrogated with the analysis methods commonly applied to neural recordings, examining two representative cases. First, to see how the representation evolves as it is propagated through the decoder, we visualized the first 3 principal components (PCs) of the activations of each hidden layer for all reaching tasks of the mouse arm (\hyperref[fig6]{Fig. 6a}). The motor intention space encoded a compressed representation with 43.4\% variance along its first principal component (79.9\% across the first three PCs), indicating a structured low-dimensional space. In the decoder, it was relatively higher dimensional in the early layers (54.6\% explained variance in the first 3 PCs); however, by the final layer, it became lower-dimensional (66.2\% variance explained in the first 3 PCs), consistent with the idea that motor activity resides in a low-dimensional subspace, potentially reflecting ``motor synergies'' \cite{Shenoy2013-jo}. We saw that a topological structure emerged in the later hidden layers of the network, where a toroidal structure with a noticeable topological invariant represents the reaching cycle.

We also found that the learned motor intention space can encode low-dimensional kinematic features. We visualized 3- and 2-dimensional PCA embeddings of motor intention vectors during fly walking behavior, showing the low-dimensional, cyclic manifold of the gait cycle (\hyperref[fig6]{Fig. 6b top}). Similar to previous work \cite{DeAngelis2019-fa}, each point represents the PCA embedding of a 30-timestep window of motor intention vectors, representing 0.06 seconds. In addition to the salient cycles along PCs 2 and 3, this manifold reflects a linear representation of gait velocity along the orthogonal PC1 axis, as shown by the color gradient denoting the average velocity of each clip smoothly changing along this axis. Quantifying this structure, the mean position along intention-latent PC1 and the frequency of the PC2/PC3 gait cycle were both strong predictors of the fly's locomotor speed (\hyperref[fig6]{Fig. 6b bottom}).

Rather than providing a complete account of the geometric structure of the controller latent space, these examples illustrate that such analyses can be applied to a controller actively performing complex closed-loop feedback control of a biomechanical body---a promising direction for understanding the underlying computations and relating them to their biological counterparts \cite{Pandarinath2018-ua, Keshtkaran2022-en}.

\section{Discussion}

Here we present MIMIC-MJX, a fast and generalizable open-source platform for simulating the neural control of naturalistic movements across a range of behaviors and species. Leveraging GPU-parallelized physics simulation and deep reinforcement learning, MIMIC-MJX registers markerless motion-tracking data to biomechanical animal models, then trains ANN controllers to imitate the movements via biomechanical actuation. Low-level controllers trained in this way can serve as starting points for virtual animals facing novel tasks, thereby enabling us to probe questions of motor learning and hierarchical control. While prior studies have demonstrated the utility of this approach \cite{Aldarondo2024-nc, Vaxenburg2025-mb}, MIMIC-MJX democratizes it, making it broadly available to academic labs with limited compute resources or expertise in computer simulation.

Neuromechanical simulations, such as those enabled by our pipeline, provide a means to address a range of outstanding questions in motor neuroscience. Having a fully configurable, observable, and manipulable neural system controlling a complex body in a physics simulation allows researchers to probe the roles of neural mechanisms, circuit motifs, learning rules, and learning algorithms on different aspects of motor control. Access to essential system features that are hard or impossible to measure and manipulate in real animals, such as detailed sensory feedback and physical properties, motivates virtual experiments. For example, researchers can repeatedly manipulate the same network in precisely defined ways, ablate specific pathways or units, or retrain an identical network under systematically varied conditions.

The modularity, flexibility, and configurability of MIMIC-MJX enable models to match the granularity of the research question. Network architectures can have varying degrees of biological realism and complexity, from generic ANNs \cite{Merel2019-mh} to connectome-inspired ones \cite{Lappalainen2024-oa, Pugliese2025-qz}. Similarly, biomechanical models can be furnished with increasing detail of the musculoskeletal system \cite{Ramalingasetty2021-rq, Gilmer2025-jj, DeWolf2024-xg}, including muscles and their detailed properties \cite{caillet2025hill}.

While MIMIC-MJX facilitates neuromechanical simulations, its current implementation has limitations.

\textbf{Body models remain hard to build.} Constructing an accurate biomechanical body model remains a major undertaking. MIMIC-MJX provides a continuously expanding set of models (currently: rat, fly, worm, stick insect, detailed mouse arm), but researchers working on other organisms, or whose research questions demand greater biomechanical detail, must furnish their own. Thankfully, emerging tools help make this more tractable~\cite{Plum2021-qo, Bolanos2021-rr, Gilmer2025-jj}.

\textbf{The control solution is underdetermined.} The control signals we recover are not uniquely determined by the kinematics; many combinations of torques or muscle activations can produce the same movement (Bernstein's degrees-of-freedom problem \cite{Bernstein1967-ez}). We constrained the solution space using energy costs, the latent bottleneck, and the body model itself, but while the learned controllers are biomechanically valid solutions consistent with the observed kinematics, they do not necessarily reflect the animal's actual motor commands. Measurements such as EMG or ground-reaction forces would be needed to further constrain the solution space and could be integrated as imitation reward terms.

\textbf{Biological detail is traded for computational efficiency.} There is an inherent trade-off between computational efficiency and the biological detail of the model. We have employed models of varying degrees of simplification for muscles, tendons, and mechano- and tactile receptors. We have further limited the number of possible contacts to minimize the computational cost of each iteration of the physics simulator by computing contact forces only between extremities and the ground, impacting behaviors with self-contact like grooming. Though these are notable limitations, as the MuJoCo simulator becomes more efficient and capable, it will be straightforward to further extend MIMIC-MJX. For example, a new physics backend compatible with MJX is in development, offering both vastly improved computational efficiency for rich contact forces and batch-rendering of parallel simulation environments, necessary to endow the agents with vision. In the future, features supporting soft-body physics would make self-collisions more realistic and better capture the elasticity of the skeletons themselves. It may also be possible to integrate surrogate fluid dynamics models~\cite{Ohana2025-hi}, enabling the inclusion of aerodynamics, hydrodynamics, acoustics, and olfaction without making the simulations prohibitively expensive.

Because MIMIC-MJX is built on an active open-source simulation ecosystem, we expect many of these technical limitations to ease as that ecosystem matures, extending the platform's utility with new features.

We see this work as a critical step toward a paradigm in which in silico simulation becomes an indispensable counterpart to in vivo experimentation. By providing a framework for training biomechanically grounded control models, MIMIC-MJX makes the virtual animal a practical experimental subject---a system whose neural and physical variables are fully observable and manipulable. As the community further develops this “virtual neuroscience” platform, these virtual animals can become a common foundation for studying how nervous systems control the bodies they evolved to move.

\section{Methods}\label{Methods}
\subsection{Body Models}
We used biomechanically validated models of the organisms we implemented in MIMIC-MJX:
\begin{itemize}
  \item Rat (\textit{Rattus norvegicus}): We use a biomechanical rat model developed in previous work \cite{Merel2019-mh}, matching the structure of Long Evans rats. The model has 74 degrees of freedom (DoF), of which the Cartesian position of the model is represented by 3 degrees and the root orientation is represented by a 4-DoF quaternion. The remaining 67 DoF represent the joint angles relative to the parent's body frame in the kinematic tree. The model has 38 controllable torque actuators, and we use a subset of the included sensors, which comprises (1) a velocimeter, (2) an accelerometer, (3) a gyroscope, and (4) force, torque, and touch sensors on its end effectors. The end effectors are chosen to be the four paws and the skull.
  \item Fruit fly (\textit{Drosophila melanogaster}): We use a fruit fly model developed in previous work \cite{Vaxenburg2025-mb}. The model consists of 67 rigid body segments linked at 66 joints, yielding 102 degrees of freedom, and was reconstructed from high-resolution confocal microscopy of a female fly. Following that work, we remove the actuators for the wings, proboscis, and antennae when training for the walking imitation task, resulting in 61 torque actuators to drive the joints. The end effectors are chosen to be the pretarsal claws.
  \item Mouse arm (\textit{Mus musculus}): A skeletal model of the mouse forelimb and estimations of muscle attachment points were provided based on light sheet microscopy data \cite{Gilmer2025-jj}. The model has four DoF in total: three in the shoulder (elevation, rotation, and extension) and one in the elbow. Our simplified model has nine controllable Hill-type muscle actuators: triceps (long), triceps (lateral), biceps (long), brachialis, pectoralis (clavicular), latissimus, posterior deltoid, anterior deltoid, and medial deltoid. Muscle attachment points were refined, and muscle parameters were found to produce forces within the range of real mouse forelimb muscles (0.2-1.2 N) \cite{Ramalingasetty2021-rq}.
  \item Worm (\textit{Caenorhabditis elegans}): We use a body model developed in previous work \cite{Chen2023-zs}. The model consists of 25 body segments linked by 24 joints. Each joint is actuated by 4 body wall muscles, dorsal left/right and ventral left/right – except for the tail, which lacks its ventral right muscle – making 95 body wall muscles. The model is constrained to 2D motion, i.e., only displacement in the XY plane and rotation along the Z axis. We also add anisotropic friction contact between the body model and the floor, which has been shown to be used by the worm to propel itself in agar \cite{Chung2023-lg}. The end-effectors were chosen to be the head and tail.
  \item Stick insect (\textit{Sungaya aeta}): We used a model consisting of 43 rigid body segments linked at 42 joints (41 revolute joints and a 6-DoF floating base), giving 41 actuated degrees of freedom driven by 41 torque actuators. The model was constructed from a 3D scan obtained with the open photogrammetry platform \textit{scAnt} \cite{Plum2021-qo}. The 3D mesh was preprocessed in Blender (Blender Foundation, Amsterdam, Netherlands; version 3.0) \cite{blender} and converted into a URDF model using \textit{Phobos} \cite{von2020phobos}; the URDF was then converted to XML and extended to an MJCF file using custom scripts. Each segment's mass and inertia are derived from its textured triangular mesh under a uniform density (with hand-tuned inertia on the thin distal leg segments), for a total body mass of $\approx$50\,mg. Contact is restricted to six floor--claw pairs, each a 0.1\,mm-radius sphere at a claw tip; all other geometry, including self-collision, is disabled.

\end{itemize}

For each of these models, we limit contacts between the model and the environment (a single ground plane) to only the end effectors to improve the throughput of MJX, which scales poorly with the number of contacts in the simulation. Because our imitation tasks only involve contact with the end effectors and implicitly constrain behavior to avoid self-collisions, these changes do not negatively impact the training results and are merely an optimization measure. In the case of the worm model, the whole body is in contact with the floor but still without self-collisions. The biomechanical body models and their associated simulation environments are maintained in a separate repository, \texttt{vnl-playground}.

\subsection{Kinematic Data}
Markerless pose estimation was performed on multicamera video data to provide the necessary pose tracking data for the subsequent fit of inverse kinematics. The data are publicly available from their respective sources:

\begin{itemize}
  \item Rat: We use a dataset collected in previous work \cite{Aldarondo2024-nc} comprised of 842 5-second clips of freely behaving kinematic data, represented by 23 keypoints captured at 50 Hz. It was collected using DANNCE v1.3 \cite{Dunn2021-zc} from multicamera video. It represents a wide range of behaviors, including various kinds of walking, rearing, and grooming.
  \item Fruit fly: We use a dataset collected in previous work \cite{Pratt2024-cd}, which uses a spherical treadmill to enable 3D kinematic tracking of the fruit fly at different locomotor speeds. This data was sampled at 300 Hz. The pose estimation was done with DeepLabCut (DLC) \cite{Mathis2018-vm} and Anipose \cite{Karashchuk2021-ys}. First DLC was used to compute the single-camera 2D keypoint trajectory, then Anipose was used to triangulate multiple camera views of 2D trajectories into a single 3D trajectory. To align the keypoint data (which uses arbitrary coordinate axes) to the body model using Cartesian coordinates, we first performed a Procrustes transformation to scale and rotate the average position of the keypoints in a single clip to the rest position of the fly body model. We then applied the transformation to all frames in the clip. Due to the data being collected on a spherical treadmill, the data was not exactly aligned to the ground plane. To address this, we then performed a second transform that aligned the tarsal-claw keypoints to the ground plane. Finally, we used the rotational velocity of the ball as a proxy for the fly's linear and angular velocity in the xy-plane. The data was then run through \texttt{stac-mjx}. Then, the result was linearly interpolated to be sampled at 500 Hz so that there are 10 physics steps per control step.
  \item Mouse arm: A three-camera dataset was labeled using the CVAT annotation tool to generate $\sim$10,000 labeled frames per camera. A calibration video was recorded by placing a ChArUco board in the view of all cameras. Calibration was performed using the SLEAP-Anipose package, which is a wrapper around Anipose to allow for easy integration of 2D pose data from SLEAP. SLEAP-Anipose uses OpenCV to perform iterative sparse-bundle adjustment. It locates the 2D points on the ChArUco board and then triangulates them into 3D. Then the reprojection error is calculated by projecting back into 2D to evaluate the 3D estimation. This reprojection error is then minimized iteratively using L-BFGS to improve the 3D estimates and provide reliable triangulation. This network was then run on a video consisting of 46 successful water reaching trials \cite{Thanawalla2025-zn}. Predictions of the pose of the right shoulder, elbow, and wrist were generated with a single animal pose estimation network using the U-Net architecture in SLEAP \cite{Pereira2022-mo}.
  \item Nematode worm: We utilized the full multiclip training dataset from \cite{Atanas2023} of wild-type worms spontaneously behaving on a plate with no food which includes locomotion, turns and reversals. We train on both forward and backward locomotion as well as minor turns, and do not use bouts of omega turns due to current segmentation pipelines having low accuracy on omega poses.
  \item Stick insect: We used a dataset collected on a flat and smooth treadmill that perpetually kept \textit{S. aeta} inside the imaging volume of a five camera array (ORX-10GS-51S5C-C, Teledyne, USA; Lens: 2/3" 8mm f/2.8 16 Megapixel Lens, Computar, Japan); active 2D motion compensation enabled the recording of untethered 3D kinematics at 75 Hz. Marker-less pose estimation of 50 keypoints was achieved across all camera views using a single U-net architecture in SLEAP \cite{Pereira2022-mo}. The network was pre-trained using synthetic data \cite{plum2023replicant}, and iteratively refined using 800 hand-annotated frames, selected from all five views. A calibration video was recorded by placing a checkerboard in view of all cameras. Camera calibration and triangulation were performed using Anipose \cite{Karashchuk2021-ys}.

    The extracted 3D coordinates were translated and rotated to align the XY plane of a body-fixed coordinate system with the treadmill plane. The 3D kinematic data was offset to account for the movement of the treadmill, estimated by tracking ArUco markers printed onto the belt surface using OpenCV, to obtain the absolute walking speed \cite{garrido2014automatic, bradski2000opencv}.

\end{itemize}

\subsection{Joint Marker Registration and Inverse Kinematics}

\newcommand{\norm}[1]{\left\lVert #1 \right\rVert}

\texttt{stac-mjx} adapts the STAC algorithm \cite{Wu2013-ci}, originally developed for robotic system identification, to estimate time-varying body configurations and time-invariant, body-local marker offsets by alternating pose and offset optimization. Let $v_t^m \in \mathbb{R}^3$ be the observed position of keypoint $m$ at frame $t$, and let $p_m(q_t,\theta_m)$ be its position predicted by MuJoCo forward kinematics from configuration $q_t$ and body-local offset $\theta_m$. The registration objective is
\begin{equation}
  \left(\hat{q}_{1:T}, \hat{\theta}\right) =  \arg \min_{q_{1:T}, \, \theta} \sum_{t=1}^{T}\sum_{m=1}^{M} \norm{p_m(q_t,\theta_m)-v_t^m}^2_2 ,
  \label{eq:stac_registration}
\end{equation}
where $q_t$ contains the joint coordinates and, where present, the root translation and orientation. Equation~\ref{eq:stac_registration} defines the maximum-likelihood estimator under independent, zero-mean, and isotropic Gaussian keypoint noise. The pose solve augments this term with joint-limit constraints and, when enabled, the temporal smoothness residual. \texttt{stac-mjx} implements this in JAX and MJX to solve many frames together on GPU.

For free-root models, we warm-start the root pose on a reference frame using the trunk keypoints. Holding the marker offsets fixed, we solve a constrained nonlinear least-squares problem over the pose trajectory, representing the free root on $\mathrm{SE}(3)$, the joint coordinates in Euclidean space, and joint limits as inequality constraints. Jointly solved frames may also be coupled by the temporal residual
\begin{equation}\label{eq:stac_temporal_residual}
  r^{\mathrm{smooth}}_t = w_{\mathrm{smooth}}
  \begin{bmatrix}
    \operatorname{Log}(G_{t-1}^{-1} G_t)^\vee \\
    h_t-h_{t-1}
  \end{bmatrix},
  \qquad t=2,\ldots,T ,
\end{equation}
where $G_t \in \mathrm{SE}(3)$ is the free-root pose and $h_t$ contains the non-root joint coordinates. Because the residual omits division by the inter-frame interval, it regularizes frame-to-frame configuration changes rather than physical velocity. JAX-LS \cite{Yi2024-jaxls} handles the joint-limit inequalities using an augmented-Lagrangian formulation and solves the resulting least-squares subproblems using Levenberg--Marquardt with conjugate-gradient linear solves. In Algorithm~\ref{alg:stac_mjx}, \(\textsc{PoseSolve}(V,\theta,q)\) denotes this constrained pose optimization, warm-started from the current $q$.

Holding the pose trajectory fixed, each predicted marker position is affine in its body-local offset, making offset calibration a separable quadratic problem with a closed-form solution. At alternating iteration $i$, we compute the update from a random subset $\mathcal{S}$ of frames with optional regularization of selected marker offsets:
\begin{equation}\label{eq:stac_offset_calibration}
  \theta^{(i+1)} = \arg \min _{\theta} \sum_{t \in \mathcal{S}} \sum_{m=1}^{M}
  \norm{p_m(q_t^{(i+1)},\theta_m) - v_t^m}_2^2 + \lambda \norm{ D(\theta-\theta^{(i)})}_2^2 ,
\end{equation}
where $D$ is a binary diagonal mask selecting the offset coordinates to regularize. Writing the predicted position as $p_m(q_t^{(i+1)},\theta_m) = b_{tm}^{(i+1)} + R_{tm}^{(i+1)}\theta_m$, with body position $b_{tm}^{(i+1)}$ and orientation $R_{tm}^{(i+1)}$ from forward kinematics, and using the orthonormality of $R_{tm}^{(i+1)}$, the minimizer is
\begin{equation}
  \theta_m^{(i+1)} = \left( \lvert\mathcal{S}\rvert I_3 + \lambda D_m^\mathsf{T}D_m \right)^{-1}
  \left[
    \sum_{t\in\mathcal{S}} (R_{tm}^{(i+1)})^\mathsf{T} (v_t^m-b_{tm}^{(i+1)}) + \lambda D_m^\mathsf{T}D_m\theta_m^{(i)}
  \right],
  \label{eq:stac_offset_closed_form}
\end{equation}
where $D_m$ is the $3 \times 3$ block of $D$ for marker $m$; being diagonal and binary, it decouples the update across coordinates. In Algorithm~\ref{alg:stac_mjx}, $\textsc{OffsetSolve}(V,q,\theta)$ denotes this closed-form update, and pose and offset solves alternate $K$ times before a final pose solve with the calibrated offsets.

\begin{algorithm}[!ht]
  \footnotesize
  \caption{\texttt{stac-mjx}: marker calibration and inverse kinematics}
  \label{alg:stac_mjx}
  \begin{algorithmic}[1]
    \Require MuJoCo model $\mathcal{M}$; calibration keypoints
    $V^{\mathrm{cal}}$; recording keypoints $V_{1:T}$
    \Require Initial marker offsets $\theta^{(0)}$; calibration iterations $K$
    \Ensure Calibrated offsets $\theta^{(K)}$ and registered poses $q_{1:T}$

    \State Initialize the MJX model at its default pose $q^{(0)}$
    \If{$\mathcal{M}$ has a free root}
    \State Warm-start the root of $q^{(0)}$ from one calibration frame using
    the trunk keypoints
    \EndIf
    \State Repeat $q^{(0)}$ across the calibration frames

    \Statex
    \Statex \textbf{Calibrate marker offsets}
    \For{$i=0,\ldots,K-1$}
    \State $q^{(i+1)} \gets
    \textsc{PoseSolve}\!\left(
      V^{\mathrm{cal}},\theta^{(i)},q^{(i)}
    \right)$
    \State $\theta^{(i+1)} \gets
    \textsc{OffsetSolve}\!\left(
      V^{\mathrm{cal}},q^{(i+1)},\theta^{(i)}
    \right)$
    \EndFor
    \State $q^{\mathrm{cal}} \gets
    \textsc{PoseSolve}\!\left(
      V^{\mathrm{cal}},\theta^{(K)},q^{(K)}
    \right)$
    \Comment{Final calibration pose solve}

    \Statex
    \Statex \textbf{Register the full recording}
    \State Split $V_{1:T}$ into overlapping context windows
    \State $q_{1:T} \gets \varnothing$; $q^{\mathrm{prev}} \gets \varnothing$
    \For{each window $W$ in temporal order}
    \State Initialize $q_W^{(0)}$ from the default pose or
    $q^{\mathrm{prev}}$ using sparse pose solves
    \State $q_W \gets
    \textsc{PoseSolve}\!\left(
      V_W,\theta^{(K)},q_W^{(0)}
    \right)$
    \State Append the central poses from $q_W$ to $q_{1:T}$
    \State $q^{\mathrm{prev}} \gets$ the final context poses from $q_W$
    \EndFor
    \State \Return $\theta^{(K)},q_{1:T}$
  \end{algorithmic}
\end{algorithm}

\subsubsection{Handling Long Continuous Sessions}
We used \texttt{stac-mjx} on a 3-hour continuous session to evaluate our trained policy on long unseen clips. Splitting such recordings into disjoint segments to fit in GPU memory introduces small discontinuities at segment boundaries. Instead, \texttt{stac-mjx} processes long recordings in overlapping context windows: the first window starts from the default pose, and each subsequent window is warm-started from the preceding window's overlap, extended and refined by a sparse keyframe solve interpolated across the window (root poses on $\mathrm{SE}(3)$, joint coordinates linearly). All frames in a window are solved jointly under the temporal-smoothness penalty and only the central frames are retained, yielding continuous trajectories with negligible edge effects and no explicit post-hoc blending. The context length and sparse-frame spacing are configuration parameters.

\FloatBarrier 
\subsection{Deep Reinforcement Learning for Motion Imitation}

\subsubsection{Parallelized training}
On-policy reinforcement learning algorithms are generally divided into two phases: data collection and policy updates. Policy updates consist of performing backpropagation on the neural networks, which are housed on the GPU. Data collection from physics environments consists of policy inference, simulating steps, and reward/observation calculation. This phase is most commonly performed on the CPU \cite{keller_autonomous_2025,chiappa_arnold_2025,xu_open-source_2024,akki_benchmarking_2025,henaff_scalable_2025} as popular physics engines such as MuJoCo are CPU-bound.
By leveraging MJX, we overcome the data transfer problem, and also substantially improve simulation speed through parallel environments---thousands of environments can run on a single GPU workstation/cluster node. Parallelizing CPU-based simulation is also possible, but typically requires provisioning additional cluster nodes, high-bandwidth interconnects, and a nontrivial distributed-training setup (job schedulers, networking and firewall configuration, synchronization, and data movement). Achieving the necessary throughput with CPUs would ordinarily require multiple server-grade machines, whereas a single GPU can do the same job with far less operational overhead.

\subsubsection{Imitation Task Environment}

\paragraph{Input Observations}
The proprioceptive observations are sampled at each timestep from the physics environment and are defined as $\boldsymbol{s}_t^p\triangleq(\boldsymbol{\theta}_t,\boldsymbol{\omega}_t,\boldsymbol{\tau}_t,z_t,\boldsymbol{\phi}_t,\boldsymbol{e}_t,\boldsymbol{k}_t,\boldsymbol{c}_t,\boldsymbol{a}_{t-1})$, where $\boldsymbol{\theta}_t$ are joint angles, $\boldsymbol{\omega}_t$ are joint angular velocities, $\boldsymbol{\tau}_t$ are applied actuator forces, $z_t$ is the root or torso height, $\boldsymbol{\phi}_t$ is the direction of the global z-axis relative to the agent (akin to a vestibular sense), $\boldsymbol{e}_t$ are egocentric end-effector positions, $\boldsymbol{k}_t$ are available kinematic sensors (accelerometer, velocimeter, and gyroscope), $\boldsymbol{c}_t$ are available touch sensors, and $\boldsymbol{a}_{t-1}$ is the previous action.

The reference observations are kinematic features of future frames in the target trajectory and are supplied as input to the policy's encoder. For free-body imitation tasks, the reference observations are defined as $\boldsymbol{s}^g_t\triangleq(\boldsymbol{\Delta r}_{t:t'}, \boldsymbol{\Delta \varphi}_{t:t'},\boldsymbol{\Delta \theta}_{t:t'},\boldsymbol{\Delta x}_{t:t'})$ where $\boldsymbol{\Delta r}$ is the difference between the agent and target root position rotated to the agent's egocentric frame, $\boldsymbol{\Delta \varphi}$ is the relative root quaternion, $\boldsymbol{\Delta \theta}$ is the difference between target and current joint angles, and $\boldsymbol{\Delta x}$ is the difference between target and current body positions rotated to the agent's egocentric frame. We computed the reference observations for 5 time steps into the future by default. The mouse arm imitation environment used a reduced fixed-base reference observation consisting of future joint-angle deltas and wrist-position deltas. We found that the choice of trajectory length affected the model's ability to generalize to longer motion clips, with a trajectory length of 5 giving the best results (\hyperref[suppfig:trajectory-length]{Supplementary Fig.~\ref*{suppfig:trajectory-length}}).

Both reference and proprioceptive observations are normalized to have $\mu \approx 0$ and $\sigma^2 \approx 1$: $\hat{o}_t = (o_t- \mu)/\sigma$. Running statistics are maintained throughout training and updated after each data-collection round. During training, multiplicative Gaussian noise is applied to the decoder's proprioceptive input, $\tilde{\boldsymbol{s}}_t^p=\boldsymbol{s}_t^p\odot(1+\sigma_p\epsilon)$ with $\epsilon\sim\mathcal{N}(0,\mathbf{I})$ in order to reduce overfitting to proprioceptive inputs; this noise is disabled during deterministic evaluation and inference.

\paragraph{Taking actions}
Each environment step takes an action vector which maps to a control signal for each actuator in the animal model. The simulation is driven forward for a fixed timestep using this control signal to actuate the body. The duration of this timestep determines the policy's action rate, which can be different for each animal model; this is defined in the environment configuration.

\paragraph{Reward}
Similar to previous work \cite{Aldarondo2024-nc,Vaxenburg2025-mb, Peng2018-di}, we decompose the motion-imitation objective into a configurable sum of tracking rewards and control costs. Together, they provide a dense reward signal representing the correct pose of the animal while discouraging energetically or dynamically excessive control. At time $t$, the instantaneous reward is
\[
  \begin{aligned}
    r_t \;=\;& \sum_{i \in \mathcal{I}} \lambda_i\,r^i_t
    + \lambda_z\,\mathbf{1}\{z_{\min} \le z_t \le z_{\max}\} \\
    &- \lambda_{\mathrm{ctrl}}\lVert a_t\rVert_2^2
    - \lambda_{\Delta\mathrm{ctrl}}\lVert a_t-a_{t-1}\rVert_2^2
    - \lambda_E c^E_t,
  \end{aligned}
\]
where $\mathcal{I}$ is the set of enabled imitation terms in the environment configuration and the three subtracted terms are the control costs.

Each imitation reward $r^i_t\!\in[0,1]$ increases as the learner better matches the registered reference trajectory. We define these terms as Gaussian kernels using L2 distance between the agent state and the reference data:
\[
  r_{t}^{i} = \exp\left(-\frac{1}{2\sigma_i^2} \, \| y_i(s_t) - y_i(s^{\mathrm{ref}}_t) \|_2^2 \right).
\]
The $\sigma_i$ term is configured for each reward term based on the distance scales.

\begin{enumerate}
  \item \textbf{Position} ($r^{\mathrm{pos}}$) measures the Cartesian displacement between the learner's root/global position and the reference, encoding \emph{where} the body is in the arena.
  \item \textbf{Quaternion / orientation} ($r^{\mathrm{quat}}$) measures angular misalignment of headings, encoding \emph{which way} the body is pointing.
  \item \textbf{Joint} ($r^{\mathrm{joint}}$) measures per-joint angle error, shaping local body configurations.
  \item \textbf{Joint velocity} ($r^{\mathrm{joint\_vel}}$) measures per-joint velocity error.
  \item \textbf{Body position} ($r^{\mathrm{body}}$) measures distances between configured body positions and their registered references.
  \item \textbf{End effector} ($r^{\mathrm{ee}}$) measures distances between the learner's distal effectors (e.g., hands, pretarsal claws) and the registered references, encouraging precise placement.
\end{enumerate}

\noindent Control costs are implemented as negative reward components:
\begin{enumerate}
  \item \textbf{Control cost} penalizes squared action magnitude, $\lVert a_t\rVert_2^2$.
  \item \textbf{Control-difference cost} penalizes squared action changes, $\lVert a_t-a_{t-1}\rVert_2^2$.
  \item \textbf{Energy cost} penalizes high-force movements at high speeds; $c^E_t=\min(P_t,E_{\max})$, where $P_t=\lVert \dot{\boldsymbol{q}}_t \odot \boldsymbol{\tau}_t\rVert_1$ is a mechanical-power proxy; $\dot{\boldsymbol{q}}_t$ is the generalized velocity vector, $\boldsymbol{\tau}_t$ is the generalized actuator-force vector, and $E_{\max}$ is the clipping threshold.
\end{enumerate}

These terms are configured per animal model. Effective defaults are reported in \hyperref[supptable:imitation-training-config]{Supplementary Table~\ref*{supptable:imitation-training-config}}. The kernel widths $\sigma_i$ were chosen heuristically to control the steepness of each reward's falloff: wide enough that a pose held momentarily stationary against a moving reference does not collapse the reward over only a few frames, yet narrow enough to preserve a discriminative gradient toward accurate tracking.

\paragraph{Termination criteria}
Early termination criteria improve sample efficiency during training by avoiding training data in unrecoverable states. The environment terminates (leading to an automatic reset during training) if any of the following conditions are met:
\begin{itemize}
  \item \textit{Maximum cartesian position error.} The agent's root position in global cartesian coordinates sufficiently diverges from the reference.
  \item \textit{Maximum orientation error.} The root quaternion of the agent sufficiently diverges from the reference.
  \item \textit{Maximum joint pose error.} The joint angles of the agent in aggregate sufficiently diverge from the reference.
  \item \textit{Numerical instability.} The simulator state contains NaN values.
\end{itemize}
The free-body rat, fruit fly, and stick insect imitation environments use the position, orientation, joint-pose, and NaN criteria above. The fixed-base mouse arm task terminates on joint-pose error or NaN values. The environment is also reset once it reaches the end of the valid reference motion trajectory clip.

\subsubsection{Training with deep reinforcement learning}
We employ Proximal Policy Optimization \cite{Schulman2017-rt} as our reinforcement learning algorithm, with several modifications tailored for trajectory imitation learning. Our PPO implementation draws heavily from the Brax framework \cite{Freeman2021-fb} while using our encoder--decoder policy architecture to compress motor intentions into a regularized stochastic latent space.

\paragraph{Problem Formalization}
We can formalize the tracking objective as a Markov Decision Process (MDP) defined by the tuple $\mathcal{M}=\left< \mathcal{S}, \mathcal{A}, \mathcal{T}, \mathcal{R}, \gamma \right>$ containing states, actions, transition dynamics, reward function, and discount factor. The physics environment determines the state $s^p_t\in \mathcal{S}$ and transition dynamics $\mathcal{T}$. The policy $\pi_{track}$ computes the per-step action $a_t\in\mathcal{A}$. The reward function computes the imitation reward, $r_t$, for the policy as a function of the simulation state, $s^p_t$ and the reference motion, $s_t^{mocap}$: $r_t=\mathcal{R}(s^p_t,s^{mocap}_t)$. The objective of the policy is to maximize the discounted return $\mathbb{E}\left[ \sum_{t=1}^T \gamma^{t-1}r_t\right]$.

\paragraph{Policy network}
Similar to prior work \cite{Aldarondo2024-nc}, our policy has an encoder--decoder architecture with a variational bottleneck in between, akin to a Variational Autoencoder \cite{kingma_auto-encoding_2022}. The encoder, $\mathcal{E}(z_t|s_{t:t'}^g)=\mathcal{N}(z_t|\mu_t^e,\sigma_t^e)$, receives a context window of reference observations as input and outputs the mean and logvariance of a diagonal Gaussian distribution. The reparameterization trick \cite{kingma_auto-encoding_2022} is used to sample this distribution. The sampled latent vector, $z_t$, is passed as input to the decoder along with the proprioceptive observation from the environment, and the decoder outputs the parameters of a normal action distribution followed by a hyperbolic tangent transformation to bound the actions within $[-1,1]$. Body models with muscle actuators use a sigmoidal transformation to bound the actions within $[0,1]$. At inference, the mean motor intention vector and the mode of the action distribution are used to generate deterministic actions.

The default \texttt{track-mjx} architecture uses an MLP encoder and an MLP decoder. Feed-forward encoder and decoder layers use LeCun Uniform initialization \cite{lecun_efficient_1998}, SiLU activations \cite{elfwing_sigmoid-weighted_2017}, and layer normalization \cite{ba_layer_2016}. Default network layer configurations are detailed in the supplement.

\paragraph{Motor intention regularization}
In order to regularize the latent variational bottleneck, $\mathcal{E}(z)$, we use two configurable losses. First, we regularize each encoded distribution toward a standard Gaussian prior:
\[
  L_{\mathrm{KL}} = KL\!\left(\mathcal{E}(z_t) \,\|\, \mathcal{N}(0,\mathbf{I})\right).
\]
Second, we optionally apply an AR(1)-inspired smoothness penalty to the latent means:
\[
  L_{\mathrm{AR}} =
  \frac{\sum_t m_t \lVert \mu^e_{t+1}-\mu^e_t\rVert_2^2}
  {\sum_t m_t},
\]
where $m_t$ masks timestep pairs that cross episode termination or truncation boundaries. This penalty was found to be helpful in prior work \cite{luo_universal_2024} for structuring the motor intention space for exploration. The regularization terms are added to the standard PPO policy objective, resulting in the minimization objective
\[
  \begin{aligned}
    L_{\pi+\mathrm{latent}}(\theta)
    &= L_{\pi}^{\mathrm{PPO}}(\theta)
    + \beta_{\mathrm{KL}}L_{\mathrm{KL}}
    + \beta_{\mathrm{temp}}L_{\mathrm{AR}}. \\
  \end{aligned}
\]
The KL and temporal weights can use custom schedules, with a linear ramping schedule being the default.

\paragraph{Value network}
We employ a standard feed-forward MLP network as our critic. The network receives the complete observation (proprioceptive and reference observations) as input. The output is a single scalar value representing the estimated value of the current state. This network uses the LeCun Uniform initialization and employs ReLU activations.

\paragraph{Environment initialization and data collection}\label{par:env_init}
We use reference motion clips with lengths of $L_{clip}$ frames sampled at a rate of $f_{mocap}$. The reference clips are split into training and held-out evaluation subsets. We initialize $N_{envs}$ environments in parallel using randomly chosen training clips. Each environment begins at a random frame chosen from a configured start-frame range, and the simulated pose is initialized to the corresponding reference pose. We define the length of a full episode to be $T_{episode}$. For the rat environment, a small amount of domain randomization is applied over joint friction loss, armature, center of mass, link masses, torso mass, and the model's default pose to improve robustness of the learned policy to different conditions.

After initialization, we define one iteration of the data collection phase as the collection of $N_{minibatch}$ number of mini-batches, each containing $B_{batch}$ trajectories spanning $L_{unroll}$ frames. We define one round of data collection as collecting $L_{unroll}$ frames of data from each of the $N_{envs}$ environments. $L_{unroll}$ must be large enough to capture coherent temporal dynamics, but short enough to ensure training stability. If $L_{unroll}$ is too small, Generalized Advantage Estimation (GAE) cannot leverage rewards from enough timesteps to be effective. If $L_{unroll}$ is too large, we obtain less stable gradient estimates, risking training destabilization. As such, multiple rounds of data collection may be needed to fill the mini-batch quota before performing a gradient update. The data collection process is outlined in \hyperref[alg:motion_imitation_pipeline]{algorithm \ref*{alg:motion_imitation_pipeline}}.

\begin{algorithm}[!htbp]
  \scriptsize
  \caption{Motion Imitation Training Pipeline}
  \label{alg:motion_imitation_pipeline}

  \textbf{Hyperparameters:} $N_{envs}$, $T_{episode}$, $L_{unroll}$, $B_{batch}$, $N_{minibatch}$, $N_{updates}$, $qvel_{\mathrm{init}}$, $num\_timesteps$

  \begin{algorithmic}[1]

    \State \textbf{// Environment Initialization}
    \For{environment $i = 1$ to $N_{envs}$}
    \State Sample clip $c_i$ and start frame $f_i$ randomly
    \State $\mathbf{q} \leftarrow \text{reference\_pose}(c_i, f_i)$
    \State $\dot{\mathbf{q}} \leftarrow \text{velocity\_init}(qvel_{\mathrm{init}}, c_i, f_i)$
    \State Initialize physics simulation with $(\mathbf{q},\dot{\mathbf{q}})$
    \State Optionally randomize physics parameters for the environment
    \EndFor

    \State
    \State \textbf{// Training Loop}
    \State $n_{\mathrm{steps}} \leftarrow 0$ \Comment{Initialize total environment timestep counter}
    \While{$n_{\mathrm{steps}} < num\_timesteps$}
    \State $\mathcal{T} \leftarrow \emptyset$ \Comment{Initialize trajectory buffer}

    \State \textbf{// Data Collection Phase}
    \For{round $r = 1$ to $\lceil (B_{batch} \times N_{minibatch}) / N_{envs} \rceil$}

    \State \textbf{// Parallel Trajectory Collection}
    \For{environment $i = 1$ to $N_{envs}$ in parallel}
    \State $\tau_i \leftarrow \emptyset$ \Comment{Initialize trajectory for env $i$}

    \For{timestep $t = 1$ to $L_{unroll}$}

    \State $\mathbf{o}_i \leftarrow [\text{reference\_window}(c_i, f_{i}:f_{i+t}); \text{proprioception}(s_i)]$
    \State $\mathbf{a}_i \leftarrow \pi_\theta(\mathbf{o}_i)$ \Comment{Policy inference}
    \State $s_i' \leftarrow \text{physics\_step}(s_i, \mathbf{a}_i)$
    \State $n_{\mathrm{steps}} \leftarrow n_{\mathrm{steps}} + 1$ \Comment{Increment total environment timestep counter}
    \State $r_i \leftarrow \text{tracking\_reward}(s_i', \text{reference}(c_i, f_i))$
    \State $\tau_i \leftarrow \tau_i \cup \{(\mathbf{o}_i, \mathbf{a}_i, r_i, s_i')\}$

    \State \textbf{// Auto-reset On Termination}
    \If{$t \geq T_{episode}$ \textbf{or} safety\_violation$(s_i')$}
    \State Sample new $(c_i, f_i)$ and reset to the corresponding reference state
    \Else
    \State $s_i \leftarrow s_i'$
    \EndIf
    \EndFor

    \State $\mathcal{T} \leftarrow \mathcal{T} \cup \{\tau_i\}$ \Comment{Update trajectory buffer}
    \EndFor
    \EndFor

    \State \textbf{// Training Phase}
    \State Update observation running statistics if normalization is enabled
    \State Compute advantages using GAE on all trajectories in $\mathcal{T}$

    \For{update $u = 1$ to $N_{updates}$}
    \State $\mathcal{T}_{shuffled} \leftarrow \text{random\_permutation}(\mathcal{T})$
    \State Split $\mathcal{T}_{shuffled}$ into $N_{minibatch}$ chunks of size $B_{batch}$
    \For{minibatch $m = 1$ to $N_{minibatch}$}
    \State $\mathcal{T}_m \leftarrow \text{minibatch}_m(\mathcal{T}_{shuffled})$
    \State Perform PPO gradient update on $\mathcal{T}_m$
    \Comment{Gradient update}
    \EndFor
    \EndFor

    \EndWhile

  \end{algorithmic}
\end{algorithm}

\subsubsection{Decoder transfer for downstream task learning}


To reuse a trained motion imitation controller on new tasks, we load the decoder parameters as a frozen low-level controller (\hyperref[fig5]{Fig. 5a, bottom}) that maps motor intentions to actions, and initialize a high-level task policy \(\pi_{\psi}\) that maps task observations to motor intentions:
\[
  z_t=\pi_{\psi}(o_t),\qquad a_t = p_{\theta}(a_t \mid z_t, s_t^{p}),
\]
where \(o_t\) is any task-specific observation (e.g., goals, terrain state, or other task variables). This modular interface leverages the task-agnostic motor intention space \(z_t\) to structure actions based on the motion imitation training, while enabling a trainable task-specific high-level policy. The high-level task policy is treated as the ``agent" in the PPO training loop, while the low-level decoder is treated as a part of the environment-- a post-processing step converting the high-level actions in motor intention space into actuator controls. With this framing, the RL action exploration of the high-level policy is done in the MIMIC-MJX motor intention space, which is more amenable to solving complex and ethological tasks than exploring in raw actuator space. The PPO, network, and simulation hyperparameters used to train the high-level policies for both downstream tasks are listed in \hyperref[supptable:downstream-task-config]{Supp. Table~\ref*{supptable:downstream-task-config}}.


\subsection{Maintain Velocity Task Environment}

In the maintain velocity environment, the agent must locomote on flat ground at a target forward speed. The reward is deliberately minimal: it specifies only the desired velocity and does not encourage any particular gait or posture.

\subsubsection{State Space and Task Overview}
The agent is initialized at the origin in a neutral standing pose facing the \(+x\) direction (identity orientation quaternion, \(z\)-offset \(\texttt{init\_z}=0\)) on a flat arena. It is rewarded for matching its forward (\(+x\)) torso velocity to a target speed \(v^\star\). Episodes run for a fixed horizon and terminate early only if the agent falls.

\subsubsection{Observations}
At time \(t\), the policy input concatenates a task--specific vector \(\mathbf{o}^{\text{task}}_t\) and a proprioceptive vector \(\mathbf{o}^{\text{prop}}_t\), following the same convention as in Bowl Escape:
\[
  \mathbf{o}_t=\big[\ \mathbf{o}^{\text{task}}_t\ \Vert\ \mathbf{o}^{\text{prop}}_t\ \big].
\]
The task--specific observation
\begin{equation}
  \mathbf{o}^{\text{task}}_t=\big[\ \mathbf{a}_{t-1},\ \mathbf{o}^{\text{kin}},\ \mathbf{o}^{\text{touch}},\ \mathbf{o}^{\text{orig}}\ \big]
\end{equation}
collects the previous action \(\mathbf{a}_{t-1}\), body--frame inertial sensors \(\mathbf{o}^{\text{kin}}\) (accelerometer, velocimeter, and gyroscope), tactile/contact readings \(\mathbf{o}^{\text{touch}}\), and the agent pose relative to the origin \(\mathbf{o}^{\text{orig}}\). The proprioceptive observation \(\mathbf{o}^{\text{prop}}_t\) is defined as for the bowl escape task. Notably, the target speed itself is not exposed to the policy; the given target speed is implicit to the task environment through the reward function.

\subsubsection{Reward and Termination}
Let \(v_t\) be the torso forward (\(+x\)) velocity and \(v^\star>0\) the target speed. Reusing the piecewise--linear tolerance \(\operatorname{tol}_{\text{lin}}\) defined for Bowl Escape, the reward is a triangular peak at \(v^\star\),
\begin{equation}
  r_t=\operatorname{tol}_{\text{lin}}\!\big(v_t;\,[v^\star,v^\star],\,v^\star\big),
\end{equation}
which equals \(1\) at \(v_t=v^\star\) and decreases linearly to \(0\) at \(v_t\in\{0,2v^\star\}\). Optional penalties on lateral velocity and yaw rate are available to discourage sideways drift and turning, but are disabled (zero weight) by default.

\paragraph{Termination}
The episode fails if the agent falls, i.e.\ if the torso drops below a minimum height \(z_t<z_{\min}\) or tilts beyond a maximum angle from vertical, \(\langle \mathbf{z}_{\text{torso}},\zhat\rangle<\cos\theta_{\max}\). Episodes also terminate on numerical failure (NaN in the simulation state).

\paragraph{Environment Parameters}
\begin{equation*}
  \begin{aligned}
    &\texttt{sim\_dt}=0.002\ \text{s},\quad \texttt{ctrl\_dt}=0.01\ \text{s},\\
    &v^\star=0.5\ \text{m/s (default; swept across }0.1\text{--}2.0\ \text{m/s)},\\
    &z_{\min}=0.03\ \text{m},\quad \theta_{\max}=60^\circ,\quad \text{horizon}=2000\ \text{control steps}.
  \end{aligned}
\end{equation*}

\subsection{Bowl Escape Task Environment}

The overall objective is for the agent to learn a way to escape from the center of a bumpy bowl-shaped arena as far as possible. Notably, the reward design itself is not concerned about a specific behavior that you use to escape the bowl, since the reward function only cares about the location of the agent.

\subsubsection{State Space and Task Overview}
The agent starts near the center of a concave heightfield (the ``bowl'') and is rewarded for reaching the rim while staying upright, with an optional term for moving near a target speed. Because MJX does not support collisions between height fields and ellipsoid geometries, we replaced the rat's ellipsoid foot geoms with boxes for this transfer task with negligible effect on the performance; the recently developed MjWarp backend resolves this issue. The ground is a heightfield \(H\in[0,1]^{N\times N}\) scaled in world coordinates by horizontal scale \(h>0\) (half--width) and vertical scale \(v_z>0\).

\paragraph{World/grid mapping.}
Let world planar coordinates be \((x,y)\in[-h,h]^2\). Define normalized texture coordinates
\[
  u=\frac{x+h}{2h},\qquad v=\frac{y+h}{2h}\qquad (u,v\in[0,1]),
\]
and corresponding grid indices
\[
  j=\big\lfloor u\,(N-1)\big\rfloor,\qquad i=\big\lfloor v\,(N-1)\big\rfloor,\qquad (i,j)\in\{0,\dots,N-1\}^2.
\]
The world's ground height is
\begin{equation}
  z_{\text{ground}}(x,y)=v_z\,H[i,j].
\end{equation}
The grid spacing in world units is \(\Delta=\frac{2h}{N-1}\).

\subsubsection{Bowl Geometry}
We construct a smooth concave ``Gaussian bowl'' and add controllable roughness via 2D Perlin noise. A radial gate blends out the noise near the origin to ensure a clean launch zone.

\paragraph{Base Gaussian Depression}
On a normalized square \((\xi,\eta)\in[-1,1]^2\) (linearly mapped from grid indices), define the radially symmetric depression
\begin{equation}
  B(\xi,\eta;\sigma,A)=A\,\exp\!\left(-\frac{\xi^2+\eta^2}{2\sigma^2}\right),
  \qquad \sigma>0,\; A<0,
\end{equation}
sampled at \(N\times N\) points to obtain \(B\in\RR^{N\times N}\).
Larger \(|A|\) deepens the bowl; smaller \(\sigma\) sharpens it.

\paragraph{2D Perlin Noise}
Let \(R_x,R_y\in\mathbb{N}\) be the desired number of periods across the domain. Construct a rectified integer lattice
\[
  \mathcal{L}=\{(p,q)\mid p\in\{0,\dots,R_x\},\ q\in\{0,\dots,R_y\}\},
\]
and assign each node \((p,q)\in\mathcal{L}\) a unit gradient \(\mathbf{g}_{pq}\in\RR^2\).

For a continuous coordinate \((\tilde u,\tilde v)\in[0,1]^2\), scale by periods,
\[
  \bar u=R_x\,\tilde u,\qquad \bar v=R_y\,\tilde v,
\]
and let \(p=\lfloor \bar u\rfloor,\ q=\lfloor \bar v\rfloor\), with local cell coordinates
\[
  t_x=\bar u-p,\qquad t_y=\bar v-q\qquad (t_x,t_y\in[0,1]).
\]
Define corner dot products
\begin{align}
  n_{00}&=\langle \mathbf{g}_{p,q},\ (t_x,\,t_y)\rangle, &
  n_{10}&=\langle \mathbf{g}_{p+1,q},\ (t_x-1,\,t_y)\rangle,\\
  n_{01}&=\langle \mathbf{g}_{p,q+1},\ (t_x,\,t_y-1)\rangle, &
  n_{11}&=\langle \mathbf{g}_{p+1,q+1},\ (t_x-1,\,t_y-1)\rangle.
\end{align}
Using the quintic fade \(f(s)=6s^5-15s^4+10s^3\), interpolate
\begin{align}
  \tilde n_0 &= (1-f(t_x))\,n_{00}+f(t_x)\,n_{10},\\
  \tilde n_1 &= (1-f(t_x))\,n_{01}+f(t_x)\,n_{11},\\
  P(\tilde u,\tilde v) &= (1-f(t_y))\,\tilde n_0+f(t_y)\,\tilde n_1.
\end{align}
The raw Perlin field \(P\) has zero mean and bounded variance; we apply a fixed affine map to center it near \([0,1]\):
\begin{equation}
  \widetilde P=\tfrac{1}{2}\,(2P+1)=P+\tfrac{1}{2}.
\end{equation}

\paragraph{Noise Gating and Normalization}
Let \(c=\tfrac{N-1}{2}\) be the grid center and \(r_{ij}=\sqrt{(i-c)^2+(j-c)^2}\) a discrete radius. Choose inner/outer gating radii \(0\le r_{\text{in}}<r_{\text{out}}\le \sqrt{2}\,\tfrac{N-1}{2}\). Define a smooth radial weight
\begin{equation}
  w_{ij}=\operatorname{smoothstep}\!\left(\frac{r_{ij}-r_{\text{in}}}{\,r_{\text{out}}-r_{\text{in}}\,}\right),
  \quad
  \operatorname{smoothstep}(s)=
  \begin{cases}
    0,& s\le 0,\\
    3s^2-2s^3,& 0<s<1,\\
    1,& s\ge 1.
  \end{cases}
\end{equation}
The raw height blends the pure bowl with the noisy bowl,
\begin{equation}
  H_0=(1-w)\odot B + w\odot\big(B+\widetilde P\big),
\end{equation}
where \(\odot\) denotes element-wise multiplication and \(w\) the matrix of \(w_{ij}\). Finally, normalize to \([0,1]\):
\begin{equation}
  H=\frac{H_0-\min(H_0)}{\max(H_0)-\min(H_0)}\;\in\;[0,1]^{N\times N}.
\end{equation}

\subsubsection{Observations}
At time \(t\), the policy input concatenates a task--specific vector \(\mathbf{o}^{\text{task}}_t\) and a proprioceptive vector \(\mathbf{o}^{\text{prop}}_t\):
\[
  \mathbf{o}_t=\big[\ \mathbf{o}^{\text{task}}_t\ \Vert\ \mathbf{o}^{\text{prop}}_t\ \big].
\]

\paragraph{Task--specific observations.}
\begin{equation}
  \mathbf{o}^{\text{task}}_t=\big[\ \mathbf{a}_{t-1},\ \mathbf{o}^{\text{kin}},\ \mathbf{o}^{\text{touch}},\ \mathbf{o}^{\text{orig}}\ \big],
\end{equation}
where \(\mathbf{a}_{t-1}\) is the previous action; \(\mathbf{o}^{\text{kin}}\) collects body--frame inertial sensors (accelerometer, velocimeter, and gyroscope); \(\mathbf{o}^{\text{touch}}\) encodes contact/tactile readings; and \(\mathbf{o}^{\text{orig}}\) provides the agent pose relative to the bowl origin (e.g., planar displacement and heading in the torso frame).

\paragraph{Proprioceptive observations.}
Let \(q\) and \(\dot q\) be generalized coordinates and velocities, excluding the floating base entries. With actuator signals \(\bm{\tau}_{\text{act}}\), torso height \(z_{\text{torso}}\), the world up-axis in the torso frame \(\hat z_{\text{torso}}\), selected end--effector positions \(\mathbf{p}_{\text{EE}}\) expressed in the torso frame, inertial sensors \(\mathbf{o}^{\text{kin}}\), tactile readings \(\mathbf{o}^{\text{touch}}\), and the previous action \(\mathbf{a}_{t-1}\),
\begin{equation}
  \mathbf{o}^{\text{prop}}_t=\big[\, q_{7:},\ \dot q_{6:},\ \bm{\tau}_{\text{act}},\ z_{\text{torso}},\ \hat z_{\text{torso}},\ \mathbf{p}_{\text{EE}},\ \mathbf{o}^{\text{kin}},\ \mathbf{o}^{\text{touch}},\ \mathbf{a}_{t-1}\,\big].
\end{equation}

\subsubsection{Reward and Termination}
Let the torso position be \(\mathbf{x}_t=(x_t,y_t,z_t)\) and \(\rho_t=\|\mathbf{x}_t\|\) its distance from the origin. The bowl half--width is \(h\).

\paragraph{Reward Scaling}
For a scalar \(s\) and target interval \([a,b]\) (with \(a\le b\)), define a piecewise--linear tolerance:
\begin{subequations}
  \begin{align}
    \operatorname{tol}_{\text{lin}}(s;\,[a,b],\,m) &=
    \begin{cases}
      0,& s\le 0,\\
      \min\!\left(1,\ \dfrac{s}{m}\right),& a=b= m,\\[6pt]
      clip\!\left(\dfrac{s-a}{b-a},\,0,\,1\right),& a<b,
    \end{cases}\\
    clip(x,0,1) &= \min(1,\max(0,x)).
  \end{align}
\end{subequations}

\paragraph{Escape shaping}
A radius--to--rim shaping that ramps from \(0\) at the center to \(1\) by the rim:
\begin{equation}
  r_{\text{esc}}(t)=\min\!\left(1,\ \frac{\rho_t}{h}\right).
\end{equation}

\paragraph{Uprightness}
Let \(\zhat\) be the world up-axis, and let \(\mathbf{z}_{\text{torso}},\mathbf{z}_{\text{head}}\) be the local \(z\)-axes (third rotation columns) of torso and head. With allowable tilt \(\theta=30^\circ\),
\begin{equation}
  r_{\text{up}}(t)=\min\!\Big(\operatorname{tol}_{\text{lin}}\!\big(\langle \mathbf{z}_{\text{torso}},\zhat\rangle;\,[\cos\theta,\infty),\,1+\cos\theta\big),\;
  \operatorname{tol}_{\text{lin}}\!\big(\langle \mathbf{z}_{\text{head}},\zhat\rangle;\,[\cos\theta,\infty),\,1+\cos\theta\big)\Big).
\end{equation}

\paragraph{Speed Setpoint}
Let \(v_t=\|\vec v_{\text{torso}}(t)\|\) be the torso linear speed and \(v^\star>0\) a target speed. A triangular peak at \(v^\star\) is
\begin{equation}
  r_{\text{spd}}(t)=\operatorname{tol}_{\text{lin}}\!\big(v_t;\,[v^\star,v^\star],\,v^\star\big),
\end{equation}
which equals \(1\) at \(v_t=v^\star\) and decreases linearly to \(0\) at \(v_t\in\{0,2v^\star\}\).

\paragraph{Total Reward}
\begin{equation}
  r_t=\underbrace{r_{\text{esc}}(t)\cdot r_{\text{up}}(t)}_{\text{reach rim while upright}},
\end{equation}
with the optional speed term \(r_{\text{spd}}(t)\) added when enabled (disabled by default).

\paragraph{Termination}
Let \(z_t\) be the torso height and \(z_{\text{ground}}(x_t,y_t)\) the terrain height below the torso. The episode fails if
\begin{equation}
  z_t\le z_{\text{ground}}(x_t,y_t)+\varepsilon,\qquad \varepsilon\approx 0.03\ \text{m}.
\end{equation}
The episode also terminates on numerical failure (a NaN in the simulation state).

\paragraph{Environment Parameters}
\begin{equation*}
  \begin{aligned}
    &N=128,\quad h=2.0,\quad v_z=0.2,\\
    &\sigma=1.25,\quad A=-10,\\
    &R_x=R_y=8,\quad r_{\text{in}}=0.05N,\quad r_{\text{out}}=0.25N,\\
    &v^\star=1.0\ \text{m/s},\quad \text{horizon}=2000\ \text{control steps}.
  \end{aligned}
\end{equation*}

\subsubsection{Kinematic manifold analysis.}
Each pose is represented by the $91$ pairwise Euclidean distances between $14$ body sites (head, three axial spine sites, two tail sites, and the proximal and distal segment of each limb), computed by forward kinematics on the recorded or simulated joint configuration through a single shared walker model. This representation is invariant to global translation and rotation, so no root or heading alignment is required. The reference distribution comprises $105{,}250$ poses subsampled at $25$\,Hz from the stac-mjx retargeted recordings. For each policy we collect $64$ stochastic rollouts of the bowl-escape task ($1{,}001$ poses per episode at a $100$\,Hz control rate, $64{,}064$ poses per condition), truncated at the first frame in which the torso crosses the rim (planar radius $\geq 2.0$\,m) so that all conditions are compared over the in-bowl phase. Features are standardized and reduced to two dimensions by PCA fit once on a class-balanced pool of the reference and all rollout conditions, then held fixed and applied to every condition; PC1 and PC2 explain $31\%$ and $25\%$ of the pooled variance for the torque conditions, and $43\%$ and $15\%$ for the position conditions, which are fit on separate axes. Contours show the $50\%$ and $90\%$ highest-density regions of a Gaussian kernel density estimate (bandwidth factor $0.30$, up to $40{,}000$ poses per condition) evaluated on a $240 \times 240$ grid and normalised to unit mass. For each policy we report its overlap with the reference, defined as the fraction of its poses falling inside the reference manifold, where that manifold is the union of balls centred on each reference pose with radius equal to the distance to its fifth nearest neighbour \citep{kynkaanniemi2019improved}.

\subsection{Gait Analysis}
We segment gait cycles by employing swing-stance transitions, detected as collisions between the rat's forepaws and the environment's ground. A gait cycle is defined from the beginning of a stance until the beginning of the next stance. The left forepaw was chosen to determine the boundaries of the gait cycles. We achieve phase normalization of joint angles by resampling all phase-locked gait cycles to 100 time points to represent the percentage of the gait cycle progression using cubic interpolation.

\subsection{Motor Intention Space Analysis}

\subsubsection{Mouse arm reaching}
To analyze the representational geometry of the network, we applied principal component analysis (PCA) to the motor intention bottleneck and to each of the three hidden layers of the decoder. For each layer, activations with dimensionality $(\text{clips}, \text{timesteps}, 512)$ were reshaped into a two-dimensional matrix of shape $(\text{clips}\times \text{timesteps}, 512)$ and PCA was performed. The first three principal components were retained, and the transformed data were reshaped back into $(\text{clips}, \text{timesteps}, 3)$ to preserve temporal and clip structure. The same procedure was applied to the motor intention representations, resulting in the top three principal components. For each analysis, the proportion of variance explained by the first three components was recorded to quantify the representational compression achieved by PCA. To enable joint visualization of neural and behavioral dynamics, the three components were concatenated with a behavioral covariate (the joint angle $q_{pose}$), producing arrays of shape $(\text{clips}, \text{timesteps}, 4)$. The 3D visualization of the top 3 PCs was visualized for the motor intention bottleneck and each layer of the decoder (\hyperref[fig6]{Fig. 6a}).

\subsubsection{Fly walking}
Using a trained policy on a dataset of fly walking, with a motor intention dimensionality of 60, we recorded network activations during each clip of walking at various speeds. We represented each trajectory frame by sampling the motor intention space activity with a time-delay embedding with a half-window of 30ms, resulting in a 31 frame window of motor intention space activations with a total of 1860 dimensions ($60 * 31$). To calculate the speed of each clip in \hyperref[fig6]{Fig. 6b}, we calculated the instantaneous speed between each frame, averaging over the clip to get the final value.

We then applied PCA to these time-delay embedded motor intention vectors and retained the top three principal components for visualization, where each point represents the embedding of a single 31-frame window. To characterize how this geometry encodes locomotion, we colored each point by the average locomotor speed of its clip. This revealed that the cyclic gait structure was spanned by PCs 2 and 3, while the orthogonal PC1 axis carried a smooth, near-linear velocity gradient, indicating that the fly's locomotor speed is encoded as a linear dimension of the motor intention manifold.

\backmatter

\beginsupplement

\clearpage

\section{Supplementary Information}

The online version of the article contains supplementary material added below.

\bmhead{Supplementary Figures}

\begin{center}
  \includegraphics[width=1.0\textwidth]{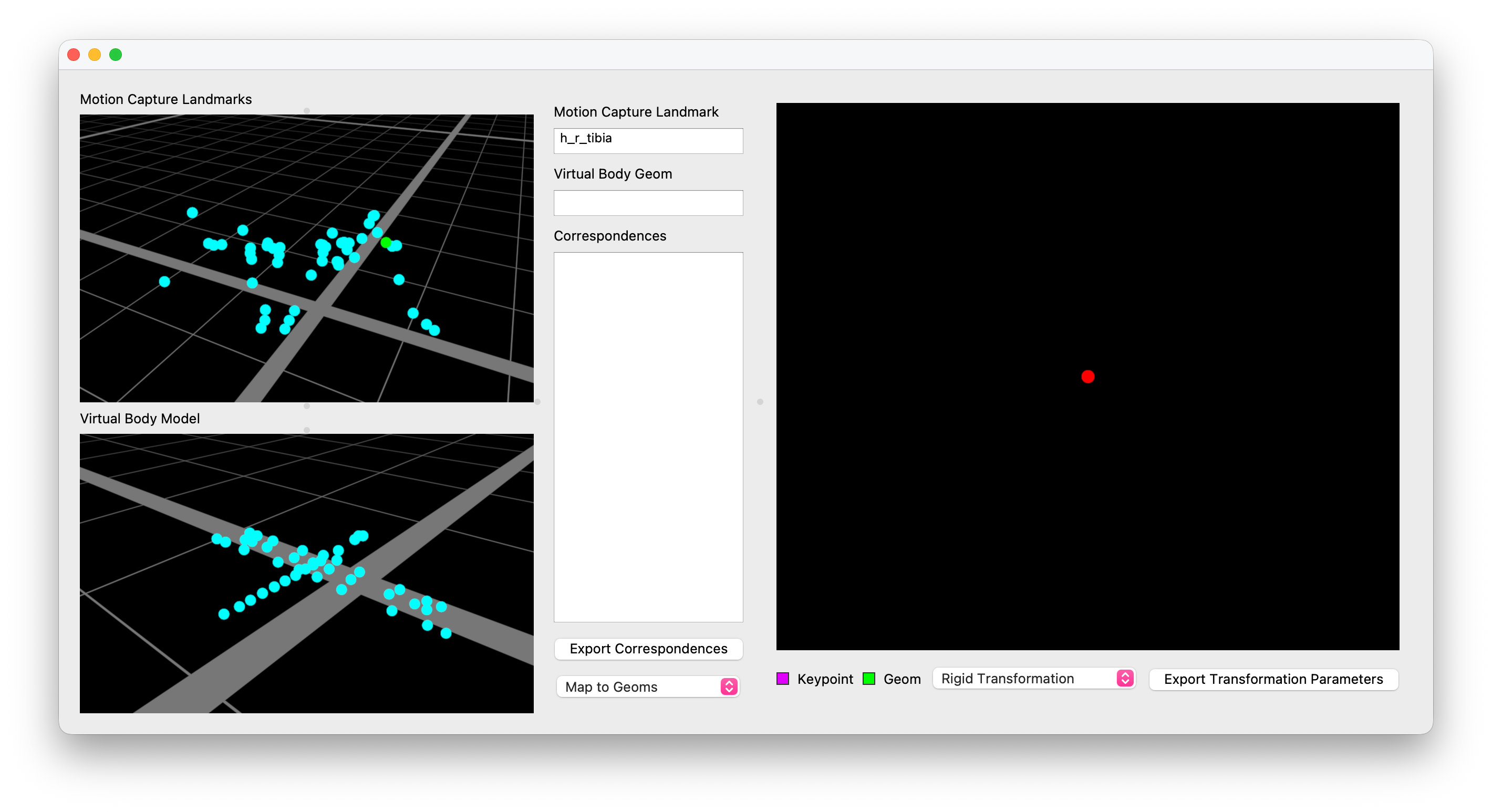}\\
  \vspace{0.2cm} 
  \includegraphics[width=1.0\textwidth]{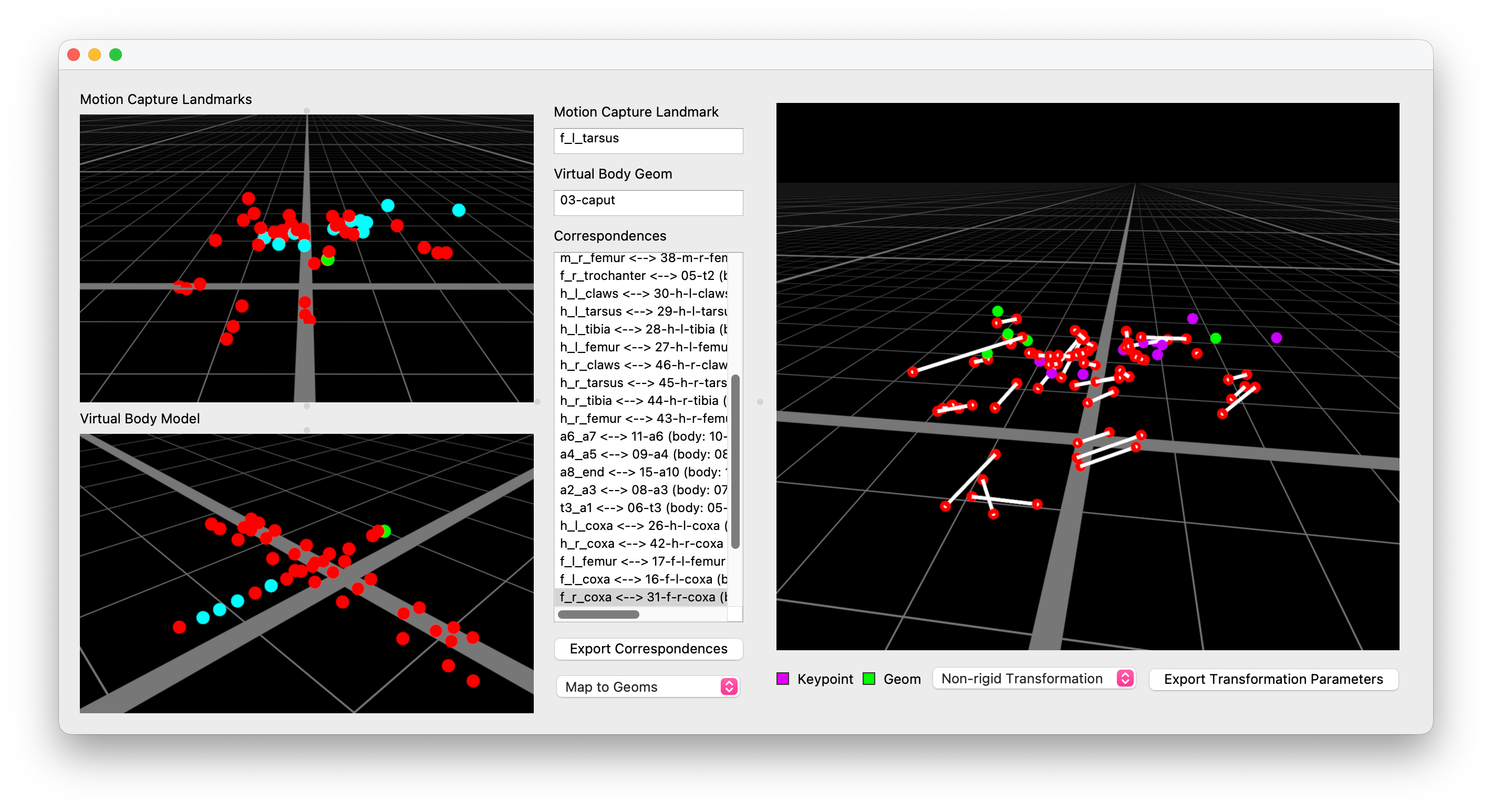}
  \captionof{figure}{Screenshots of the STAC Keypoint Correspondence user interface. Top: Initial interface displaying pose tracking landmarks (blue) and the virtual body model. Bottom: Interface after establishing correspondences. Red dots indicate established correspondences; blue dots indicate unmatched points.}
  \label{suppfig:stac-ui}
\end{center}
\vspace{1em} 

\begin{center}
  \includegraphics{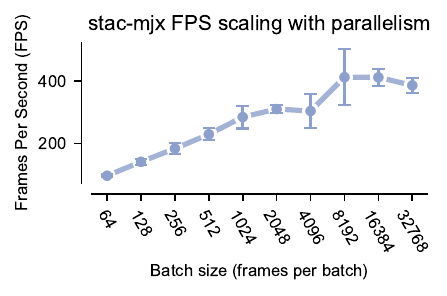}
  \captionof{figure}{\texttt{stac-mjx} frame processing speed as a function of the number of parallel batches. The plot shows frames per second (FPS) when processing 360,000 total frames across different batch configurations.}
  \label{suppfig:stac-mjx-fps-scaling}
\end{center}
\vspace{1em}

\begin{center}
  \includegraphics[width=0.85\textwidth]{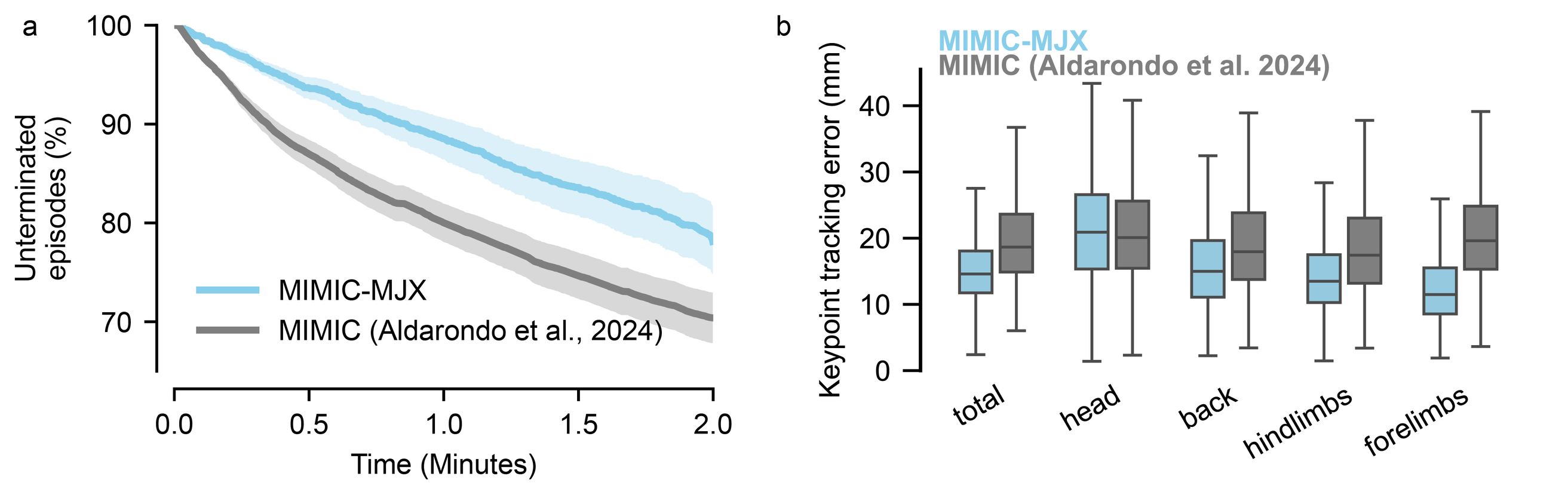}
  \captionof{figure}{\textbf{a}, Comparison of long-duration tracking stability between MIMIC-MJX and MIMIC \cite{Aldarondo2024-nc}. Unterminated episode percentage (solid line, mean; shaded band, $\pm 1$ standard error of the mean) is shown over two minutes across 22 held-out sessions from 7 rats. \textbf{b}, Comparison of keypoint tracking error between MIMIC-MJX and MIMIC, pooling keypoint errors over a continuous 3-hour held-out session, grouped by body region. Box: median and IQR; whiskers: $1.5\times$IQR (Tukey).}
  \label{suppfig:mimic-comparison}
\end{center}
\vspace{1em}

\begin{center}
  \includegraphics[width=1.0\textwidth]{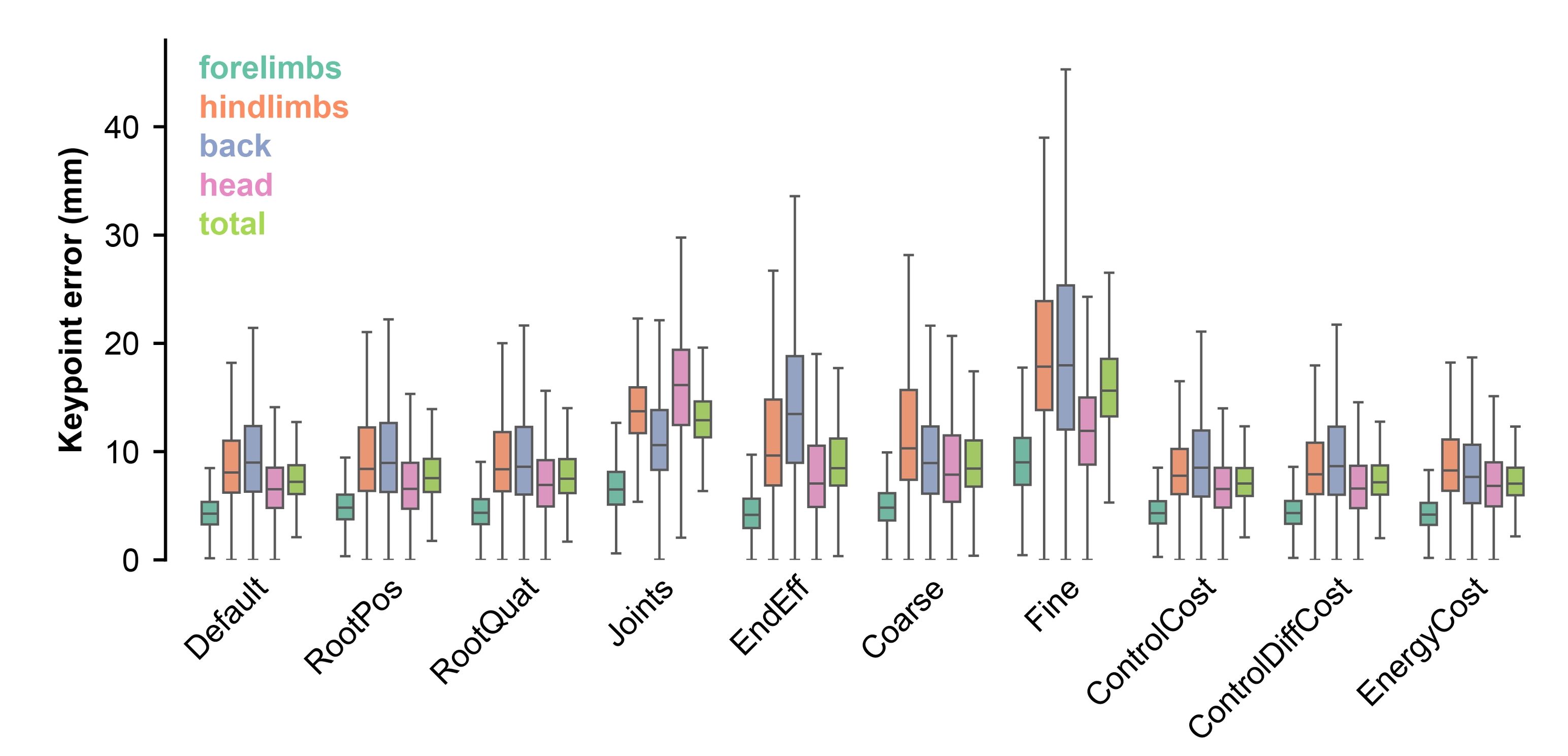}
  \captionof{figure}{Ablation studies on various reward terms of the rat. Each label denotes the reward term (or group of terms) removed; Default is the full reward Keypoint error distributions are shown between \texttt{track-mjx}-reproduced trajectories and \texttt{stac-mjx}-registered reference. Each distribution pools keypoint errors across 22 held-out 3-hour sessions (7 rats), grouped by body region. Box: median and IQR; whiskers: $1.5\times$IQR (Tukey).}
  \label{suppfig:reward-ablation}
\end{center}
\vspace{1em}

\begin{center}
  \includegraphics[width=1.0\textwidth,height=0.9\textheight,keepaspectratio]{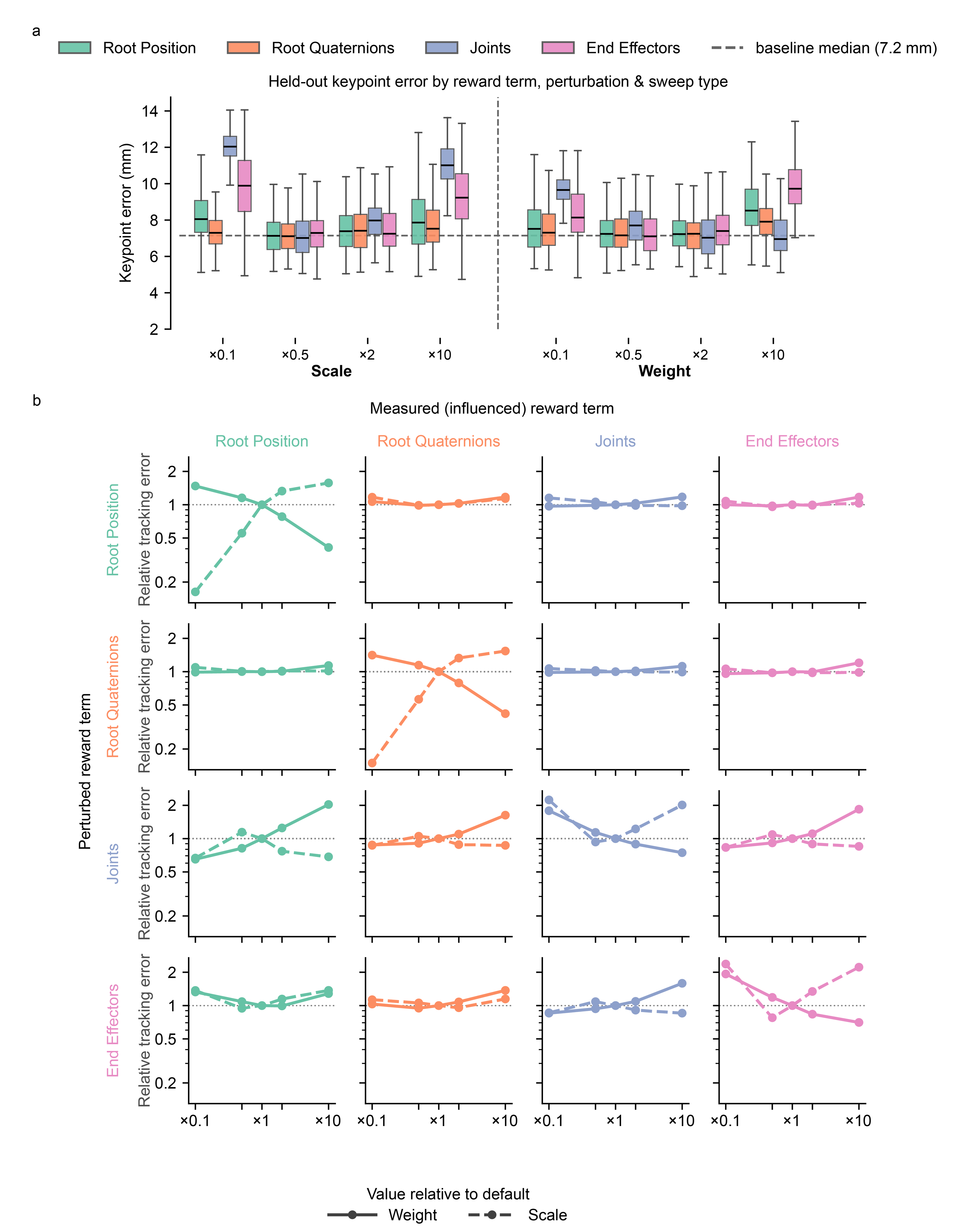}
  \captionof{figure}{Reward perturbation analysis for rat tracking. \textbf{a}, Held-out keypoint error after varying reward-term scales or weights for root position, root quaternion, joint, and end-effector terms. Each distribution pools keypoint errors over a continuous 3-hour held-out session; box: median and IQR, whiskers: $1.5\times$IQR (Tukey); the dashed line marks the baseline (unperturbed) median (7.2 mm). \textbf{b}, Relative tracking error of each measured reward term as each reward component is perturbed across values relative to the default. Solid and dashed lines indicate weight and scale perturbations, respectively.}
  \label{suppfig:reward-perturbation}
\end{center}
\vspace{1em}

\begin{center}
  \includegraphics[width=1.0\textwidth]{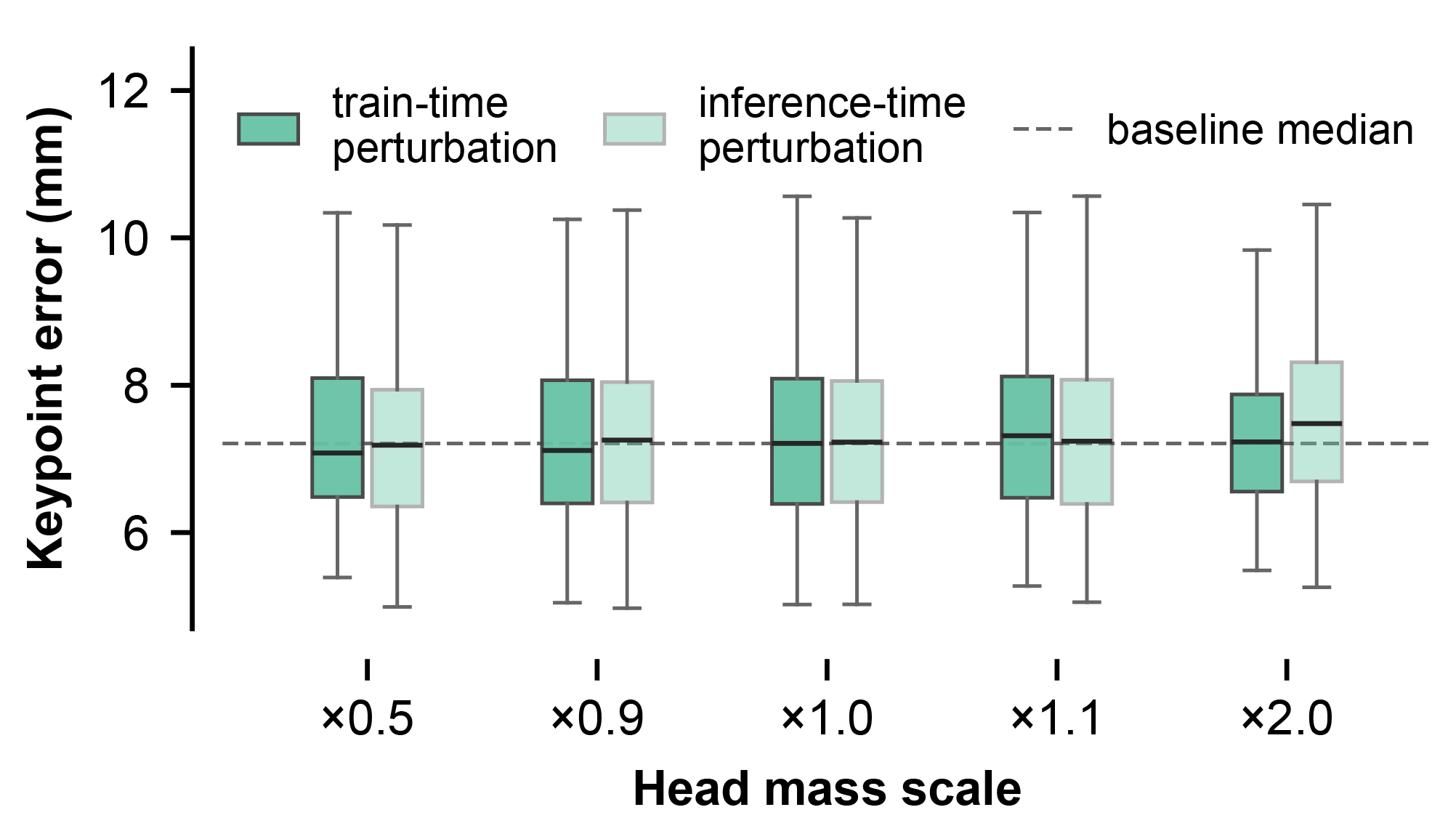}
  \captionof{figure}{Sensitivity of tracking performance to head-mass perturbations. Held-out keypoint error for train-time and inference-time head-mass perturbations across mass scale factors. Each box pools keypoint errors over a continuous 3-hour held-out session (box: median and IQR; whiskers: $1.5\times$IQR, Tukey); the dashed line marks the baseline median.}
  \label{suppfig:head-mass-sensitivity}
\end{center}
\vspace{1em}

\begin{center}
  \includegraphics[width=0.8\textwidth]{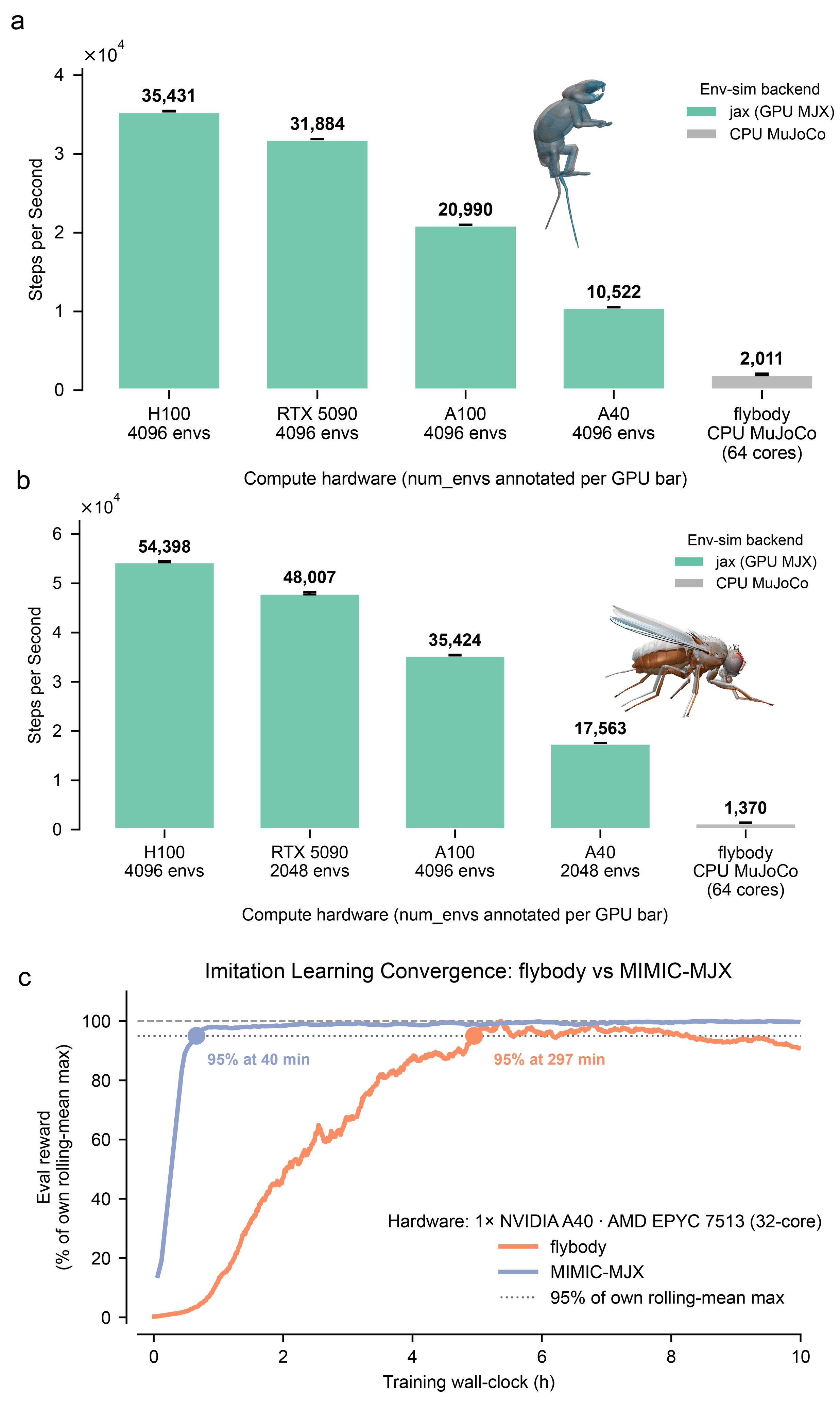}
  \captionof{figure}{\textbf{Hardware throughput and imitation learning convergence.}
    \textbf{a}, Steps per second (SPS) for the rat environment across different compute hardware (H100, RTX 5090, A100, A40, CPU); number of parallel environments (envs) annotated per bar, comparing the MJX JAX (GPU-based) backend against a MuJoCo (CPU-based) baseline (\texttt{flybody}, 64 cores).
    \textbf{b}, SPS for the fly environment on the same hardware as (a).
  \textbf{c}, Imitation learning convergence comparing \texttt{flybody} training using DMPO and \texttt{track-mjx} using PPO. Curves show eval reward as a percentage of each method's own rolling-mean max; markers indicate time to reach 95\% of that maximum (40 min for \texttt{track-mjx} vs.\ 297 min for \texttt{flybody}).}
  \label{suppfig:hardware-throughput}
\end{center}
\vspace{1em}

\begin{center}
  \includegraphics[width=1.0\textwidth]{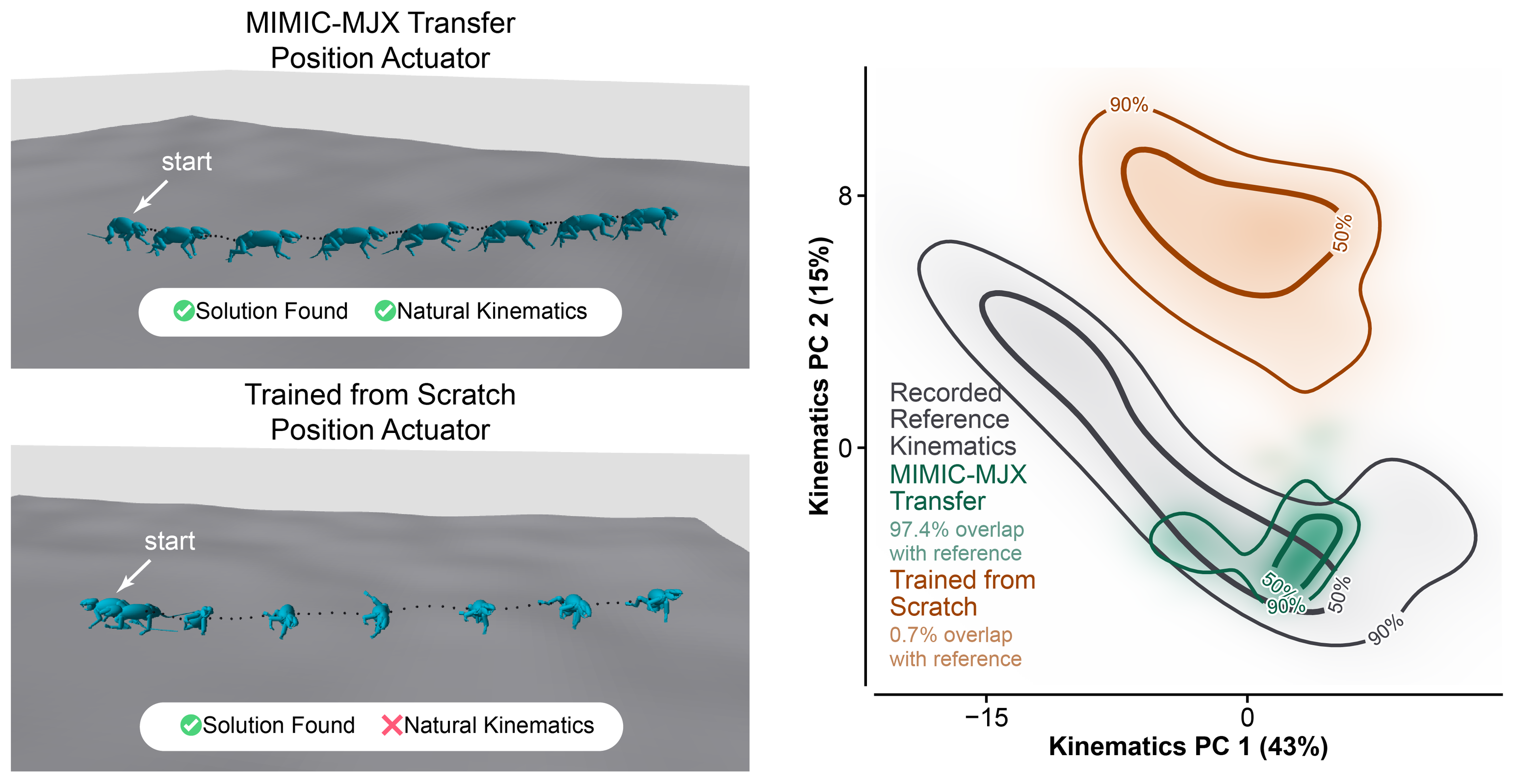}
  \captionof{figure}{\textbf{Position-actuator Bowl Escape rollouts and their kinematic occupancy relative to recorded reference kinematics.} Left, overlaid pose montages of one representative escape episode for the MIMIC-MJX transfer policy (top) and a policy trained from scratch (bottom), both using position actuators; arrow, start pose; dotted line, torso path; badges denote task success (``Solution Found'') and natural-kinematics designation (``Natural Kinematics''). Right, two-dimensional kinematic manifold: kernel-density occupancy of per-frame pose features (91 pairwise landmark distances) on position-native PC axes (variance in labels), for recorded reference kinematics (grey), transfer (green) and from-scratch (orange); contours enclose 50\% and 90\% of probability mass, pooled over 64 rollouts per policy (in-bowl frames). Percentages, fraction of each policy's frames inside the reference k-nearest-neighbour manifold (k = 5): 97.4\% (transfer), 0.7\% (from scratch).}
  \label{suppfig:bowl-position-actuators}
\end{center}
\vspace{1em}

\begin{center}
  \includegraphics[width=1.0\textwidth]{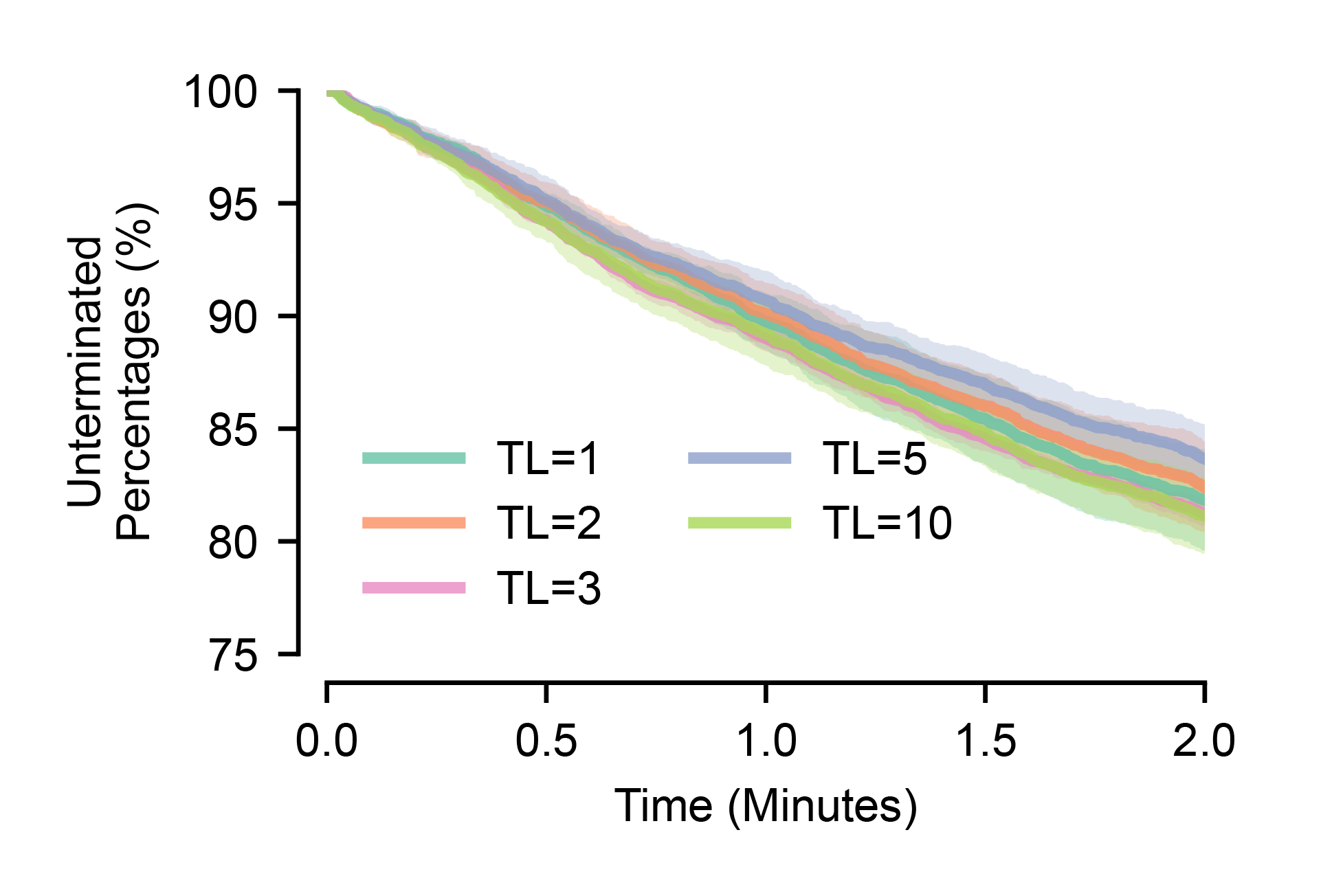}
  \captionof{figure}{Evaluation of continuous tracking performance for variable trajectory length (TL) on a held-out dataset.}
  \label{suppfig:trajectory-length}
\end{center}

\begin{center}
  \includegraphics[width=1.0\textwidth]{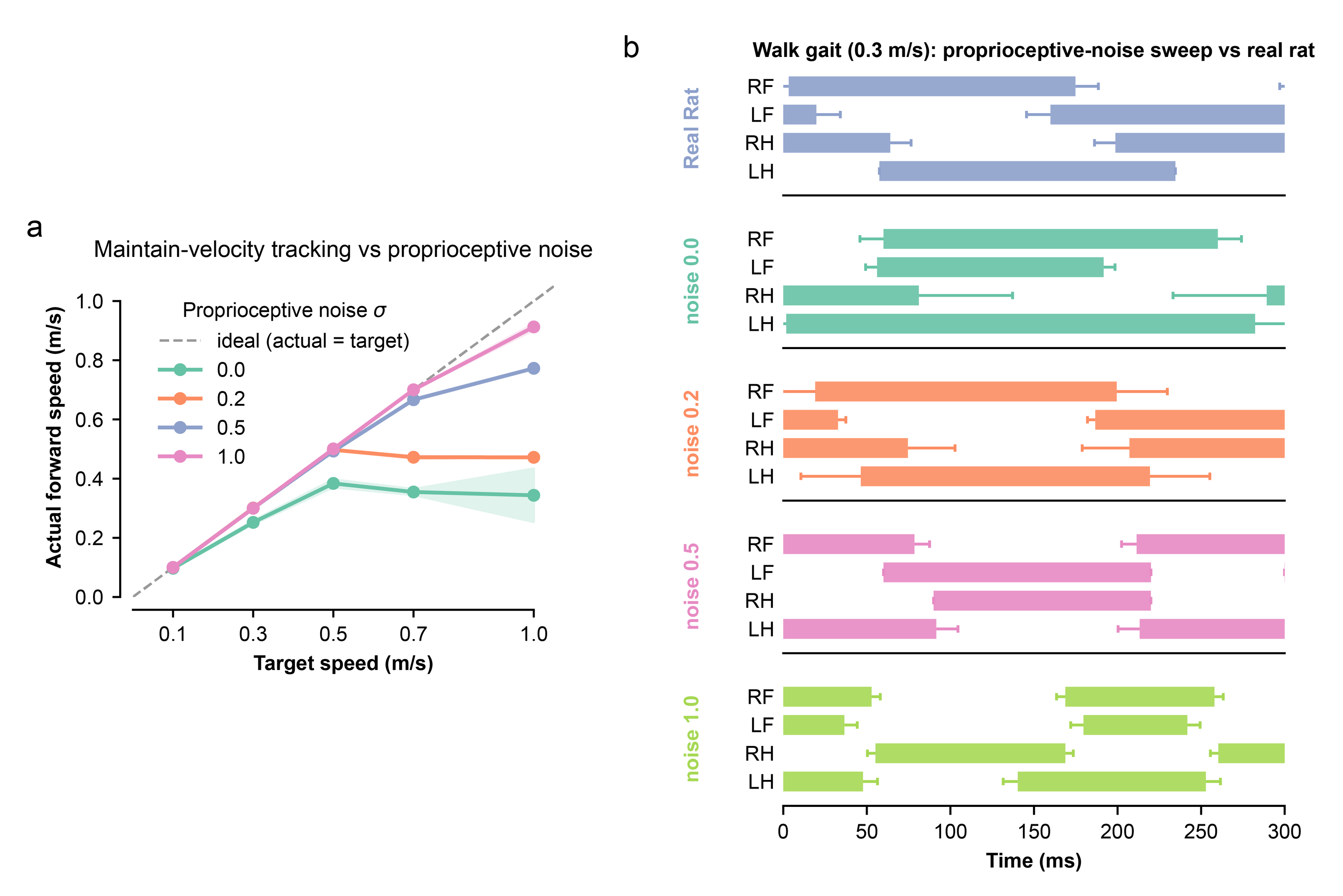}
  \captionof{figure}{Maintain velocity task. \textbf{a}, maintain velocity task performance with different proprioceptive noise level. \textbf{b}, for 0.3 m/s target velocity, the average gait contact pattern for each proprioceptive noise level and the real rat.}
  \label{suppfig:maintain_vel_noise}
\end{center}

\clearpage 
\bmhead{Supplementary Tables}

\begin{center}
  \captionof{table}{Comparison of MIMIC-MJX to other simulation frameworks. Entries reflect native, documented support in the primary workflow.}
  \label{supptable:framework-comparison}

  \vspace{0.4em}
  \scriptsize
  \setlength{\tabcolsep}{2pt}
  \renewcommand{\arraystretch}{1.2}

  \begin{tabularx}{0.99\textwidth}{
      >{\raggedright\arraybackslash}p{2.2cm}
      >{\centering\arraybackslash}p{1.7cm}
      >{\centering\arraybackslash}p{1.2cm}
      >{\centering\arraybackslash}p{1.45cm}
      >{\centering\arraybackslash}p{1.65cm}
      >{\raggedright\arraybackslash\hspace{0pt}}X
    }
    \toprule
    \textbf{Framework} &
    \makecell{\textbf{Model}\\\textbf{registration}} &
    \makecell{\textbf{Parallel}\\\textbf{sim.}} &
    \makecell{\textbf{Deep RL}\\\textbf{workflow}} &
    \makecell{\textbf{Animal}\\\textbf{model}\\\textbf{support}} &
    \textbf{Primary emphasis} \\
    \midrule
    \textbf{\mbox{MIMIC-MJX}} & \yes & \yes & \yes & \yes & Neuromechanical emulation from measured behavior \\
    MyoSuite                  & \no  & \limited & \yes & \no & Musculoskeletal control benchmarks \\
    OpenSim + Moco            & \yes & \limited & \limited & \yes & Musculoskeletal modeling, inverse methods, optimal control \\
    SCONE                     & \no  & \limited & \limited & \yes & Predictive simulation with neuromuscular controllers \\
    FARMS                     & \no  & \limited & \limited & \yes & Modular animal/robot neuromechanical simulation \\
    AnimatLab                 & \no  & \no  & \no  & \yes & Interactive neuromechanical simulation and biological circuits \\
    Isaac Lab                 & \yes  & \yes & \yes & \no & Robot learning with GPU accelerated simulation workflows \\
    \bottomrule
  \end{tabularx}
\end{center}

\begin{center}
  \captionof{table}{Imitation training config parameters for each animal body model.}
  \label{supptable:imitation-training-config}

  \vspace{0.5em} 

  \resizebox{\textwidth}{!}{%
    \begin{tabular}{lccccc}
      \hline
      \textbf{PPO Params} & \textbf{Rat} & \textbf{Fly} & \textbf{Worm} & \textbf{Mouse Arm} & \textbf{Stick Insect} \\
      \hline
      num envs & 4096 & 4096 & 8192 & 4096 & 4096 \\
      batch size & 1024 & 1024 & 1024 & 1024 & 1024 \\
      num minibatches & 16 & 16 & 16 & 8 & 16 \\
      num updates per batch & 4 & 4 & 4 & 4 & 4 \\
      learning rate & 0.0001 & 0.0001 & 0.0001 & 0.0001 & 0.0001 \\
      clipping epsilon & 0.2 & 0.2 & 0.2 & 0.2 & 0.2 \\
      discounting & 0.95 & 0.95 & 0.95 & 0.95 & 0.97 \\
      entropy cost & 0.01 & 0.01 & 0.002 & 0.001 & 0.0001 \\
      unroll length & 20 & 20 & 20 & 20 & 20 \\
      kl weight & 0.1 & 0.1 & 0.0001 & 0.00001 & 0.1 \\
      \hline
      \textbf{Network Params} & & & & & \\
      \hline
      encoder layer sizes & [512,256,256] & [256,256] & [512,512,512] & [512,512,512] & [512,512,512] \\
      decoder layer sizes & [512,512,256,256] & [256,256] & [512,512,512] & [512,512,512] & [512,512,512,256] \\
      critic layer sizes & [512,512,256,256] & [256,256] & [512,512,512] & [512,512,512] & [512,512,256] \\
      intention size & 16 & 60 & 8 & 4 & 60 \\
      \hline
      \textbf{Sim Params} & & & & & \\
      \hline
      sim dt & 0.002 & 0.0002 & 0.01 & 0.00125 & 0.002 \\
      ctrl dt (sim\_dt * steps...) & 0.01 & 0.002 & 0.1 & 0.0025 & 0.01 \\
      solver & CG & CG & CG & CG & CG \\
      iterations & 5 & 4 & 4 & 6 & 6 \\
      ls iterations & 5 & 4 & 4 & 6 & 6 \\
      \hline
      \textbf{Reward Params} & & & & & \\
      \hline
      pos exp scale & 400 & 400 & 0.01 & 0.0 & 40000 \\
      quat exp scale & 4 & 4 & 30 & 0.0 & 4 \\
      joint exp scale & 0.25 & 0.25 & 1 & 0.2 & 0.25 \\
      end eff exp scale & 500 & 100 & 0.01 & 0.0 & 50000 \\
      body pos exp scale & 0.0 & 8 & 0.0 & 0.0 & 8 \\
      joint vel exp scale & 0.0 & 0.5 & 0.0 & 0.0 & 0.5 \\
      pos weight & 1 & 1 & 1 & 0.0 & 1 \\
      quat weight & 1 & 1 & 1 & 0.0 & 1 \\
      joint weight & 1 & 1 & 2 & 5 & 1 \\
      end eff weight & 1 & 1 & 1 & 0.0 & 1 \\
      body pos weight & 0.0 & 0.0 & 0.0 & 0.0 & 0.0 \\
      joint vel weight & 0.0 & 0.0 & 0.0 & 0.0 & 0.0 \\
      control cost & 0.02 & 0.02 & 0.02 & 0.15 & 0.02 \\
      control difference cost & 0.02 & 0.1 & 1.0 & 0.0 & 0.02 \\
      energy cost & 0.01 & 0.001 & 0.0 & 0.01 & 0.001 \\
      \hline
    \end{tabular}
  }
\end{center}

\begin{center}
  \captionof{table}{Breakdown of average wall-clock times of one control timestep for the
    MIMIC-MJX rodent imitation policy. The table is generated with MJX on a single NVIDIA~RTX 5090
    GPU, as a function of the number of parallel environments~$N$. Component times (Policy, RL env,
    MuJoCo) are each timed \emph{separately} and divided by~$N$, in microseconds; ``Component sum''
    adds them (a cross-check, not a measured end-to-end step time). ``Real time sim.''\ is the
    simulated time advanced per step ($\Delta t_{\mathrm{ctrl}}$). $^\dagger$Batched wall-clock per
    step, from a separate scanned run: $8.66$, $16.13$, $33.91$, $111.79$~ms for $N=1$, $256$,
    $1024$, $4096$ --- not the component sum~$\times\,N$. ``1-env real time''~$=\Delta
    t_{\mathrm{ctrl}}/(\text{batched wall per step})$ is the wall-clock speed of a \emph{single}
    environment; ``Throughput'' is the aggregate simulated seconds produced per wall-clock second
    across all $N$ environments. No domain randomization; solver: newton, 5 iterations, 5 substeps
  ($\Delta t_{\mathrm{sim}}=2\,\mathrm{ms}$).}
  \label{tab:steptime-rodent}

  \vspace{0.4em}
  \footnotesize
  \setlength{\tabcolsep}{3pt}
  \renewcommand{\arraystretch}{1.2}

  \begin{tabular}{r rrr r r r r}
    \toprule
    & \multicolumn{3}{c}{Component ($\mu$s/env)} & Component & Real time & 1-env$^\dagger$ & Throughput$^\dagger$ \\
    \cmidrule(lr){2-4}
    $N$ & Policy & RL env & MuJoCo & sum ($\mu$s/env) & sim.\ (ms) & real time & (sim\,s/wall\,s) \\
    \midrule
    1    & 66.520 & 143.130 & 7586.981 & 7796.631 & 10.0 & 115.4\% & 1.2   \\
    256  & 0.489  & 0.598   & 66.465   & 67.552   & 10.0 & 62.0\%  & 158.8 \\
    1024 & 0.252  & 0.146   & 33.318   & 33.717   & 10.0 & 29.5\%  & 302.0 \\
    4096 & 0.185  & 0.040   & 27.433   & 27.658   & 10.0 & 8.9\%   & 366.4 \\
    \bottomrule
  \end{tabular}
\end{center}

\begin{center}
  \captionof{table}{Breakdown of average wall-clock times of one control timestep for the MIMIC-MJX fruit fly (fly) imitation policy simulated with MJX on a single NVIDIA~RTX 5090 GPU, as a function of the number of parallel environments~$N$. Component times (Policy, RL env, MuJoCo) and Total are amortized per environment. ``Real time sim.''\ is the simulated time advanced per step ($\Delta t_{\mathrm{ctrl}}$). ``1-env real time''~$=\Delta t_{\mathrm{ctrl}}/\text{wall-per-step}$ is the wall-clock speed of a single environment; ``Throughput'' is the aggregate simulated seconds produced per wall-clock second across all $N$ environments. no domain randomization; solver: cg, 5 iterations, 10 substeps ($\Delta t_{\mathrm{sim}}=0.2\,\mathrm{ms}$).}
  \label{supptable:fruitfly-step-time}

  \vspace{0.4em}
  \footnotesize
  \setlength{\tabcolsep}{3pt}
  \renewcommand{\arraystretch}{1.2}

  \begin{tabular}{r rrr r r r r}
    \toprule
    & \multicolumn{3}{c}{Component ($\mu$s/env)} & Total & Real time & 1-env & Throughput \\
    \cmidrule(lr){2-4}
    $N$ & Policy & RL env & MuJoCo & ($\mu$s/env) & sim.\ (ms) & real time & (sim\,s/wall\,s) \\
    \midrule
    1    & 78.375 & 143.395 & 8281.323 & 8503.093 & 2.0 & 22.6\% & 0.2   \\
    256  & 0.481  & 0.609   & 76.745   & 77.835   & 2.0 & 10.5\% & 26.8  \\
    1024 & 0.174  & 0.157   & 32.177   & 32.508   & 2.0 & 6.1\%  & 62.4  \\
    4096 & 0.095  & 0.045   & 16.481   & 16.621   & 2.0 & 2.9\%  & 119.4 \\
    \bottomrule
  \end{tabular}
\end{center}

\begin{center}
  \captionof{table}{\textbf{Comparison with \texttt{flybody} Walking imitation.} \texttt{flybody} numbers were taken from Supplementary Table 5 of the paper \cite{Vaxenburg2025-mb}, which was run on a single core of an Intel Xeon~E5-2697~v3 @ 2.60\,GHz using the CPU MuJoCo physics backend and Distributional Maximum a posteriori Policy Optimization (DMPO) as the RL algorithm (a variant of MPO \cite{Abdolmaleki2018-xg}). Walking uses the same $500$\,Hz control / $5000$\,Hz physics ($\Delta t_{\mathrm{ctrl}}=2$\,ms, 10 substeps) as our simulation. MIMIC-MJX rows: NVIDIA RTX~5090, N = number of parallel envs during training. ``Real time'' is the wall-clock speed of a single environment vs.\ the clock ($>100\%$ = faster than real time). ``Throughput'' is the aggregate simulated seconds produced per wall-clock second.
  }
  \label{supptable:flybody-head-to-head}

  \vspace{0.4em}
  \footnotesize
  \setlength{\tabcolsep}{2pt}
  \renewcommand{\arraystretch}{1.2}

  \begin{tabular}{>{\raggedright\arraybackslash}p{2.4cm} rrrr rrr}
    \toprule
    & \multicolumn{4}{c}{Per-control-step time (ms)} & \multicolumn{3}{c}{Real time \& throughput} \\
    \cmidrule(lr){2-5}\cmidrule(lr){6-8}
    Configuration & Policy & RL env & MuJoCo & Total & RT sim & Real & Thrpt. \\
    & & & & & (ms) & time & (sim\,s/wall\,s) \\
    \midrule
    \texttt{flybody} Walking (1 CPU core)    & 4.31   & 4.49    & 5.10   & 13.73  & 2.0 & 14.6\% & 0.15  \\
    \midrule
    MIMIC-MJX, $N{=}1$   & 0.078  & 0.143   & 8.281  & 8.503  & 2.0 & 22.6\% & 0.23  \\
    \bottomrule
  \end{tabular}
\end{center}

\begin{center}
  \captionof{table}{Downstream task training configuration for the high-level policies in the Bowl Escape and Maintain Velocity transfer tasks. Both tasks reuse a frozen MIMIC-MJX rat imitation decoder as the low-level controller (\hyperref[supptable:imitation-training-config]{Supp. Table~\ref*{supptable:imitation-training-config}}) and train only the high-level policy in motor-intention space with PPO.}
  \label{supptable:downstream-task-config}

  \vspace{0.5em} 

  \resizebox{0.7\textwidth}{!}{%
    \begin{tabular}{lcc}
      \hline
      \textbf{PPO Params} & \textbf{Bowl Escape} & \textbf{Maintain Velocity} \\
      \hline
      num timesteps & $3\times10^{8}$ & $3\times10^{8}$ \\
      num envs & 4096 & 4096 \\
      batch size & 1024 & 1024 \\
      num minibatches & 16 & 16 \\
      num updates per batch & 4 & 4 \\
      unroll length & 20 & 20 \\
      episode length & 2000 & 2000 \\
      learning rate & 0.0001 & 0.0001 \\
      clipping epsilon & 0.1 & 0.3 \\
      discounting & 0.98 & 0.99 \\
      entropy cost & 0.0001 & 0.01 \\
      reward scaling & 1 & 1 \\
      action repeat & 1 & 1 \\
      normalize observations & True & True \\
      seed & 0 & 0 \\
      \hline
      \textbf{Network Params} & & \\
      \hline
      high-level policy layer sizes & [512,512,256,256] & [1024,512,256] \\
      value (critic) layer sizes & [512,512,512,256,256] & [1024,512,256] \\
      motor intention size & 16 & 16 \\
      \hline
      \textbf{Sim Params} & & \\
      \hline
      sim dt & 0.002 & 0.002 \\
      ctrl dt & 0.01 & 0.01 \\
      solver & Newton & Newton \\
      iterations & 10 & 5 \\
      ls iterations & 5 & 5 \\
      torque actuators & True & True \\
      rescale factor & 0.9 & 0.9 \\
      \hline
    \end{tabular}
  }
\end{center}

\bmhead{Supplementary Movies}

Available online at: \url{https://mimic-mjx.talmolab.org}

\begin{enumerate}
  \item \textbf{Supp. Movie 1}: The video shows MIMIC-MJX motion capture registration and replay demonstrated across diverse morphologies: rat, fly, mouse arm, worm, and stick insect.
  \item \textbf{Supp. Movie 2}: The video shows transfer learning performance in the rat Bowl Escape task, comparing agents with and without pretrained decoder initialization.
\end{enumerate}




\section{Acknowledgments}
We acknowledge G. Sean Escola, Yuval Tassa and Josh Merel for helpful discussions throughout the development of this project.

This work was supported by a L.I.F.E. Foundation grant to TDP, a Salk Innovation Award to EA and TDP, an NIH BRAIN Initiative grant U01NS136507 to BP\"O, TDP, BWB, EA, a SFARI Investigator Award to BP\"O, an ARNI award to BP\"O and BR, an AFOSR award FA9550-19-1-0386 to BWB, and the Richard \& Joan Komen University Chair to BWB. ETTA was supported by a Swartz Theory Postdoctoral Fellowship, and the AFOSR FA9550-19-1-0386. EW was supported by the Swedish Foundation for International Cooperation in Research and Higher Education (STINT). FP and DL were supported by a Biotechnology and Biological Sciences Research Council award (grant number BB/Y513787/1). Additional support includes National Institutes of Health grant F32 NS120998 and the Salk Pioneer Fund Postdoctoral Scholar Award to AN, the Salk Alumni Fellowship Award and Salk Women \& Science Research Award to AT, National Institutes of Health grant F32 NS126231 and the Salk Alumni Fellowship Award to KWH, and National Institutes of Health grants R01 NS128898 and R01 NS111479 to EA. BAR and AS were supported by the National Science Foundation and DoD OUSD (R\&E) under Cooperative Agreement PHY-2229929 (The NSF AI Institute for Artificial and Natural Intelligence). BAR and AS were also supported by the Natural Sciences and Engineering Research Council of Canada (Discovery Grant: RGPIN-2020-05105; Discovery Accelerator Supplement: RGPAS-2020-00031) and the Canadian Institute for Advanced Research (Canada AI Chair; Learning in Machine and Brains Fellowship). AS was also supported by the FRQNT Master's Scholarship. The research was enabled, in part, by support provided by (Calcul Québec) (https://www.calculquebec.ca/en/), the Digital Research Alliance of Canada (https://alliancecan.ca/en), and NVIDIA, in the form of computational resources. SWF acknowledges funding from the Howard Hughes Medical Institute and the Freedom Together Foundation.

\bmhead{Data Availability}
We have open-sourced all the data used in this work. This includes data that is already publicly available, as well as data collected by our team. Data that was already publicly available has been cited in this work. Model weights are available on Hugging Face at \url{https://huggingface.co/talmolab/MIMIC-MJX} and sample configuration files and datasets are available at \url{https://huggingface.co/datasets/talmolab/MIMIC-MJX}.

\bmhead{Code Availability} Our code is open-sourced and available at \url{https://github.com/talmolab/stac-mjx} and \url{https://github.com/talmolab/track-mjx}. The biomechanical body models and simulation environments are available at \url{https://github.com/talmolab/vnl-playground}. Documentation, tutorials, and usage guides are available on the documentation page at \url{https://mimic-mjx.talmolab.org}.

\begin{appendices}




\end{appendices}


\bibliography{references} 

@ARTICLE{Richards2019-om,
  title     = "A deep learning framework for neuroscience",
  author    = "Richards, Blake A and Lillicrap, Timothy P and Beaudoin, Philippe
               and Bengio, Yoshua and Bogacz, Rafal and Christensen, Amelia and
               Clopath, Claudia and Costa, Rui Ponte and de Berker, Archy and
               Ganguli, Surya and Gillon, Colleen J and Hafner, Danijar and
               Kepecs, Adam and Kriegeskorte, Nikolaus and Latham, Peter and
               Lindsay, Grace W and Miller, Kenneth D and Naud, Richard and
               Pack, Christopher C and Poirazi, Panayiota and Roelfsema, Pieter
               and Sacramento, João and Saxe, Andrew and Scellier, Benjamin and
               Schapiro, Anna C and Senn, Walter and Wayne, Greg and Yamins,
               Daniel and Zenke, Friedemann and Zylberberg, Joel and Therien,
               Denis and Kording, Konrad P",
  journal   = "Nature Neuroscience",
  publisher = "Nature Publishing Group",
  volume    =  22,
  number    =  11,
  pages     = "1761--1770",
  month     =  oct,
  year      =  2019,
  language  = "en"
}

@article{Rudin2021-ma,
  title         = {Learning to walk in minutes using massively parallel deep
                   reinforcement learning},
  author        = {Rudin, Nikita and Hoeller, David and Reist, Philipp and
                   Hutter, Marco},
  journal       = {arXiv [cs.RO]},
  month         = sep,
  year          = 2021,
  archiveprefix = {arXiv},
  primaryclass  = {cs.RO}
}

@article{Abdolmaleki2018-xg,
  title         = {Maximum a Posteriori Policy Optimisation},
  author        = {Abdolmaleki, Abbas and Springenberg, Jost Tobias and Tassa,
                   Yuval and Munos, Remi and Heess, Nicolas and Riedmiller,
                   Martin},
  journal       = {arXiv [cs.LG]},
  month         = jun,
  year          = 2018,
  archiveprefix = {arXiv},
  primaryclass  = {cs.LG}
}

@article{Cook2019-hl,
  title     = {Whole-animal connectomes of both Caenorhabditis elegans sexes},
  author    = {Cook, Steven J and Jarrell, Travis A and Brittin, Christopher A
               and Wang, Yi and Bloniarz, Adam E and Yakovlev, Maksim A and
               Nguyen, Ken C Q and Tang, Leo T-H and Bayer, Emily A and Duerr,
               Janet S and Bülow, Hannes E and Hobert, Oliver and Hall, David H
               and Emmons, Scott W},
  journal   = {Nature},
  publisher = {Springer Science and Business Media LLC},
  volume    = 571,
  number    = 7763,
  pages     = {63--71},
  month     = jul,
  year      = 2019,
  language  = {en}
}

@article{Dorkenwald2024-ju,
  title     = {Neuronal wiring diagram of an adult brain},
  author    = {Dorkenwald, Sven and Matsliah, Arie and Sterling, Amy R and
               Schlegel, Philipp and Yu, Szi-Chieh and McKellar, Claire E and
               Lin, Albert and Costa, Marta and Eichler, Katharina and Yin,
               Yijie and Silversmith, Will and Schneider-Mizell, Casey and
               Jordan, Chris S and Brittain, Derrick and Halageri, Akhilesh and
               Kuehner, Kai and Ogedengbe, Oluwaseun and Morey, Ryan and Gager,
               Jay and Kruk, Krzysztof and Perlman, Eric and Yang, Runzhe and
               Deutsch, David and Bland, Doug and Sorek, Marissa and Lu, Ran and
               Macrina, Thomas and Lee, Kisuk and Bae, J Alexander and Mu, Shang
               and Nehoran, Barak and Mitchell, Eric and Popovych, Sergiy and
               Wu, Jingpeng and Jia, Zhen and Castro, Manuel A and Kemnitz, Nico
               and Ih, Dodam and Bates, Alexander Shakeel and Eckstein, Nils and
               Funke, Jan and Collman, Forrest and Bock, Davi D and Jefferis,
               Gregory S X E and Seung, H Sebastian and Murthy, Mala and
               {FlyWire Consortium}},
  journal   = {Nature},
  publisher = {Springer Science and Business Media LLC},
  volume    = 634,
  number    = 8032,
  pages     = {124--138},
  month     = oct,
  year      = 2024,
  language  = {en}
}

@article{DeWolf2024-xg,
  title    = {Neuro-musculoskeletal modeling reveals muscle-level neural
              dynamics of adaptive learning in sensorimotor cortex},
  author   = {DeWolf, Travis and Schneider, Steffen and Soubiran, Paul and
              Roggenbach, Adrian and Mathis, Mackenzie},
  journal  = {bioRxiv},
  pages    = {2024.09.11.612513},
  month    = sep,
  year     = 2024,
  language = {en}
}

@article{Pugliese2025-qz,
  title    = {Connectome simulations identify a central pattern generator
              circuit for fly walking},
  author   = {Pugliese, Sarah M and Chou, Grant M and Abe, Elliott T T and
              Turcu, Denis and Lancaster, Jackson K and Tuthill, John C and
              Brunton, Bingni W},
  journal  = {bioRxiv},
  pages    = {2025.09.12.675944},
  month    = sep,
  year     = 2025,
  language = {en}
}

@article{Thanawalla2025-zn,
  title       = {Cerebellar outputs for rapid directional refinement of forelimb
                 movement},
  author      = {Thanawalla, Ayesha R and Wilcox, Oren and Rhee, Eliza and
                 Jiang, Juan and Huang, Kee Wui and Yusufi, Raihana and
                 Saklaway, Dalia and Nagamori, Akira and Conner, James M and
                 Chen, Albert I and Azim, Eiman},
  journal     = {Neuroscience},
  institution = {bioRxiv},
  number      = {biorxiv;2025.10.01.679895v1},
  month       = oct,
  year        = 2025
}

@article{von2020phobos,
  title   = {Phobos: A tool for creating complex robot models},
  author  = {von Szadkowski, Kai and Reichel, Simon},
  journal = {Journal of Open Source Software},
  volume  = {5},
  number  = {45},
  pages   = {1326},
  year    = {2020}
}

@manual{blender,
  title        = {Blender: A 3D Modelling and Rendering Package},
  author       = {{Blender Online Community}},
  organization = {Blender Foundation},
  address      = {Amsterdam, Netherlands},
  year         = {2024},
  note         = {Version 4.1},
  url          = {https://www.blender.org}
}

@article{plum2023replicant,
  title     = {replicAnt: a pipeline for generating annotated images of animals in complex environments using Unreal Engine},
  author    = {Plum, Fabian and Bulla, Ren{\'e} and Beck, Hendrik K and Imirzian, Natalie and Labonte, David},
  journal   = {Nature Communications},
  volume    = {14},
  number    = {1},
  pages     = {7195},
  year      = {2023},
  publisher = {Nature Publishing Group UK London}
}

@article{garrido2014automatic,
  title     = {Automatic generation and detection of highly reliable fiducial markers under occlusion},
  author    = {Garrido-Jurado, S and Mu{\~n}oz-Salinas, R and Madrid-Cuevas, FJ and Mar{\'\i}n-Jim{\'e}nez, MJ},
  journal   = {Pattern Recognition},
  volume    = {47},
  number    = {6},
  pages     = {2280--2292},
  year      = {2014},
  publisher = {Elsevier}
}

@article{bradski2000opencv,
  title     = {The opencv library.},
  author    = {Bradski, Gary},
  journal   = {Dr. Dobb's Journal: Software Tools for the Professional Programmer},
  volume    = {25},
  number    = {11},
  pages     = {120--123},
  year      = {2000},
  publisher = {Miller Freeman Inc.}
}

@article{Shenoy2013-jo,
  title     = {Cortical control of arm movements: a dynamical systems
               perspective},
  author    = {Shenoy, Krishna V and Sahani, Maneesh and Churchland, Mark M},
  journal   = {Annu. Rev. Neurosci.},
  publisher = {Annual Reviews},
  volume    = 36,
  number    = 1,
  pages     = {337--359},
  month     = jul,
  year      = 2013,
  language  = {en}
}

@article{Cotton2025-ni,
  title         = {{KinTwin}: Imitation learning with torque and muscle driven
                   biomechanical models enables precise replication of
                   able-bodied and impaired movement from markerless motion
                   capture},
  author        = {Cotton, R James},
  journal       = {arXiv [cs.CV]},
  month         = may,
  year          = 2025,
  archiveprefix = {arXiv},
  primaryclass  = {cs.CV}
}

@misc{Yadan2019Hydra,
  author       = {Omry Yadan},
  title        = {Hydra - A framework for elegantly configuring complex applications},
  howpublished = {Github},
  year         = {2019},
  url          = {https://github.com/facebookresearch/hydra}
}

@article{OpenWorm-Sarma2018-bk,
  title     = {{OpenWorm}: overview and recent advances in integrative
               biological simulation of Caenorhabditis elegans},
  author    = {Sarma, Gopal P and Lee, Chee Wai and Portegys, Tom and Ghayoomie,
               Vahid and Jacobs, Travis and Alicea, Bradly and Cantarelli,
               Matteo and Currie, Michael and Gerkin, Richard C and Gingell,
               Shane and Gleeson, Padraig and Gordon, Richard and Hasani, Ramin
               M and Idili, Giovanni and Khayrulin, Sergey and Lung, David and
               Palyanov, Andrey and Watts, Mark and Larson, Stephen D},
  journal   = {Philos. Trans. R. Soc. Lond. B Biol. Sci.},
  publisher = {The Royal Society},
  volume    = 373,
  number    = 1758,
  pages     = 20170382,
  month     = sep,
  year      = 2018,
  language  = {en}
}

@article{ModWorm-Kim2019-wr,
  title     = {Modular integration of neural connectomics, dynamics and
               biomechanics for identification of behavioral sensorimotor
               pathways in Caenorhabditis elegans},
  author    = {Kim, Jimin and Florman, Jeremy T and Santos, Julia A and Alkema,
               Mark J and Shlizerman, Eli},
  journal   = {bioRxiv},
  publisher = {Cold Spring Harbor Laboratory},
  pages     = 724328,
  year      = 2019
}

@article{BAAIWorm-Zhao2024-wi,
  title     = {An integrative data-driven model simulating {C}. elegans brain,
               body and environment interactions},
  author    = {Zhao, Mengdi and Wang, Ning and Jiang, Xinrui and Ma, Xiaoyang
               and Ma, Haixin and He, Gan and Du, Kai and Ma, Lei and Huang,
               Tiejun},
  journal   = {Nat. Comput. Sci.},
  publisher = {Springer Science and Business Media LLC},
  volume    = 4,
  number    = 12,
  pages     = {978--990},
  month     = dec,
  year      = 2024,
  language  = {en}
}

@article{NeuroSimWorm-Wang2025-yp,
  title     = {{NeuroSimWorm}: A multisensory framework for modeling and
               simulating neural circuits of Caenorhabditis elegans},
  author    = {Wang, Jiaxin and Zhang, Mengxiao and Wang, Kejun and Kang, Lijun
               and Feng, Liang and Pan, Gang and Tang, Huajin},
  journal   = {Neurocomputing},
  publisher = {Elsevier BV},
  volume    = 638,
  number    = 130055,
  pages     = 130055,
  month     = jul,
  year      = 2025,
  language  = {en}
}

@misc{Yi2024-jaxls,
  title        = {jaxls: Nonlinear least squares in JAX},
  author       = {Brent Yi},
  year         = {2024},
  howpublished = {\url{https://github.com/brentyi/jaxls}}
}

@misc{keller_autonomous_2025,
  title     = {Autonomous {Behavior} and {Whole}-{Brain} {Dynamics} {Emerge} in {Embodied} {Zebrafish} {Agents} with {Model}-based {Intrinsic} {Motivation}},
  url       = {http://arxiv.org/abs/2506.00138},
  doi       = {10.48550/arXiv.2506.00138},
  urldate   = {2025-06-06},
  publisher = {arXiv},
  author    = {Keller, Reece and Tornell, Alyn and Pei, Felix and Pitkow, Xaq and Kozachkov, Leo and Nayebi, Aran},
  month     = may,
  year      = {2025},
  note      = {arXiv:2506.00138 [q-bio]}
}

@misc{chiappa_arnold_2025,
  title      = {Arnold: a generalist muscle transformer policy},
  shorttitle = {Arnold},
  url        = {http://arxiv.org/abs/2508.18066},
  doi        = {10.48550/arXiv.2508.18066},
  urldate    = {2025-09-09},
  publisher  = {arXiv},
  author     = {Chiappa, Alberto Silvio and An, Boshi and Simos, Merkourios and Li, Chengkun and Mathis, Alexander},
  month      = aug,
  year       = {2025},
  note       = {arXiv:2508.18066 [cs]}
}

@misc{xu_open-source_2024,
  title     = {Open-{Source} {Reinforcement} {Learning} {Environments} {Implemented} in {MuJoCo} with {Franka} {Manipulator}},
  url       = {http://arxiv.org/abs/2312.13788},
  doi       = {10.48550/arXiv.2312.13788},
  urldate   = {2025-09-12},
  publisher = {arXiv},
  author    = {Xu, Zichun and Li, Yuntao and Yang, Xiaohang and Zhao, Zhiyuan and Zhuang, Lei and Zhao, Jingdong},
  month     = jul,
  year      = {2024},
  note      = {arXiv:2312.13788 [cs]
               version: 3}
}

@misc{akki_benchmarking_2025,
  title     = {Benchmarking {Model} {Predictive} {Control} and {Reinforcement} {Learning} {Based} {Control} for {Legged} {Robot} {Locomotion} in {MuJoCo} {Simulation}},
  url       = {http://arxiv.org/abs/2501.16590},
  doi       = {10.48550/arXiv.2501.16590},
  urldate   = {2025-09-12},
  publisher = {arXiv},
  author    = {Akki, Shivayogi and Chen, Tan},
  month     = jan,
  year      = {2025},
  note      = {arXiv:2501.16590 [cs]}
}

@misc{henaff_scalable_2025,
  title     = {Scalable {Option} {Learning} in {High}-{Throughput} {Environments}},
  url       = {http://arxiv.org/abs/2509.00338},
  doi       = {10.48550/arXiv.2509.00338},
  urldate   = {2025-09-12},
  publisher = {arXiv},
  author    = {Henaff, Mikael and Fujimoto, Scott and Rabbat, Michael},
  month     = aug,
  year      = {2025},
  note      = {arXiv:2509.00338 [cs]}
}

@incollection{lecun_efficient_1998,
  address   = {Berlin, Heidelberg},
  title     = {Efficient {BackProp}},
  isbn      = {978-3-540-49430-0},
  url       = {https://doi.org/10.1007/3-540-49430-8_2},
  language  = {en},
  urldate   = {2025-09-19},
  booktitle = {Neural {Networks}: {Tricks} of the {Trade}},
  publisher = {Springer},
  author    = {LeCun, Yann and Bottou, Leon and Orr, Genevieve B. and Müller, Klaus -Robert},
  editor    = {Orr, Genevieve B. and Müller, Klaus-Robert},
  year      = {1998},
  doi       = {10.1007/3-540-49430-8_2},
  pages     = {9--50}
}

@misc{elfwing_sigmoid-weighted_2017,
  title     = {Sigmoid-{Weighted} {Linear} {Units} for {Neural} {Network} {Function} {Approximation} in {Reinforcement} {Learning}},
  url       = {http://arxiv.org/abs/1702.03118},
  doi       = {10.48550/arXiv.1702.03118},
  urldate   = {2025-09-19},
  publisher = {arXiv},
  author    = {Elfwing, Stefan and Uchibe, Eiji and Doya, Kenji},
  month     = nov,
  year      = {2017},
  note      = {arXiv:1702.03118 [cs]}
}

@misc{ba_layer_2016,
  title     = {Layer {Normalization}},
  url       = {http://arxiv.org/abs/1607.06450},
  doi       = {10.48550/arXiv.1607.06450},
  urldate   = {2025-09-19},
  publisher = {arXiv},
  author    = {Ba, Jimmy Lei and Kiros, Jamie Ryan and Hinton, Geoffrey E.},
  month     = jul,
  year      = {2016},
  note      = {arXiv:1607.06450 [stat]}
}

@misc{kingma_auto-encoding_2022,
  title     = {Auto-{Encoding} {Variational} {Bayes}},
  url       = {http://arxiv.org/abs/1312.6114},
  doi       = {10.48550/arXiv.1312.6114},
  urldate   = {2025-09-19},
  publisher = {arXiv},
  author    = {Kingma, Diederik P. and Welling, Max},
  month     = dec,
  year      = {2022},
  note      = {arXiv:1312.6114 [stat]}
}

@misc{luo_universal_2024,
  title     = {Universal {Humanoid} {Motion} {Representations} for {Physics}-{Based} {Control}},
  url       = {http://arxiv.org/abs/2310.04582},
  doi       = {10.48550/arXiv.2310.04582},
  urldate   = {2025-05-30},
  publisher = {arXiv},
  author    = {Luo, Zhengyi and Cao, Jinkun and Merel, Josh and Winkler, Alexander and Huang, Jing and Kitani, Kris and Xu, Weipeng},
  month     = apr,
  year      = {2024},
  note      = {arXiv:2310.04582 [cs]}
}

@article{Chung2023-lg,
  title    = {{ElegansBot}: Development of equation of motion deciphering
              locomotion including omega turns of Caenorhabditis elegans},
  author   = {Chung, Taegon and Chang, Iksoo and Kim, Sangyeol},
  journal  = {eLife},
  month    = dec,
  year     = 2023,
  language = {en}
}

@article{Chen2023-zs,
  title     = {A connectome-based digital twin Caenorhabditis elegans capable of
               intelligent sensorimotor behavior},
  author    = {Chen, Zhongyu and Yu, Yuguo and Xue, Xiangyang},
  journal   = {Mathematics},
  publisher = {MDPI AG},
  volume    = 11,
  number    = 11,
  pages     = 2442,
  month     = may,
  year      = 2023,
  language  = {en}
}

@misc{Biewald2020-mg,
  title  = {Experiment Tracking with Weights and Biases},
  author = {Biewald, Lukas},
  year   = 2020
}

@article{Karashchuk2021-ys,
  title     = {Anipose: A toolkit for robust markerless {3D} pose estimation},
  author    = {Karashchuk, Pierre and Rupp, Katie L and Dickinson, Evyn S and
               Walling-Bell, Sarah and Sanders, Elischa and Azim, Eiman and
               Brunton, Bingni W and Tuthill, John C},
  journal   = {Cell Rep.},
  publisher = {Elsevier BV},
  volume    = 36,
  number    = 13,
  pages     = 109730,
  month     = sep,
  year      = 2021,
  language  = {en}
}

@article{Dickinson2000-xc,
  title     = {How animals move: an integrative view},
  author    = {Dickinson, M H and Farley, C T and Full, R J and Koehl, M A and
               Kram, R and Lehman, S},
  journal   = {Science},
  publisher = {American Association for the Advancement of Science (AAAS)},
  volume    = 288,
  number    = 5463,
  pages     = {100--106},
  month     = apr,
  year      = 2000,
  language  = {en}
}

@article{Ramalingasetty2021-rq,
  title     = {A whole-body musculoskeletal model of the mouse},
  author    = {Ramalingasetty, Shravan Tata and Danner, Simon M and Arreguit,
               Jonathan and Markin, Sergey N and Rodarie, Dimitri and Kathe,
               Claudia and Courtine, Grégoire and Rybak, Ilya A and Ijspeert,
               Auke Jan},
  journal   = {IEEE Access},
  publisher = {Institute of Electrical and Electronics Engineers (IEEE)},
  volume    = 9,
  pages     = {163861--163881},
  month     = dec,
  year      = 2021,
  language  = {en}
}

@article{Gilmer2025-jj,
  title     = {A novel biomechanical model of the proximal mouse forelimb
               predicts muscle activity in optimal control simulations of
               reaching movements},
  author    = {Gilmer, Jesse I and Coltman, Susan K and Cuenu, Geraldine and
               Hutchinson, John R and Huber, Daniel and Person, Abigail L and Al
               Borno, Mazen},
  journal   = {J. Neurophysiol.},
  publisher = {American Physiological Society Rockville, MD},
  volume    = 133,
  number    = 4,
  pages     = {1266--1278},
  month     = apr,
  year      = 2025,
  language  = {en}
}

@article{Pratt2024-cd,
  title     = {Miniature linear and split-belt treadmills reveal mechanisms of
               adaptive motor control in walking Drosophila},
  author    = {Pratt, Brandon G and Lee, Su-Yee J and Chou, Grant M and Tuthill,
               John C},
  journal   = {Curr. Biol.},
  publisher = {Elsevier BV},
  volume    = 34,
  number    = 19,
  pages     = {4368--4381.e5},
  month     = oct,
  year      = 2024,
  language  = {en}
}

@misc{MuJoCo-XLA-AuthorsUnknown-il,
  title        = {{MuJoCo} {XLA} ({MJX})},
  author       = {{MuJoCo XLA Authors}},
  howpublished = {\url{https://mujoco.readthedocs.io/en/stable/mjx.html}}
}

@misc{Bradbury2018-ge,
  title   = {{JAX}: composable transformations of {Python+NumPy} programs},
  author  = {Bradbury, James and Frostig, Roy and Hawkins, Peter and Johnson,
             Matthew James and Leary, Chris and Maclaurin, Dougal and Necula,
             George and Paszke, Adam and VanderPlas, Jake and Wanderman-Milne,
             Skye and Zhang, Qiao},
  year    = 2018,
  version = {0.3.13}
}

@article{Freeman2021-fb,
  title         = {Brax -- A differentiable physics engine for large scale rigid
                   body simulation},
  author        = {Freeman, C Daniel and Frey, Erik and Raichuk, Anton and
                   Girgin, Sertan and Mordatch, Igor and Bachem, Olivier},
  journal       = {arXiv [cs.RO]},
  month         = jun,
  year          = 2021,
  archiveprefix = {arXiv},
  primaryclass  = {cs.RO}
}

@article{Wang-Chen2024-uc,
  title     = {{NeuroMechFly} {v2}: simulating embodied sensorimotor control in
               adult Drosophila},
  author    = {Wang-Chen, Sibo and Stimpfling, Victor Alfred and Lam, Thomas Ka
               Chung and Özdil, Pembe Gizem and Genoud, Louise and Hurtak, Femke
               and Ramdya, Pavan},
  journal   = {Nat. Methods},
  publisher = {Springer Science and Business Media LLC},
  volume    = 21,
  number    = 12,
  pages     = {2353--2362},
  month     = dec,
  year      = 2024,
  language  = {en}
}

@article{Zador2023-oj,
  title     = {Catalyzing next-generation Artificial Intelligence through
               {NeuroAI}},
  author    = {Zador, Anthony and Escola, Sean and Richards, Blake and Ölveczky,
               Bence and Bengio, Yoshua and Boahen, Kwabena and Botvinick,
               Matthew and Chklovskii, Dmitri and Churchland, Anne and Clopath,
               Claudia and DiCarlo, James and Ganguli, Surya and Hawkins, Jeff
               and Körding, Konrad and Koulakov, Alexei and LeCun, Yann and
               Lillicrap, Timothy and Marblestone, Adam and Olshausen, Bruno and
               Pouget, Alexandre and Savin, Cristina and Sejnowski, Terrence and
               Simoncelli, Eero and Solla, Sara and Sussillo, David and Tolias,
               Andreas S and Tsao, Doris},
  journal   = {Nat. Commun.},
  publisher = {Springer Science and Business Media LLC},
  volume    = 14,
  number    = 1,
  pages     = 1597,
  month     = mar,
  year      = 2023,
  language  = {en}
}

@article{Wu2013-ci,
  title     = {{STAC}: Simultaneous tracking and calibration},
  author    = {Wu, Tingfan and Tassa, Yuval and Kumar, Vikash and Movellan,
               Javier and Todorov, Emanuel},
  journal   = {2013 13th IEEE-RAS International Conference on Humanoid Robots
               (Humanoids)},
  publisher = {IEEE},
  pages     = {469--476},
  month     = oct,
  year      = 2013
}

@article{Peng2018-di,
  title         = {{DeepMimic}: Example-guided deep reinforcement learning of
                   physics-based character skills},
  author        = {Peng, Xue Bin and Abbeel, Pieter and Levine, Sergey and van
                   de Panne, Michiel},
  journal       = {arXiv [cs.GR]},
  month         = apr,
  year          = 2018,
  archiveprefix = {arXiv},
  primaryclass  = {cs.GR}
}

@article{Merel2018-my,
  title         = {Neural probabilistic motor primitives for humanoid control},
  author        = {Merel, Josh and Hasenclever, Leonard and Galashov, Alexandre
                   and Ahuja, Arun and Pham, Vu and Wayne, Greg and Teh, Yee
                   Whye and Heess, Nicolas},
  journal       = {arXiv [cs.LG]},
  month         = nov,
  year          = 2018,
  archiveprefix = {arXiv},
  primaryclass  = {cs.LG}
}

@article{Merel2019-mh,
  title         = {Deep neuroethology of a virtual rodent},
  author        = {Merel, Josh and Aldarondo, Diego and Marshall, Jesse and
                   Tassa, Yuval and Wayne, Greg and Ölveczky, Bence},
  journal       = {arXiv [q-bio.NC]},
  month         = nov,
  year          = 2019,
  archiveprefix = {arXiv},
  primaryclass  = {q-bio.NC}
}

@article{Vaxenburg2025-mb,
  title     = {Whole-body physics simulation of fruit fly locomotion},
  author    = {Vaxenburg, Roman and Siwanowicz, Igor and Merel, Josh and Robie,
               Alice A and Morrow, Carmen and Novati, Guido and Stefanidi,
               Zinovia and Both, Gert-Jan and Card, Gwyneth M and Reiser,
               Michael B and Botvinick, Matthew M and Branson, Kristin M and
               Tassa, Yuval and Turaga, Srinivas C},
  journal   = {Nature},
  publisher = {Springer Science and Business Media LLC},
  volume    = 643,
  number    = 8074,
  pages     = {1312--1320},
  month     = jul,
  year      = 2025,
  language  = {en}
}

@article{Mathis2018-vm,
  title     = {{DeepLabCut}: markerless pose estimation of user-defined body
               parts with deep learning},
  author    = {Mathis, Alexander and Mamidanna, Pranav and Cury, Kevin M and
               Abe, Taiga and Murthy, Venkatesh N and Mathis, Mackenzie Weygandt
               and Bethge, Matthias},
  journal   = {Nat. Neurosci.},
  publisher = {Springer Science and Business Media LLC},
  volume    = 21,
  number    = 9,
  pages     = {1281--1289},
  month     = sep,
  year      = 2018,
  language  = {en}
}

@article{Dunn2021-zc,
  title     = {Geometric deep learning enables {3D} kinematic profiling across
               species and environments},
  author    = {Dunn, Timothy W and Marshall, Jesse D and Severson, Kyle S and
               Aldarondo, Diego E and Hildebrand, David G C and Chettih, Selmaan
               N and Wang, William L and Gellis, Amanda J and Carlson, David E
               and Aronov, Dmitriy and Freiwald, Winrich A and Wang, Fan and
               Ölveczky, Bence P},
  journal   = {Nat. Methods},
  publisher = {Springer Science and Business Media LLC},
  volume    = 18,
  number    = 5,
  pages     = {564--573},
  month     = may,
  year      = 2021,
  language  = {en}
}

@article{Pereira2022-mo,
  title     = {{SLEAP}: A deep learning system for multi-animal pose tracking},
  author    = {Pereira, Talmo D and Tabris, Nathaniel and Matsliah, Arie and
               Turner, David M and Li, Junyu and Ravindranath, Shruthi and
               Papadoyannis, Eleni S and Normand, Edna and Deutsch, David S and
               Wang, Z Yan and McKenzie-Smith, Grace C and Mitelut, Catalin C
               and Castro, Marielisa Diez and D'Uva, John and Kislin, Mikhail
               and Sanes, Dan H and Kocher, Sarah D and Wang, Samuel S-H and
               Falkner, Annegret L and Shaevitz, Joshua W and Murthy, Mala},
  journal   = {Nat. Methods},
  publisher = {Springer Science and Business Media LLC},
  volume    = 19,
  number    = 4,
  pages     = {486--495},
  month     = apr,
  year      = 2022,
  language  = {en}
}

@article{Schulman2017-rt,
  title         = {Proximal Policy Optimization Algorithms},
  author        = {Schulman, John and Wolski, Filip and Dhariwal, Prafulla and
                   Radford, Alec and Klimov, Oleg},
  journal       = {arXiv [cs.LG]},
  month         = jul,
  year          = 2017,
  archiveprefix = {arXiv},
  primaryclass  = {cs.LG}
}

@article{Lobato-Rios2022-om,
  title     = {{NeuroMechFly}, a neuromechanical model of adult Drosophila
               melanogaster},
  author    = {Lobato-Rios, Victor and Ramalingasetty, Shravan Tata and Özdil,
               Pembe Gizem and Arreguit, Jonathan and Ijspeert, Auke Jan and
               Ramdya, Pavan},
  journal   = {Nat. Methods},
  publisher = {Springer Science and Business Media LLC},
  volume    = 19,
  number    = 5,
  pages     = {620--627},
  month     = may,
  year      = 2022,
  language  = {en}
}

@article{Aldarondo2024-nc,
  title     = {A virtual rodent predicts the structure of neural activity across
               behaviours},
  author    = {Aldarondo, Diego and Merel, Josh and Marshall, Jesse D and
               Hasenclever, Leonard and Klibaite, Ugne and Gellis, Amanda and
               Tassa, Yuval and Wayne, Greg and Botvinick, Matthew and Ölveczky,
               Bence P},
  journal   = {Nature},
  publisher = {Springer Science and Business Media LLC},
  volume    = 632,
  number    = 8025,
  pages     = {594--602},
  month     = aug,
  year      = 2024,
  language  = {en}
}

@article{DeAngelis2019-fa,
  title     = {The manifold structure of limb coordination in walking Drosophila},
  author    = {DeAngelis, Brian D and Zavatone-Veth, Jacob A and Clark, Damon A},
  journal   = {Elife},
  publisher = {eLife Sciences Publications, Ltd},
  volume    = 8,
  month     = jun,
  year      = 2019,
  language  = {en}
}

@article{Anderson2014-cq,
  title    = {Toward a science of computational ethology},
  author   = {Anderson, David J and Perona, Pietro},
  journal  = {Neuron},
  volume   = 84,
  number   = 1,
  pages    = {18--31},
  month    = oct,
  year     = 2014,
  language = {en}
}

@article{Datta2019-qs,
  title    = {Computational Neuroethology: A Call to Action},
  author   = {Datta, Sandeep Robert and Anderson, David J and Branson, Kristin
              and Perona, Pietro and Leifer, Andrew},
  journal  = {Neuron},
  volume   = 104,
  number   = 1,
  pages    = {11--24},
  month    = oct,
  year     = 2019,
  language = {en}
}

@article{Pereira2020-zz,
  title     = {Quantifying behavior to understand the brain},
  author    = {Pereira, Talmo D and Shaevitz, Joshua W and Murthy, Mala},
  journal   = {Nat. Neurosci.},
  publisher = {Nature Publishing Group},
  pages     = {1--13},
  month     = nov,
  year      = 2020,
  language  = {en}
}

@article{Pereira2019-np,
  title    = {Fast animal pose estimation using deep neural networks},
  author   = {Pereira, Talmo D and Aldarondo, Diego E and Willmore, Lindsay and
              Kislin, Mikhail and Wang, Samuel S-H and Murthy, Mala and
              Shaevitz, Joshua W},
  journal  = {Nat. Methods},
  volume   = 16,
  number   = 1,
  pages    = {117--125},
  month    = jan,
  year     = 2019,
  language = {en}
}

@article{Biderman2024-hp,
  title     = {Lightning Pose: improved animal pose estimation via
               semi-supervised learning, Bayesian ensembling and cloud-native
               open-source tools},
  author    = {Biderman, Dan and Whiteway, Matthew R and Hurwitz, Cole and
               Greenspan, Nicholas and Lee, Robert S and Vishnubhotla, Ankit and
               Warren, Richard and Pedraja, Federico and Noone, Dillon and
               Schartner, Michael M and Huntenburg, Julia M and Khanal, Anup and
               Meijer, Guido T and Noel, Jean-Paul and Pan-Vazquez, Alejandro
               and Socha, Karolina Z and Urai, Anne E and Abbot, Larry and
               Acerbi, Luigi and Aguillon-Rodriguez, Valeria and Ahmadi, Mandana
               and Amjad, Jaweria and Angelaki, Dora and Arlandis, Jaime and
               Ashwood, Zoe C and Banga, Kush and Barrell, Hailey and Bayer,
               Hannah M and Benson, Brandon and Benson, Julius and Bhagat, Jai
               and Birman, Dan and Bonacchi, Niccolò and Bougrova, Kcenia and
               Boussard, Julien and Bruijns, Sebastian A and Buchanan, E Kelly
               and Campbell, Robert and Carandini, Matteo and Catarino, Joana A
               and Cazettes, Fanny and Chapuis, Gaelle A and Churchland, Anne K
               and Dan, Yang and Davatolhagh, Felicia and Dayan, Peter and
               Denève, Sophie and DeWitt, Eric E J and Dong, Ling Liang and
               Engel, Tatiana and Fabbri, Michele and Faulkner, Mayo and Fetcho,
               Robert and Fiete, Ila and Findling, Charles and Freitas-Silva,
               Laura and Ganguli, Surya and Gercek, Berk and Ghani, Naureen and
               Gordeliy, Ivan and Haetzel, Laura M and Harris, Kenneth D and
               Hausser, Michael and Hiratani, Naoki and Hofer, Sonja and Hu, Fei
               and Huber, Felix and Hurwitz, Cole and Khanal, Anup and Krasniak,
               Christopher S and Krishnagopal, Sanjukta and Krumin, Michael and
               Kundu, Debottam and Landemard, Agnès and Langdon, Christopher and
               Langfield, Christopher and Laranjeira, Inês and Latham, Peter and
               Lau, Petrina and Lee, Hyun Dong and Liu, Ari and Mainen, Zachary
               F and Makri-Cottington, Amalia and Martinez-Vergara, Hernando and
               McMannon, Brenna and McRoberts, Isaiah and Meijer, Guido T and
               Melin, Maxwell and Meshulam, Leenoy and Miller, Kim and Miska,
               Nathaniel J and Mitelut, Catalin and Mohammadi, Zeinab and
               Mrsic-Flogel, Thomas and Murakami, Masayoshi and Noel, Jean-Paul
               and Nylund, Kai and Oloomi, Farideh and Pan-Vazquez, Alejandro
               and Paninski, Liam and Pezzotta, Alberto and Picard, Samuel and
               Pillow, Jonathan W and Pouget, Alexandre and Rau, Florian and
               Rossant, Cyrille and Roth, Noam and Roy, Nicholas A and Saniee,
               Kamron and Schaeffer, Rylan and Schartner, Michael M and Shi,
               Yanliang and Soares, Carolina and Socha, Karolina Z and Soitu,
               Cristian and Steinmetz, Nicholas A and Svoboda, Karel and Taheri,
               Marsa and Tessereau, Charline and Urai, Anne E and Varol, Erdem
               and Wells, Miles J and West, Steven J and Whiteway, Matthew R and
               Windolf, Charles and Winter, Olivier and Witten, Ilana and Wool,
               Lauren E and Xu, Zekai and Yu, Han and Zador, Anthony M and
               Zhang, Yizi and Cunningham, John P and Sawtell, Nathaniel B and
               Paninski, Liam and {The International Brain Laboratory}},
  journal   = {Nat. Methods},
  publisher = {Springer Science and Business Media LLC},
  pages     = {1--13},
  month     = jun,
  year      = 2024,
  language  = {en}
}

@book{Bernstein1967-ez,
  title     = {The coordination and regulation of movements},
  author    = {Bernstein, N},
  publisher = {Pergamon Press},
  address   = {Oxford, England},
  year      = 1967
}

@article{Doerig2023-lw,
  title    = {The neuroconnectionist research programme},
  author   = {Doerig, Adrien and Sommers, Rowan P and Seeliger, Katja and
              Richards, Blake and Ismael, Jenann and Lindsay, Grace W and
              Kording, Konrad P and Konkle, Talia and van Gerven, Marcel A J and
              Kriegeskorte, Nikolaus and Kietzmann, Tim C},
  journal  = {Nat. Rev. Neurosci.},
  volume   = 24,
  number   = 7,
  pages    = {431--450},
  month    = jul,
  year     = 2023,
  language = {en}
}

@article{Melis2024-pn,
  title     = {Machine learning reveals the control mechanics of an insect wing
               hinge},
  author    = {Melis, Johan M and Siwanowicz, Igor and Dickinson, Michael H},
  journal   = {Nature},
  publisher = {Nature Publishing Group},
  volume    = 628,
  number    = 8009,
  pages     = {795--803},
  month     = apr,
  year      = 2024,
  language  = {en}
}

@article{Ohana2025-hi,
  title     = {The well: a large-scale collection of diverse physics simulations
               for machine learning},
  author    = {Ohana, Ruben and McCabe, Michael and Meyer, Lucas and Morel, Rudy
               and Agocs, Fruzsina J and Beneitez, Miguel and Berger, Marsha and
               Burkhart, Blakesley and Dalziel, Stuart B and Fielding, Drummond
               B and Fortunato, Daniel and Goldberg, Jared A and Hirashima,
               Keiya and Jiang, Yan-Fei and Kerswell, Rich R and Maddu,
               Suryanarayana and Miller, Jonah and Mukhopadhyay, Payel and
               Nixon, Stefan S and Shen, Jeff and Watteaux, Romain and Blancard,
               Bruno Régaldo-Saint and Rozet, François and Parker, Liam H and
               Cranmer, Miles and Ho, Shirley},
  journal   = {Proceedings of the 38th International Conference on Neural
               Information Processing Systems},
  publisher = {Curran Associates Inc.},
  address   = {Red Hook, NY, USA},
  volume    = 37,
  number    = {Article 1430},
  pages     = {44989--45037},
  series    = {NIPS '24},
  month     = jun,
  year      = 2025,
  language  = {en}
}

@article{Lappalainen2024-oa,
  title     = {Connectome-constrained networks predict neural activity across
               the fly visual system},
  author    = {Lappalainen, Janne K and Tschopp, Fabian D and Prakhya, Sridhama
               and McGill, Mason and Nern, Aljoscha and Shinomiya, Kazunori and
               Takemura, Shin-Ya and Gruntman, Eyal and Macke, Jakob H and
               Turaga, Srinivas C},
  journal   = {Nature},
  publisher = {Springer Science and Business Media LLC},
  volume    = 634,
  number    = 8036,
  pages     = {1132--1140},
  month     = oct,
  year      = 2024,
  language  = {en}
}

@article{Plum2021-qo,
  title     = {{scAnt}-an open-source platform for the creation of {3D} models
               of arthropods (and other small objects)},
  author    = {Plum, Fabian and Labonte, David},
  journal   = {PeerJ},
  publisher = {PeerJ},
  volume    = 9,
  number    = {e11155},
  pages     = {e11155},
  month     = apr,
  year      = 2021,
  language  = {en}
}

@article{Bolanos2021-rr,
  title     = {A three-dimensional virtual mouse generates synthetic training
               data for behavioral analysis},
  author    = {Bolaños, Luis A and Xiao, Dongsheng and Ford, Nancy L and LeDue,
               Jeff M and Gupta, Pankaj K and Doebeli, Carlos and Hu, Hao and
               Rhodin, Helge and Murphy, Timothy H},
  journal   = {Nat. Methods},
  publisher = {Nature Publishing Group},
  pages     = {1--4},
  month     = apr,
  year      = 2021,
  language  = {en}
}

@article{Grillner2006-bc,
  title     = {Biological pattern generation: the cellular and computational
               logic of networks in motion},
  author    = {Grillner, Sten},
  journal   = {Neuron},
  publisher = {Elsevier BV},
  volume    = 52,
  number    = 5,
  pages     = {751--766},
  month     = dec,
  year      = 2006,
  language  = {en}
}

@article{Robinson1968-rl,
  title     = {The oculomotor control system: A review},
  author    = {Robinson, D A},
  journal   = {Proc. IEEE Inst. Electr. Electron. Eng.},
  publisher = {Institute of Electrical and Electronics Engineers (IEEE)},
  volume    = 56,
  number    = 6,
  pages     = {1032--1049},
  year      = 1968
}

@article{Marder2007-td,
  title     = {Understanding circuit dynamics using the stomatogastric nervous
               system of lobsters and crabs},
  author    = {Marder, Eve and Bucher, Dirk},
  journal   = {Annu. Rev. Physiol.},
  publisher = {Annual Reviews},
  volume    = 69,
  number    = 1,
  pages     = {291--316},
  month     = feb,
  year      = 2007,
  language  = {en}
}

@article{Kristan2005-zj,
  title     = {Neuronal control of leech behavior},
  author    = {Kristan, Jr, William B and Calabrese, Ronald L and Friesen, W
               Otto},
  journal   = {Prog. Neurobiol.},
  publisher = {Elsevier BV},
  volume    = 76,
  number    = 5,
  pages     = {279--327},
  month     = aug,
  year      = 2005
}

@article{Ulutas2025-ig,
  title     = {High-resolution in vivo kinematic tracking with customized
               injectable fluorescent nanoparticles},
  author    = {Ulutas, Emine Zeynep and Pradhan, Amartya and Koveal, Dorothy and
               Markowitz, Jeffrey E},
  journal   = {Sci. Adv.},
  publisher = {American Association for the Advancement of Science (AAAS)},
  volume    = 11,
  number    = 40,
  pages     = {eadu9136},
  month     = oct,
  year      = 2025,
  language  = {en}
}

@article{Monsees2022-nh,
  title     = {Estimation of skeletal kinematics in freely moving rodents},
  author    = {Monsees, Arne and Voit, Kay-Michael and Wallace, Damian J and
               Sawinski, Juergen and Charyasz, Edyta and Scheffler, Klaus and
               Macke, Jakob H and Kerr, Jason N D},
  journal   = {Nat. Methods},
  publisher = {Springer Science and Business Media LLC},
  volume    = 19,
  number    = 11,
  pages     = {1500--1509},
  month     = nov,
  year      = 2022,
  language  = {en}
}

@article{Eyjolfsdottir2016-ua,
  title         = {Learning recurrent representations for hierarchical behavior
                   modeling},
  author        = {Eyjolfsdottir, Eyrun and Branson, Kristin and Yue, Yisong and
                   Perona, Pietro},
  journal       = {arXiv [cs.AI]},
  month         = oct,
  year          = 2016,
  archiveprefix = {arXiv},
  primaryclass  = {cs.AI}
}

@article{Pandarinath2018-ua,
  title    = {Inferring single-trial neural population dynamics using sequential
              auto-encoders},
  author   = {Pandarinath, Chethan and O'Shea, Daniel J and Collins, Jasmine and
              Jozefowicz, Rafal and Stavisky, Sergey D and Kao, Jonathan C and
              Trautmann, Eric M and Kaufman, Matthew T and Ryu, Stephen I and
              Hochberg, Leigh R and Henderson, Jaimie M and Shenoy, Krishna V
              and Abbott, L F and Sussillo, David},
  journal  = {Nat. Methods},
  volume   = 15,
  number   = 10,
  pages    = {805--815},
  month    = oct,
  year     = 2018
}

@article{Keshtkaran2022-en,
  title     = {A large-scale neural network training framework for generalized
               estimation of single-trial population dynamics},
  author    = {Keshtkaran, Mohammad Reza and Sedler, Andrew R and Chowdhury,
               Raeed H and Tandon, Raghav and Basrai, Diya and Nguyen, Sarah L
               and Sohn, Hansem and Jazayeri, Mehrdad and Miller, Lee E and
               Pandarinath, Chethan},
  journal   = {Nat. Methods},
  publisher = {Springer Science and Business Media LLC},
  volume    = 19,
  number    = 12,
  pages     = {1572--1577},
  month     = dec,
  year      = 2022,
  language  = {en}
}

@article{Atanas2023,
  author    = {Atanas, Adam A.
               and Kim, Jungsoo
               and Wang, Ziyu
               and Bueno, Eric
               and Becker, McCoy
               and Kang, Di
               and Park, Jungyeon
               and Kramer, Talya S.
               and Wan, Flossie K.
               and Baskoylu, Saba
               and Dag, Ugur
               and Kalogeropoulou, Elpiniki
               and Gomes, Matthew A.
               and Estrem, Cassi
               and Cohen, Netta
               and Mansinghka, Vikash K.
               and Flavell, Steven W.},
  title     = {Brain-wide representations of behavior spanning multiple timescales and states in <em>C.{\&}{\#}xa0;elegans</em>},
  journal   = {Cell},
  year      = {2023},
  month     = {Sep},
  day       = {14},
  publisher = {Elsevier},
  volume    = {186},
  number    = {19},
  pages     = {4134-4151.e31},
  issn      = {0092-8674},
  doi       = {10.1016/j.cell.2023.07.035},
  url       = {https://doi.org/10.1016/j.cell.2023.07.035}
}

@article{danner_spinal_2023,
  title    = {Spinal control of locomotion before and after spinal cord injury},
  doi      = {10.1101/2023.03.22.533794},
  author   = {Danner, Simon and Shepard, Courtney and Hainline, Casey and Shevtsova, Natalia and Rybak, Ilya and Magnuson, David},
  journal  = {bioRxiv},
  month    = jun,
  year     = {2023}
}

@article{caillet2025hill,
  title={Hill-type models of skeletal muscle and neuromuscular actuators: a systematic review},
  author={Caillet, Arnault H and Phillips, Andrew TM and Carty, Christopher and Farina, Dario and Modenese, Luca},
  journal={IEEE Reviews in Biomedical Engineering},
  year={2025},
  publisher={IEEE}
}

@article{kynkaanniemi2019improved,
  title={Improved precision and recall metric for assessing generative models},
  author={Kynk{\"a}{\"a}nniemi, Tuomas and Karras, Tero and Laine, Samuli and Lehtinen, Jaakko and Aila, Timo},
  journal={Advances in neural information processing systems},
  volume={32},
  year={2019}
}

\pagebreak

\end{document}